\documentclass[12pt]{article}
\usepackage[T1]{fontenc}
\usepackage{amsmath}
\usepackage{amsfonts}
\usepackage{graphicx}
\usepackage[mathscr]{euscript}
\usepackage{rotating}
\usepackage{pdflscape}
\usepackage[hmargin=2.2cm,vmargin=2.75cm]{geometry} 
\usepackage{booktabs,caption}
\usepackage{natbib}
\usepackage{setspace}
\usepackage{bbm}
\usepackage{amsthm}
\usepackage{enumerate}
\usepackage{threeparttable}
\usepackage{verbatim}
\usepackage{caption}
\usepackage{url}
\usepackage{amssymb}
\usepackage{subcaption,array}
\usepackage{adjustbox}
\usepackage[titletoc,title]{appendix}
\usepackage{ulem} 
\usepackage{longtable}
\usepackage{diagbox}

\usepackage[hidelinks,breaklinks]{hyperref}
\usepackage{url}
\hypersetup{colorlinks=true,allcolors=blue}

\usepackage{framed}
\usepackage[dvipsnames]{xcolor}

\theoremstyle{definition}

\begin{document}

\title{The geographic flow of bank funding and access to credit: Branch networks, local synergies and competition\thanks{This research has benefited from the financial support of SSHRC and the National Natural Science Foundation of China. We thank Mark Egan, Carlo Regiani, and Howard Smith for insightful discussions.  Helpful comments were provided by Jason Allen, Heski Bar-Isaac, Dean Corbae, Lu Han, Ali Horta\c{c}su, Jean-Fran\c{c}ois Houde, Jakub Kastl, Nate Miller, Matt Osborne, Nicola Pavanini, Xavier Vives, and participants in
seminars and conferences at the Bank of Canada, Carey Business School, Notre Dame, Princeton, University of Toronto, Vancouver School of Economics, Washington University in St. Louis, \textit{Industrial Organization of the Financial Sector conference} at the
Becker-Friedman Institute, \textit{Jornadas de Economia Industrial}, \textit{CEPR/JIE School and Conference on Applied Industrial Organization}, {\it Drexel-PFED Conference on Credit Markets and the Macroeconomy}, {\it NBER Conference on Financial Market Regulation}, {\it Barcelona Summer Forum}. We are grateful to Paul Lim for his excellent research assistance.}}

\author{Victor Aguirregabiria\footnote{Department of Economics, University of Toronto. 150 St. George Street, Toronto, ON, M5S 3G7, Canada, \href{mailto: victor.aguirregabiria@utoronto.ca}{victor.aguirregabiria@utoronto.ca}.} \\ University of Toronto, CEPR \and Robert Clark\footnote{Department of Economics, Queen's University, Dunning Hall, 94 University Avenue, Kingston, ON, K7L 3N6, Canada, \href{mailto: clarkr@econ.queensu.ca}{clarkr@econ.queensu.ca}.} \\ Queen's University \and Hui Wang \footnote{Guanghua School of Management, Peking University, Beijing, 100871, China, \href{mailto:jackie.wang@gsm.pku.edu.cn}{jackie.wang@gsm.pku.edu.cn}.} \\ Peking University}

\maketitle
\thispagestyle{empty}

\begin{abstract}
Geographic dispersion of depositors, borrowers, and banks may prevent funding from flowing to high loan demand areas, limiting credit access. Using bank-county-year level data, we provide evidence of geographic imbalance of deposits and loans and develop a methodology for investigating the contribution to this imbalance of branch networks, market power, and scope economies. Results are based on a novel measure of imbalance and estimation of a structural model of bank competition that admits interconnections across locations and between deposit and loan markets.  Counterfactual experiments show branch networks and competition contribute importantly to credit flow but benefit more affluent markets.

\vspace{0.4cm}

\vspace{0.4cm}
\end{abstract}

\setcounter{page}{0}
\baselineskip19.5pt

\newpage
\section{Introduction}

An important determinant of credit provision is the availability of
deposits (\citeauthor{Jayaratne_Morgan_2000}, \citeyear{Jayaratne_Morgan_2000}; 
\citeauthor{Ben-David_Palvia_2017}, \citeyear{Ben-David_Palvia_2017}). However, in
any given region, the demand for loans may not always coincide with the
availability of deposits. Geographic frictions, such as asymmetric information and transaction costs, limit the flow of funds across regions such that there can arise substantial geographic heterogeneity in access to credit and possibly even \textit{credit deserts}. In turn, limited credit access can impact entrepreneurship levels, employment, wages, and economic growth (see, for instance, \citeauthor{Gine_Townsend_2004}, \citeyear{Gine_Townsend_2004}).

 
 Wholesale liquidity markets can  improve the flow of funds.  Banks can buy and sell liquidity in the interbank wholesale market, although transaction costs may arise due to bank precautionary motives and liquidity hoarding (\citeauthor{Ashcraft_2011}, \citeyear{Ashcraft_2011};
\citeauthor{Acharya_Merrouche_2012}, \citeyear{Acharya_Merrouche_2012}). Banks may also  use their branch networks to overcome geographic frictions and move liquidity from one region to another, incurring transaction costs that are likely smaller than those from the interbank market (\citeauthor{Coase_1937}, \citeyear{Coase_1937}). However, two counterbalancing forces can negatively affect a bank's willingness to transfer funds between its branches: (i) economies of scope and other synergies between deposits and loans at the branch level and (ii) local market power. Scope economies may arise because clients
 prefer to have their deposit account and mortgage in the same bank or because a bank's cost of managing a deposit account and a loan may be
smaller if they belong to the same client.\footnote{See for instance 
\citeauthor{Kashyap_Rajan_2002} (\citeyear{Kashyap_Rajan_2002}), \citeauthor{Mester_Nakamura_2006} (\citeyear{Mester_Nakamura_2006}), 
\citeauthor{Norden_Weber_2010} (\citeyear{Norden_Weber_2010}), and
\citeauthor{Egan_Lewellen_2017} (\citeyear{Egan_Lewellen_2017}).}  These and other synergies create incentives to concentrate lending activity in
 branches with high levels of deposit, and therefore limit the geographic flow of liquidity to markets more in need of credit.\footnote{Factors other than scope economies can generate synergies between deposits and loans at the branch or local market level. For instance, the Community Reinvestment Act (CRA) incentivizes banks to use local deposits to fund local loans. In this paper, we are not concerned with identifying the specific sources of synergies, either economies of scope or others. We want to study how these synergies affect the geographic imbalances between deposits and loans.} Spatial heterogeneity in local market power also negatively impacts the geographic flow of credit, affecting the extent to which  changes in the marginal cost of loans are passed through to borrowers through loan interest rates.

This paper aims to provide systematic evidence on the extent to which deposits and loans are geographically imbalanced in the US banking industry and to investigate empirically the contribution of branch networks, scope economies, and local market power to this imbalance. 
To perform our analysis, we assemble a dataset from the US banking industry for 1998-2010, merging data at the bank-county-year level from two sources. Deposit and branch-network information is collected from the  \textit{Summary of Deposits} (SOD) provided by the \textit{Federal Deposit Insurance Corporation} (FDIC). Information on lending is from the \textit{Home Mortgage Disclosure Act} (HMDA) data set, which provides information on mortgage loans. 


Our first contribution is to develop an index of the imbalance between deposits and loans that captures the degree to which a depository bank transfers funds away from where deposits are collected (henceforth Imbalance Index or \textit{II}). To do so, we adapt techniques developed in sociology and labor economics to quantify residential segregation. These measures capture the extent to which individuals from different social groups live together or apart within a given geographical area.\footnote{See for instance \cite{Jahn_Schmid_1947},
\cite{Duncan_Duncan_1955}, \cite{Atkinson_1970}, \citeauthor{White_1983} (\citeyear{White_1983}, \citeyear{White_1986}), and \cite{Cutler_Glaeser_1999}. More recently, they have been used by \cite{Gentzkow_Shapiro_2019}  to quantify the degree of polarization in political speech in the US.} Our findings suggest that, while many banks exhibit a strong home
bias with deposit and loan shares being roughly equivalent in each county where they operate, some banks transfer significant funds between geographic locations.  Furthermore, we find evidence that some regions of the country have much larger shares of total deposits than they do of loans and vice versa, implying an important geographic imbalance.

We do not take a stand on whether greater imbalance is good or bad. On the one hand, greater imbalance implies markets are integrated with deposit funding flowing to those where it is most in demand. On the other, it could mean funding is being withdrawn from markets with specific characteristics (poorer, less white) and allocated to markets with different characteristics, such that some populations have limited access to credit. Our objective is to document disparities in credit access and to gain some understanding of the role that branch networks, scope economies, and local market power play in generating these imbalances. 

Investigating the factors contributing to geographic imbalance of deposits and loans requires a model that allows for interconnections across geographic locations and between deposit and loan markets such that local shocks to deposits or loans can endogenously affect the volume of loans and deposits in every local market. The second contribution of this paper is to develop and estimate a structural
model of bank oligopoly competition for \textit{both} deposits and loans in multiple geographic markets, allowing for rich interconnections. We characterize an equilibrium of this multimarket oligopoly model and propose a simple algorithm to solve it. Our approach allows us to perform counterfactual experiments that provide evidence of the effect of branch networks, scope economies, local market power, and various public policies on the geographic diffusion of funds.  While other papers have pointed out the existence of economies of scope between deposits and loans at the bank level, our approach allows us to determine at which geographic level these synergies occur, focusing on the possibility that they may be local. 

Our model is not just a theoretical construct. It is a tool that can help us understand and potentially influence real-world concerns in banking and finance. In this model, differentiated banks sell deposit and loan products in multiple local markets (counties).  The model incorporates four main variables, which may affect the demand and costs of loans and deposits in a local market.  First, the number of branches the bank has in the local market may affect the marginal cost of managing deposits and loans and influence consumer awareness and willingness to pay. Second, the total amount of deposits the bank has at the national level may reduce the bank's risk for liquidity shortage and the need to borrow at interbank wholesale markets. This introduces an important interconnection between local markets in a bank's operation. The third factor is the amount of deposits (loans) the bank has in the local market, which may increase consumer demand for loans (deposits) and reduce the bank's marginal cost of loans (deposits) due to economies of scope in managing deposits and loans. The resulting structure bears a resemblance to models of two-sided markets. Finally, the model includes the fraction of securitized loans to capture the extent to which banks move loans off their balance sheets. Securitization is especially important for non-depository institutions, so our model also allows these {\it shadow banks} to play a role on the loan side of the market.

The structural parameters associated with branch networks, scope economies, and local market power are fundamental for the model's predictions. Estimation must address endogeneity and simultaneity of local and total deposits and
loans. Our identification approach involves controlling for a rich fixed-effects specification of the unobservables, that includes \textit{bank $\times$ county}, \textit{county $\times$ year}, and \textit{bank $\times$ year} fixed effects.
We obtain difference-in-differences transformations of the structural equations and apply instrumental variables / GMM to these transformed equations.  For a bank's local deposits and loans, we use the lagged values of the number of branches, deposits, and loans for a bank in a county as instrumental variables, along with the corresponding values for banks in counties that are neighbors of contiguous counties. For a bank's total deposits, we use the socioeconomic conditions in geographically distant counties where the bank operates branches as instruments. We then use these moment conditions to obtain a GMM estimator of the model's structural parameters. The validity of these instruments requires that unobservables are not serially or spatially correlated after controlling for fixed effects. We present evidence from tests that validate these restrictions.

It is important to point out that datasets containing information on interest rates for loans and deposits for all depository institutions at the county level are not available, and so we estimate the social surplus (i.e., consumers' willingness-to-pay net of banks' marginal costs) for the different deposit and loan products, as well as how
social surplus depends on different variables such as local
bank branches. We show that these primitives can be identified without information on prices of deposits and loans and require imposing weaker conditions than if we were trying to identify demand and marginal costs separately. Furthermore, this approach allows us to assess welfare effects as measured by changes in the social surplus.  Access to deposit and loan interest rate data would be crucial if our objective were to estimate demand and marginal cost separately. However, that is not the purpose of this paper. 

Estimation yields the following results. The number of branches in a county increases the social surplus for deposits and loans, but the effect is substantially stronger for deposits. Securitization has a strong positive impact on the amount of loans. Substantial synergies exist between deposits and loans at the bank and local market levels.  Finally, the effect of a bank's total deposits on the social surplus for loans is positive and significant both economically and statistically, implying that banks' internal liquidity reduces the cost of lending. 

Our structural approach allows us to evaluate the effect of various market features and policies on the Imbalance Index and the value of loans, deposits, and social surplus nationally and for different counties. In a first experiment, we study the contribution of branch networks to the geographic flow of credit by imposing the restriction that banks only have access to locally generated deposits and not those earned throughout their network in each market. Our second experiment evaluates the effects of eliminating local synergies between deposits and loans. A third experiment looks at the impact of removing local market power and obtains the model equilibrium under the condition that, in all local markets, banks act competitively with prices equal to marginal cost. In a fourth experiment, we remove shadow banks as competitors in loan markets and obtain a measure of the contribution of these institutions to the geographic distribution of credit and its evolution over time. We then present two policy experiments. In the first, we evaluate the effect of a regulation prohibiting banks from operating branch networks in multiple states, as was the case before the Riegle-Neal Act of 1994. We implement this counterfactual by dividing every multi-state bank into different independent banks, one for each state. In the second, we study the impact of introducing a deposit tax that approximates the effect of inflation taxing away the real value of nominal deposits.

Our findings suggest that branch networks are essential for spreading liquidity across the regions where a bank operates, especially in smaller, poorer, and more rural areas, significantly increasing bank-level imbalance index scores. On the other hand,  local synergies decrease the imbalance index by generating a home bias and preventing funds from flowing away from where they are generated. Local market power substantially negatively affects the geographic flow of credit. Limited competition in small counties is essential in constraining the amount of credit these counties receive. Shadow banks also play an important role in credit provision, particularly in large and urban markets. In terms of policy experiments, we find that Riegle-Neal affected lending and the flow of funds, but the impact was moderated by the  limitations in cross-border lending activity.  A deposit tax has a modest negative impact on lending activity, felt disproportionately by socially disadvantaged markets.

Our model builds on and extends the literature on structural models of bank competition.\footnote{For an overview of this literature, see \citeauthor{Clark_etal_2021} (\citeyear{Clark_etal_2021}).} Previous work has looked separately at the market's loan or deposit sides. \citeauthor{Corbae_DErasmo_2021} (\citeyear{Corbae_DErasmo_2019}, \citeyear{Corbae_DErasmo_2021}), 
\citeauthor{Benetton_2018} (\citeyear{Benetton_2018}), \citeauthor{Crawford_Pavanini_2018} (\citeyear{Crawford_Pavanini_2018}), and \citeauthor{Benetton_Gavazza_2021} (\citeyear{Benetton_Gavazza_2021}) all focus on the loan side. \citeauthor{Dick_2008} (\citeyear{Dick_2008}), \citeauthor{Ho_Ishii_2011} (\citeyear{Ho_Ishii_2011}), \citeauthor{Honka_Hortacsu_2017} (\citeyear{Honka_Hortacsu_2017}), \citeauthor{Egan_Lewellen_2017} (\citeyear{Egan_Lewellen_2017}), and \citeauthor{Xiao_2020} (\citeyear{Xiao_2020}) estimate differentiated demand
models for bank deposits. \citeauthor{Egan_Hortacsu_2017} (\citeyear{Egan_Hortacsu_2017}) distinguish between insured and uninsured deposits and endogenize bank defaults and
bank runs. \citeauthor{Aguirregabiria_Clark_2016} (\citeyear{Aguirregabiria_Clark_2016}) estimate a model of banks' geographic location of branches and study the
role of geographic risk diversification in the configuration of bank branch networks. \citeauthor{Wang_Whited_2020} (\citeyear{Wang_Whited_2020}) embed simple demand models for both deposits and loans into a corporate finance model to understand the impact of various financial frictions for the transmission of monetary policy. Given their focus, their models of deposits and loans are only at the national level, only for a subset of lenders, and do not allow for synergies between the two sides of the market. Similarly, \citeauthor{Drechsler_Savov_2017} (\citeyear{Drechsler_Savov_2017}) study the role of market power in the transmission of monetary policy using a Dixit-Stiglitz model of monopolistic competition. \cite{Oberfield2024} study the branch-network expansion patterns and show that banks located in markets where deposits are plentiful relative to demand for credit minimize the need for wholesale funding. We extend all of these previous studies by considering an equilibrium model for deposits and loans that allows for interconnections between these markets at the local level and for the effect of a bank's total liquidity on the costs of loans in local markets. This rich connectivity is necessary to answer the specific questions we pose here, but it is also a contribution in its own right. 
 
We are also related to a recent set of papers that take
advantage of the exogenous variation provided by the shale boom to study the extent to which banks use their branch networks to transfer funds from one local market to another (\citeauthor{Gilje_2017}, \citeyear{Gilje_2017}; \citeauthor{Gilje_etal_2016}, \citeyear{Gilje_etal_2016}; \citeauthor{Loutskina_Strahan_2015}, \citeyear{Loutskina_Strahan_2015}; \citeauthor{Petkov_2017}, \citeyear{Petkov_2017}; and \citeauthor{Cortes_Strahan_2017}, \citeyear{Cortes_Strahan_2017}). Our paper complements the empirical findings by \cite{Gilje_etal_2016} in different ways. First, our empirical analysis of the relationship between the geographic location of a bank's branches (deposits) and loans extends to all US local markets (counties). Second, we study the contribution of local market power to the geographic flow of
banks' funds. Third, our approach to identifying the effect of total deposits on local loans exploits more general sources of exogenous variation than those associated with local catastrophic events or discoveries of natural resources. Finally, our structural model allows us to identify the different sources of transaction costs for the flow of funding, and to perform counterfactual experiments to evaluate the effect on credit of reducing these costs.

In the next section, we describe the data and present descriptive evidence on the geographic dispersion of
deposits and loans. In Section \ref{sec:model}, we describe our model, and in Section \ref{sec:estimation}, we explain how we go about estimating it. Section \ref{sec:results} presents our empirical results and Section \ref{sec:cfs} describes our counterfactual experiments. Finally, Section \ref{sec:conclusions} concludes.

\section{Data and descriptive evidence}\label{sec:data}

\subsection{Data sources}

We combine two data sources at the bank-county level. Branch and deposit information is collected from the Summary of Deposit (SOD) data provided by the Federal Deposit Insurance Corporation (FDIC). Information on mortgage loans comes from the Home Mortgage Disclosure Act (HMDA) data set. 

The SOD dataset is updated on June 30th of each year and covers all depository institutions insured by the FDIC, including commercial banks and saving associations. The dataset includes information at the branch level on total
deposits, location, and bank affiliation. Based on the county identifier of
each branch, we can construct a measure of the number of branches and total
deposits for each bank in each county.  We focus on total deposits, including insured and uninsured, as in \cite{Mankart_etal_2018}.\footnote{Deposits are usually assigned based on the account holder's address, branch activity, or where the account was opened (\citeauthor{FDIC_sod_reporting_2020}, \citeyear{FDIC_sod_reporting_2020}). Note that survey evidence from \cite{Kiser_2002} suggests that most individuals open their main checking/deposit account around the time of their first home purchase.}$^,$\footnote{A small proportion of branches in the SOD data set (around 5\% of all branches) have zero recorded deposits. These might be offices in charge of loans or administrative issues. We exclude them in our analysis.}

Under HMDA, most institutions must disclose information on the mortgage loans they originate, refinance, or purchase in a given year.\footnote{There are geographic restrictions on loan reporting. According to the Community Reinvestment Act, large banks must report on all loans regardless of location. Regardless of size, lenders located in an MSA must report on loans originated in an MSA but can choose not to report loans outside MSAs. Only small lenders outside of MSAs do not have to report, but according to the US census, about $83\%$ of the population lived in an MSA  during our sample period. Therefore, HMDA captures most residential mortgage lending activity.} At the level of financial institution, county, and year, we have information on the number and volume of mortgage applications, mortgage loans issued, and mortgage loans subsequently securitized.\footnote{We use the information on securitization to capture differences in banks' cost of lending at the county level. Summary statistics on securitization are reported in Table 1 below.} The type of institutions reporting to HMDA include both depository institutions and non-depository institutions -- mainly Independent Mortgage Companies (IMCs), which are commonly described as \textit{shadow banks}.\footnote{IMCs are for-profit lenders that are neither affiliated nor subsidiaries of banks' holding companies.} 

Our analysis focuses on the depository institutions reporting to HMDA, including banks and Savings Associations, that can be matched with the SOD data.\footnote{We match banks in the SOD and HMDA datasets using their certificate number (provided by the FDIC to every insured depository institution) or/and their RSSD number (assigned by the Federal Reserve to every financial institution). We match thrifts using their docket numbers. We match financial institutions supervised by the OCC through the Call Reports, which allow us to match information from SOD and HMDA.} Focusing on depository institutions is consistent with the research questions addressed in this paper, since they rely heavily on branching and deposits to fund their loans. By contrast, shadow banks rely on wholesale funding and mortgage brokers (\citeauthor{Rosen_2011}, \citeyear{Rosen_2011}). Nevertheless, our structural model of demand and supply of mortgages includes competition from shadow banks. Together, these depository institutions and shadow banks represent the working sample that we use for estimation and counterfactual experiments. They account for $80\%$ of all deposits and $94\%$ of all mortgage loans. Other financial institutions, including the depository institutions that we cannot match with HMDA, are excluded from our analysis because we cannot assign either their deposits or loans (or both) to particular counties. Although large in number, these institutions represent a negligible fraction of lending activity and a small share of deposits. We derive bank-level characteristics from balance sheets and income-statement information in the banks' quarterly reports provided to the different regulatory bodies: the Federal Reserve Board (FRB)'s Report on Condition and Income (Call Reports) for commercial banks and the Office of Thrift Supervision's (OTS) Thrift Financial Report (TFR) for saving associations. Appendix \ref{sec:working_sample} presents a detailed description of the construction of our working sample.

County-level data on socioeconomic characteristics are obtained from various products of the US Census Bureau. Population counts by age, gender, and ethnic group are obtained from the Population Estimates. Median household income at the county level is extracted from the State and County Data Files, whereas income per capita is provided by the Bureau of Economic Analysis (BEA). We also use information on county-level house prices for 2742 counties from the Federal Housing Finance Agency (see \citeauthor{Bogin_etal_2019}, \citeyear{Bogin_etal_2019}), and county-level bankruptcy data from the U.S. Bankruptcy Courts.\footnote{More specifically, we use Table F 5A Business and Nonbusiness Bankruptcy County Cases Commenced, by Chapter of the Bankruptcy Code During the 12-Month Period Ending June 30, 2007.} 

\subsection{Data features}\label{sec:data_features}

Four features of our data and empirical approach deserve specific discussion. First, we have data on mortgage loans at the bank-county-year level but not on other forms of bank credit at the same level of disaggregation. Ideally, we would incorporate information on other types of bank loans, but, to our knowledge, such data are not publicly available at the bank-county-year level.\footnote{Some data on small-business loans are available, but, for most of our sample period, only very large banks (i.e., those with more than \$1 billion in assets) were required to reveal this lending activity.} However, mortgage loans represent the most substantial part of bank loans and even of bank assets. Using bank-level information from the 2010 Call Reports, \cite{Mankart_etal_2018} show that mortgages account for between 62\% and 72\% of all bank loans, and between 38\% and 45\% of total bank assets, depending on bank size (with larger banks typically having smaller shares). They also report that bank deposits represent between 68\% and
85\% of total bank liabilities. These patterns hold in our sample too. Therefore, our focus on deposits and mortgages, though motivated by data availability, captures a substantial fraction of total bank liabilities and assets, respectively.

Second,  our empirical focus is on stocks of deposits and flows of new loans, as opposed to either the stocks of both deposits and loans or only
new deposits and new loans.\footnote{We include refinances in our sample since borrowers can move their mortgage to a new bank when they refinance, so the refinance decision looks very similar to the initial decision to get a mortgage with a lender.}  These are the levels at which most work studying deposit- or loan markets in isolation have considered these series. See for instance \cite{Dick_2008} and \cite{Egan_Hortacsu_2017} for deposits, and \cite{Benetton_2018} and \cite{Buchak_etal_2024} for loans.
The assumption underlying the decision is that consumers
can choose in every period where to put their entire stock of deposits and where to get new loans (or where to refinance their loan). We are justified in making this assumption because switching costs are higher for loans than for deposits. While there are costs involved in moving deposits, they are typically less important than the financial penalties imposed when moving mortgage loans from one financial institution to another. That said, one problem with using stocks is that past deposit inflows,  which make up today's stock of deposits, have already been used to fund prior mortgages. To address this, flows could be constructed from the SOD dataset as the net change in deposit stocks at the bank-county level by first differencing by year. However, this represents a net change -- newly attracted deposits minus withdrawn deposits. This net change can be negative, which makes the deposit share at the county level difficult to calculate. It is also the case that the net change in deposits would underestimate the funds available to banks to create new loans since a fraction of existing loans are repaid (come due) every year. Therefore, we construct an {\it adjusted flow} measure, including net deposit changes along with those deposits freed up today as a fraction of previously funded loans is paid off. The Call Reports provide information for each bank on the fraction of loans coming due each year, and we use this to construct our measure. We have performed our analysis using new loans and adjusted deposit flows. Our findings regarding the evolution of the Imbalance Index are robust to this alternative measure. Details are provided in Appendix \ref{appendix_construction_flows}.\footnote{We have also considered the case of {\it stocks} of deposits and {\it stocks} of loans. We take advantage of the fact that, in addition to the flow of new loans at the bank-county level, we have information on stocks of loans at the bank level (i.e., aggregated across counties). To infer stock at the bank-county level, we assume that the distribution of stocks across counties is the same as the distribution of flows for the same bank. We have performed descriptive analysis using this measure, and the results are essentially unchanged. See Appendix \ref{appendix_construction_flows}.}

Third, publicly available data on interest rates of deposits and loans for all financial institutions are unavailable at the bank-county-year level or even at a more aggregate geographic level. Furthermore, the existing proprietary data on interest rates are not as clean as the quantity data on deposits and mortgage loans that we use, and they are based on geographic interpolations and, therefore, contain potentially important measurement errors. The loan-rate data, particularly, are available only for a small set of lenders. The lack of good price data at the bank-county-year level would be an essential limitation if we wanted to estimate demand and marginal cost separately, but this is not the goal of this paper. To answer all the empirical questions in this paper, we need to estimate the value of consumers' willingness-to-pay net of banks' marginal costs (i.e., the \textit{social surplus}) for the different deposit and loan products, as well as how net willingness-to-pay depends on different variables such as local bank branches and market power. We show that these primitives can be identified without information on prices of deposits and loans and require imposing weaker conditions than if we were trying to identify demand and marginal costs separately.  One might be particularly concerned that without price data, it is impossible to say anything meaningful about market power. Still, we provide evidence that the 
predictions of our structural model are consistent with theoretical implications on market power and with the stylized facts about margins and spreads presented in recent work on market power in the industry. We discuss this in more detail in Sections \ref{sec:model} and \ref{sec:decomposition}.

Finally,  we define our markets as counties, the primary administrative divisions for most states. Markets determine the set of banks competing with each other for
consumer deposits and loans within a geographic area. Although other market definitions, such as State or Metropolitan Statistical Area, have been
employed in some previous empirical studies on the US banking industry, many have considered county as their measure of geographic market (see for
instance \citeauthor{Huang_2008}, \citeyear{Huang_2008};
\citeauthor{Gowrisankaran_Krainer_2011}, \citeyear{Gowrisankaran_Krainer_2011}; \citeauthor{Uetake_Watanabe_2019}, \citeyear{Uetake_Watanabe_2019}).

\subsection{Summary statistics}
Our working sample consists of all matched depository banks along with shadow banks (cases 1, 3, and 7 from Table \ref{tab:types_banks}), in $3,146$ counties during the period 1998-2010. Table \ref{tab:summary_statistics} presents summary statistics from this sample. The top panel provides bank-level statistics based on $61,418$ bank-year observations for depository banks, and $18,552$ bank-year observations for shadow banks. The bottom panel includes county-level statistics for $40,811$ county-year observations. 

The median number of counties where a depository bank obtains deposits from its branches is $2$, while the median number where a bank sells mortgage loans is $8$.\footnote{Table \ref{tab:multistate_banks} in Appendix \ref{sec:appendix_multistate} presents information on multi-state branching and lending.} Banks' branch networks are geographically more concentrated than the networks of
counties where they provide loans. Similarly,  the median number of banks providing deposit services in a county is only $4$, but the median number of banks selling mortgages is $85$. This includes the shadow banks. The median number of counties where shadow banks actively sell loans is 31. The median Herfindahl-Hirschman index (HHI) is $3,450$ for deposit markets and $655$ for loan markets. A possible explanation for the different market structures is that the sunk cost of entry in a local market is higher for deposits than for loans.

\begin{table}[h]
\caption{Summary Statistics \label{tab:summary_statistics}}
\centering
\resizebox{\textwidth}{!}{\begin{tabular}{lccccc}
\hline\hline
\multicolumn{6}{c}{} \\ 
\multicolumn{6}{l}{Panel A: Bank Level Statistics} \\ \hline
\textbf{Variable} & \textbf{Mean} & \textbf{S. D.} & \textbf{Pctile 5} & 
\textbf{Median} & \textbf{Pctile 95} \\ \hline
\multicolumn{6}{l}{\textbf{Depository Banks: 61,418 bank-year observations}} \\ 
\multicolumn{1}{r}{\textit{Number of branches}} & {\small 15.22} & {\small 113.46} & {\small 1} & {\small 4} & {\small 34} \\ 
\multicolumn{1}{r}{\textit{Number of counties with deposits $>$ 0}} & 
{\small 4.01} & {\small 17.80} & {\small 1} & {\small 2} & {\small 10} \\ 
\multicolumn{1}{r}{\textit{Number of counties with new loans $>$ 0}} & 
{\small 29.78} & {\small 149.39} & {\small 1} & {\small 8} & {\small 68} \\ 
\multicolumn{1}{r}{\textit{Total deposits  (in million \$)}} & {\small 985}
& {\small 11,369} & {\small 35} & {\small 150} & {\small 1,720} \\
\multicolumn{1}{r}{\textit{Total new loans  (in million \$)}} & {\small 186}
& {\small 3,149} & {\small 1} & {\small 12} & {\small 253} \\ 
\multicolumn{1}{r}{\textit{Securitization rate of loans (first year) (\%)}} & {\small 20.80} & {\small 31.10} & {\small 0.00} & {\small 0.00} & {\small 88.63} \\ 
&  &  &  &  &  \\ 
\multicolumn{6}{l}{\textbf{Shadow Banks: 18,552 bank-year observations}} \\ 
\multicolumn{1}{r}{\textit{Number of counties with new loans $>$ 0}} & 
{\small 151.52} & {\small 361.40} & {\small 2} & {\small 31} & {\small 813} \\ 
\multicolumn{1}{r}{\textit{Total new loans  (in million \$)}} & {\small 742}
& {\small 5,280} & {\small 1} & {\small 79} & {\small 2088} \\ 
\multicolumn{1}{r}{\textit{Securitization rate of loans (first year) (\%)}} & {\small 76.75} & {\small 38.14} & {\small 0.00} & {\small 98.97} & {\small 100.00} \\ 
&  &  &  &  &  \\ \hline
&  &  &  &  &  \\ 
\multicolumn{6}{l}{Panel B: County Level Statistics (40,811 county-year observations)} \\ \hline
\textbf{Variable} & \textbf{Mean} & \textbf{S. D.} & \textbf{Pctile 5} & 
\textbf{Median} & \textbf{Pctile 95} \\ \hline
\textbf{Banks} &  &  &  &  &  \\ 
\multicolumn{1}{r}{\textit{Number of branches (per county)}} & {\small 22.90}
& {\small 61.85} & {\small 0} & {\small 6} & {\small 101} \\ 
\multicolumn{1}{r}{\textit{Number of banks with branches (per county)}} & {\small 6.04} & {\small 8.24} & {\small 0} & {\small 4} & {\small 20} \\ 
\multicolumn{1}{r}{\textit{Number of banks with new loans (per county)}} & {\small 113.69}
& {\small 99.70} & {\small 13} & {\small 85} & {\small 319} \\ 
\multicolumn{1}{r}{\textit{  "   -- Depository banks}} & {\small 44.82}
& {\small 39.71} & {\small 6} & {\small 34} & {\small 124} \\ 
\multicolumn{1}{r}{\textit{  "   -- Shadow banks}} & {\small 68.87}
& {\small 62.97} & {\small 6} & {\small 50} & {\small 196} \\ 
\multicolumn{1}{r}{\textit{HHI market of deposits}} & {\small 4381} & 
{\small 2899} & {\small 1202} & {\small 3450} & {\small 10000} \\
\multicolumn{1}{r}{\textit{HHI market of new loans}} & {\small 910} & 
{\small 879} & {\small 267} & {\small 655} & {\small 2344} \\ 
\multicolumn{1}{r}{\textit{Deposits per capita (in ,000 \$)}} & {\small 8.67}
& {\small 9.84} & {\small 0.00} & {\small 7.57} & {\small 19.79} \\ 
\multicolumn{1}{r}{\textit{New loans per capita  (in ,000 \$) }} & {\small 3.31} & {\small 4.08} & {\small 0.38} & {\small 2.05} & {\small 10.21} \\ 
\multicolumn{1}{r}{\textit{Securitization rate of loans (first year) (\%)}} & {\small 64.90} & {\small 13.88} & {\small 39.22} & {\small 67.09} & {\small 83.59}
\\ 
&  &  &  &  &  \\ 
\textbf{Demographics} &  &  &  &  &  \\ 
\multicolumn{1}{r}{\textit{Income per capita (in ,000 \$)}} & {\small 27.89}
& {\small 8.09} & {\small 18.09} & {\small 26.56} & {\small 41.74} \\ 
\multicolumn{1}{r}{\textit{Population (in ,000 people)}} & {\small 93.40} & 
{\small 301.24} & {\small 3.04} & {\small 25.24} & {\small 396.38} \\ 
\multicolumn{1}{r}{\textit{Share population }$\leq $ \textit{19 (in \%)}} & 
{\small 27.44} & {\small 3.47} & {\small 22.16} & {\small 27.28} & {\small 33.17}
\\ 
\multicolumn{1}{r}{\textit{Share population }$\geq $ \textit{50 (in \%)}} & 
{\small 33.27} & {\small 6.33} & {\small 23.37} & {\small 33.00} & {\small 44.23}
\\ 
\multicolumn{1}{r}{\textit{Annual change in house price index}} & {\small 3.21} & {\small 7.40} & {\small -7.48} & {\small 3.35} & {\small 13.91} \\ 
\multicolumn{1}{r}{\textit{Number of bankruptcy filings per year}} & {\small %
435} & {\small 1530} & {\small 6} & {\small 106} & {\small 1790} \\ 
&  &  &  &  &  \\ \hline\hline
\end{tabular}}
\end{table}

\clearpage

\begin{figure}[!h]
    \caption{Number of banks and branches per county}\label{fig:nbr_banks_branches}
    \centering
    \subfloat[Depository banks with deposits >0 \\per county]{\includegraphics[width=0.5\linewidth]{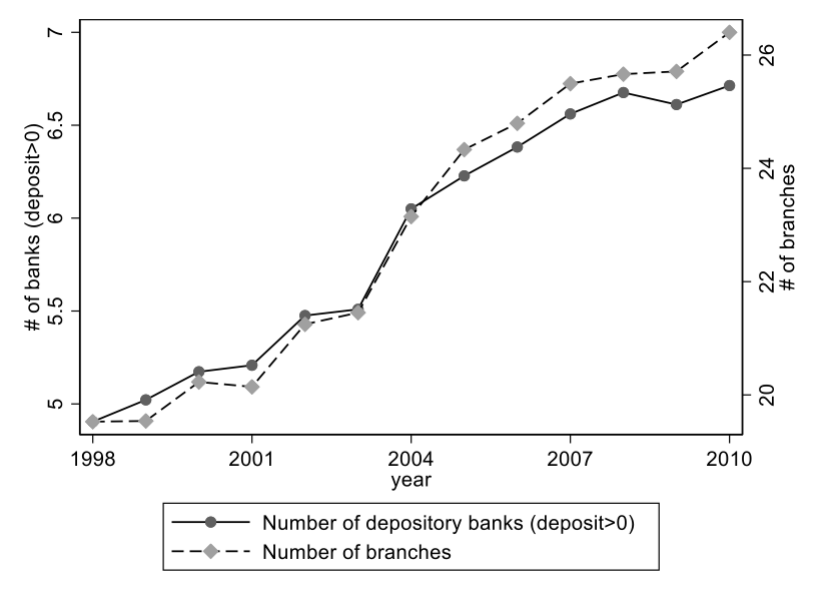}\label{fig:Figure_1_a_May_2023}}
     \centering
    \subfloat[Depository and shadow banks with \\loans >0 per county]
    {\includegraphics[width=0.5\linewidth]{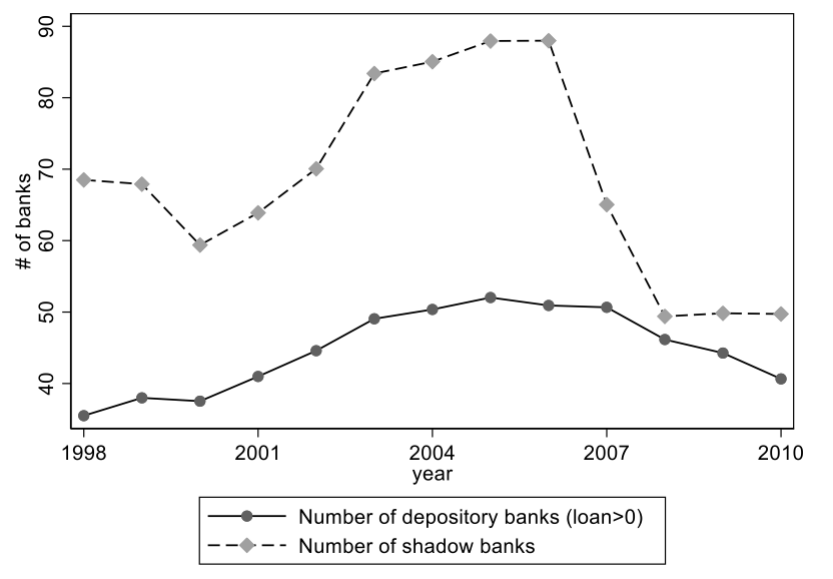}
    \label{fig:Figure_1_b_May_2023}}
\end{figure}

 Figure \ref{fig:Figure_1_a_May_2023} shows the evolution of the number of banks and branches
per county for depository institutions. At the start of our sample period, there were just under five banks and 20 branches per county taking deposits. These numbers increased steadily to nearly 7 and 26, respectively, by 2010. The increase coincides with the
rolling out of Riegle-Neal, which permitted banks to branch across state lines. Over the same period, the percentage of  banks with branches in multiple states
increased from less than 1\% to around 7\%.  Figure \ref{fig:Figure_1_b_May_2023} displays the evolution of the number of depository institutions and shadow banks offering loans per county. The number of depository banks making loans increases steadily until the crisis and then decreases by around 20\%. From 1998 to 2010 the percentage of banks making loans in multiple states increased from 40\% to 55\%. The number of shadow banks increases until the crisis after which it drops sharply.

\subsection{Geographic imbalance of deposits and loans}

In this subsection, we present evidence of the extent to which deposits and loans are geographically imbalanced.  We adapt the measures of residential segregation used in sociology and labor economics to capture the dissimilarity between the geographic distributions of deposits and loans, both for individual banks and for all the banks.

\begin{figure}[!h]
    \caption{Distribution of borrower/lender counties}\label{fig:maps}
    \centering
    \subfloat[]{\includegraphics[width=0.7\linewidth]{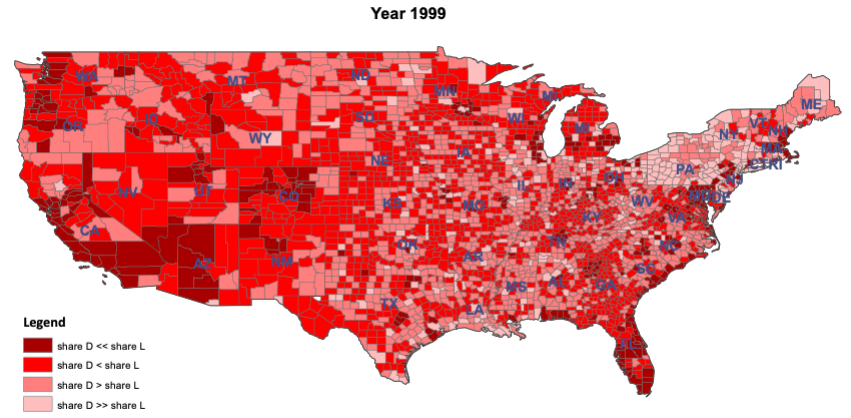}\label{fig:Map_1999}}\\
    \subfloat[]{\includegraphics[width=0.7\linewidth]{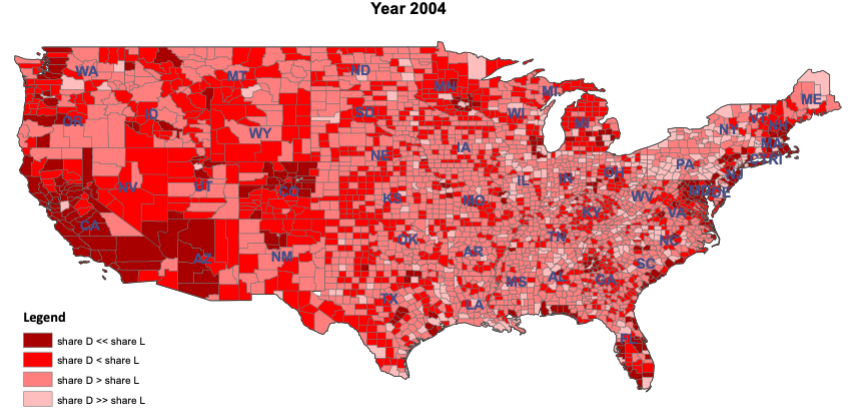}\label{fig:Map_2004}}\\
    \subfloat[]{\includegraphics[width=0.7\linewidth]{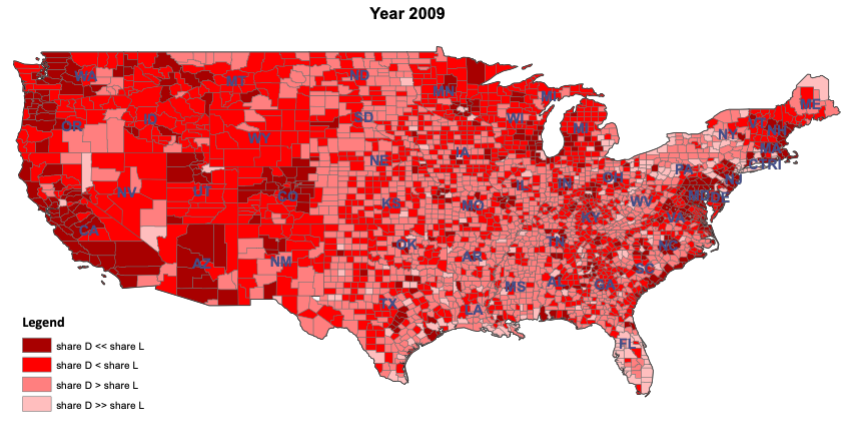}\label{fig:Map_2009}}
\end{figure}

Figure \ref{fig:maps} presents maps with the geographic distribution of counties' positions as net borrowers or net lenders. We present these maps for three
different years: 1999, 2004, and 2009.  We calculate the county's share of deposits for every county-year over aggregate national deposits.
Similarly, we calculate the county's share of new loans over the nation's aggregate amount of new loans. Based on these shares, we construct at
the county level the index $S_{L-D}$ that represents the difference between
the county's share of new mortgage loans and its share of deposits. The values of $S_{L-D}$ provide the geographic distribution of the borrowing and
lending positions of the different counties. By construction, the mean over counties of these indexes equals zero, and there are positive and negative values for net borrowing and net lending counties, respectively.  The 10th, 50th, and 90th percentiles of the distribution of the index $S_{L-D}$ are -23.8\%, -0.5\%, and 35.6\%, respectively,  as percentages of the average deposit share of a county. That is, for counties at the top 10\% (bottom 10\%), the difference between the share
of loans, and the share of deposits exceeds (is below) the national average
by more than 41 (28) percentage points. Using these cutoff points, we sort counties into four groups: (i) counties belonging to the top 10
percentiles of $S_{L-D}$ (Share Loans $>>$ Share Deposits); (ii) counties
between the 10th and 50th percentiles (Share Loans $>$ Share
Deposits); (iii) counties between the 50th and 90th percentiles 
(Share Loans $<$ Share Deposits); and (iv) counties belonging to the bottom
10 percentiles  (Share Loans $<<$ Share Deposits).

Figure \ref{fig:maps} shows clear evidence of deposit and loan imbalances: some regions have high share of deposits, but low share of loans and vice versa.
It also reveals regional patterns in net borrowing/lending positions, the most obvious of which is that counties in the interior tend to be net lenders, while those on the coasts are typically net borrowers. There have also been interesting changes related to the mortgage boom and subsequent financial crisis at the end of the decade. In 1999, several counties in California were in the bottom 10 percentiles of $S_{L-D}$, indicating that their share of deposits
was much larger than their share of total loans. By 2004, almost all counties in the state were in the top 10 percentiles, likely reflecting the build-up
of mortgage debt during the housing boom. Five years later, during the crisis, many counties had flipped again with deposit shares higher than loan shares.

Borrowing from the literature on racial geographic segregation, we consider the following index to capture the {\it imbalance} of deposits and loans for bank $j$:
\begin{equation}
    II_{jt} \text{ } = \text{ }
    \frac{1}{2}
    \sum_{m=1}^{M}
    \left\vert             
        \frac{q_{jmt}^{d}}{Q_{jt}^{d}} -
        \frac{q_{jmt}^{\ell }}{Q_{jt}^{\ell}}
    \right\vert,
\label{individual dissimilarity index}
\end{equation}
where $q_{jmt}^{d}$ and $q_{jmt}^{\ell }$ represent the amount of deposits and loans, respectively,  in county $m$ and year $t$ for bank $j$, and $Q_{jt}^{d}$
and $Q_{jt}^{\ell }$ represent the $j$'s total deposits and loans. This index is a measure of the imbalance of a bank's deposits and loans or a measure of the bank's home bias. An {\it Imbalance Index} (II) score equal to zero represents an extreme case of home bias, i.e., the bank's geographic distributions of loans and deposits are identical. At the other extreme, an II equal to one means that the bank gets all its deposits in markets where it does not provide
loans and sells loans only in markets where it does not have deposits.

\begin{figure}[!h]
\caption{Distribution of Bank-Level II between Deposits and Loans }\label{fig:bank_segregation}
\centering
\begin{minipage}{\linewidth}
\centering
\includegraphics[width=0.8\linewidth]{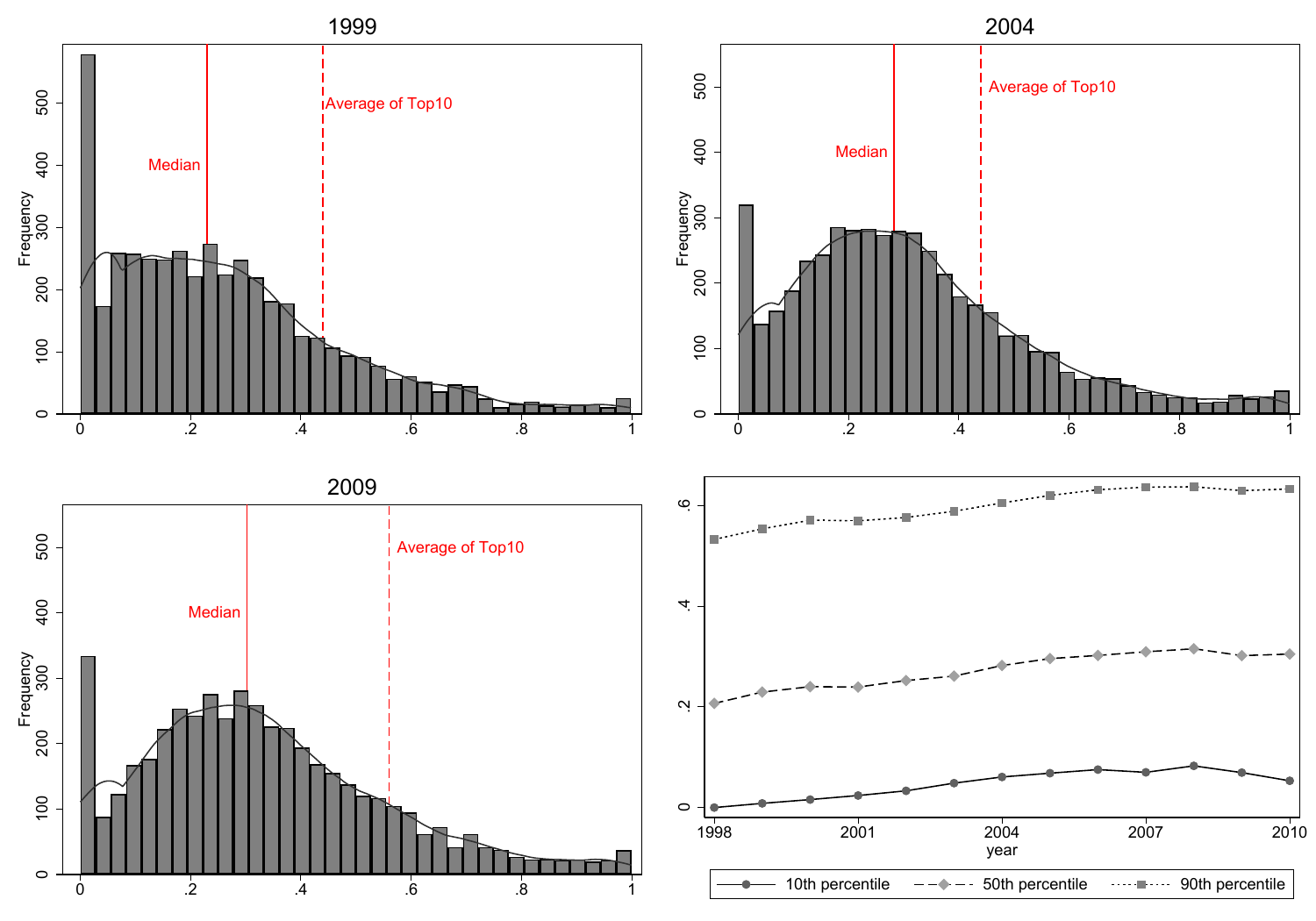}
\footnotetext{Note: The solid red vertical line represents the median of the distribution, while the dashed line shows the mean II for the Top 10 banks as measured by the banks' assets in each year. The bottom right panel shows the evolution of the 10th, 50th, and 90th percentile of the bank-level imbalance index distribution. }
\end{minipage}
\end{figure}

Figure \ref{fig:bank_segregation} presents the empirical distribution of the II calculated at the bank-year level.\footnote{Here we present the II using stocks of deposits and flows of loans. In Appendix \ref{appendix_construction_flows}, we show that the distributions of the bank-level IIs are almost identical when using adjusted deposit flows instead. The correlation between the two  IIs is $0.963$.} We can see that while most banks are involved to some degree in transferring funds across geographic locations, some have a substantial home bias. Each year, there is a mass of banks with a score equal to zero. These are exclusively banks operating in a single county (i.e., taking deposits and making loans in a single county). At the other extreme are banks with very high II scores: the II is greater than $0.5$ for
roughly 20\% of the banks. We can also see a noticeable shift to the right of the distribution over time, suggesting that more deposit funds are being distributed outside the county where they were generated (home county). A Kolmogorov–Smirnov test rejects equality of the distributions in each of the three years.  Note that the distribution shifted rightward between 2004 and 2007 until the Great Financial Crisis started, at which point the trend reversed, and the distribution shifted left. The median bank-level index increased from 0.23 to 0.32 between 1999 and 2010, peaking in 2008. 

\label{sec:imbalance_explanation}

\begin{figure}[!h]
    \caption{Time Series of the National II}\label{fig:national_segregation}
    \centering
    \subfloat[Depository \& shadow banks (Cases 1,3,7)]{\includegraphics[width=0.5\linewidth]{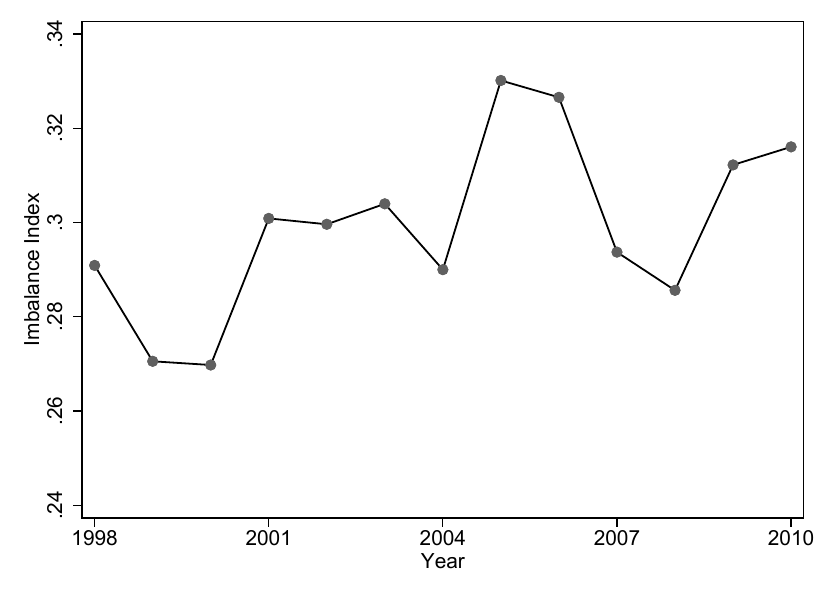}\label{fig:national_segregation_137}}
     \centering
    \subfloat[Depository banks (Cases 1,3)]
    {\includegraphics[width=0.5\linewidth]{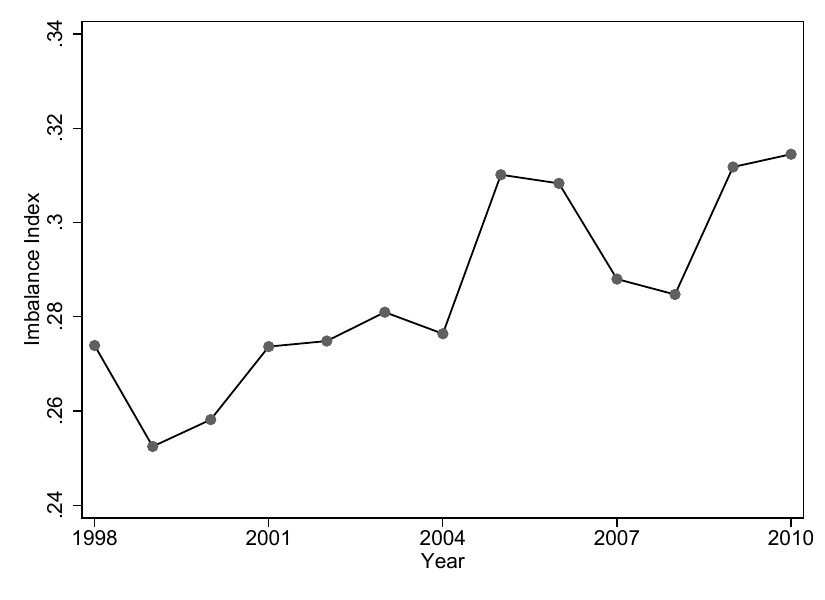}
    \label{fig:national_segregation_13}}
\end{figure}

This increase over time is also noticeable in Figure \ref{fig:national_segregation}, which presents the time series of a national-level II calculated using county-level observations:
\begin{equation}
    II_{t} \text{ } = \text{ }
    \frac{1}{2}
    \sum_{m=1}^{M}
    \left\vert 
        \frac{Q_{mt}^{d}}{Q_{t}^{d}}-
        \frac{Q_{mt}^{\ell }}{Q_{t}^{\ell}}
    \right\vert,
\label{aggregate segregation index}
\end{equation}
where $\frac{Q_{mt}^{d}}{Q_{t}^{d}}$ and $\frac{Q_{mt}^{\ell }}{Q_{t}^{\ell}}$ are the shares of county $m$ in the aggregate national amounts of deposits and new mortgage loans, respectively. It measures the imbalance of funds between geographic locations. Figure \ref{fig:national_segregation_137} presents the national II for the estimation sample (depository banks that can be matched with HMDA and shadow banks). It exhibits a positive trend, although the overall level of variation is not large (between $0.27$ and $0.33$). It peaked in 2005 before the crisis and dropped temporarily during the crisis. Figure \ref{fig:national_segregation_13} restricts attention to matched depository banks. It displays the same general trend, but is shifted down, suggesting shadow banks are important for the diffusion of credit.

\subsubsection{Bank heterogeneity in the II}
\label{sec:imbalance_explanation}

In this subsection, we investigate the effect of bank characteristics on the II.  Specifically, we study the impact of bank size, as measured by number of counties in which they operate total assets, and total deposits (Figures \ref{fig:breakdown_counties}, \ref{fig:breakdown_assets}, and \ref{fig:breakdown_deposits}); geographic location of the bank's headquarters, as measured by urban vs rural (Figure \ref{fig:breakdown_geo}), and by North, South, East or West (Figure  \ref{fig:breakdown_region}); and socioeconomic composition of the bank's headquarter county, as measured by median income (Figure \ref{fig:breakdown_wealth}) and by the percentage of non-whites (Figure  \ref{fig:breakdown_race}). In each case we report Kolmogorow-Smirnov tests for equality of the distributions.

 Whether measured by assets, deposits, or county-network, the size of a bank has a positive effect on its II, with the most significant impact coming from county presence. The rightward shift over time observed in Figure \ref{fig:bank_segregation}  is driven by the fact that the size distribution of banks is changing, with fewer small banks over time. However, even if we restrict attention to the ten largest banks in the US as measured by assets, the same pattern is present, as seen from the dashed vertical line in the figure (and from Table \ref{tab:II_top_10}).  The II scores for these banks are much higher than for the median sample (0.44 vs 0.23 in 1999), and they increased to 0.56 by 2009. For these larger banks, the rise in their II occurs because the increase in the number of counties in which they make loans outpaces the rise in the number of counties where they collect deposits.

 Although initially, there was no significant difference in the IIs of poor vs. rich or rural vs. urban counties, by 2002, banks with headquarters in poor and rural counties began to see their IIs shift rightward and remain that way until the end of our sample. These results suggest that over time, banks headquartered in poorer and more rural counties are sending more of their deposits out of counties where they are raised. In contrast, the racial composition of the headquarters county has almost no effect on the II.

\section{Model\label{sec:model}}

Consider an economy with $M$ geographic markets, indexed by $m \in \mathcal{M}=\{1,2,...,M\}$, and $J$ banks, indexed by $j \in \{1,2,...,J\}$.\footnote{For the sake of notational simplicity, we omit in this section the time subindex $t$.}   Let $\mathcal{M}_{j}^{d}$ represent the set of markets where  bank $j$ has branches and sells deposits. Similarly, $\mathcal{M}_{j}^{\ell }$ represents the set of markets where bank $j$ sells loans. For traditional banks, $\mathcal{M}_{j}^{\ell }$ includes all the markets where the bank has branches, but it may include other markets where the bank has contacts with mortgage brokers that provide clients for the bank. Therefore, $\mathcal{M}_{j}^{\ell}$ includes the set $\mathcal{M}_{j}^{d}$ but it can be larger, i.e., $\mathcal{M}_{j}^{d}\subseteq \mathcal{M}_{j}^{\ell}$. For shadow bank $j$,  $\mathcal{M}_{j}^{d}$ is, by definition, empty.  We take networks $\{\mathcal{M}_{j}^{d}\}_{j=1}^{J}$ and $\{\mathcal{M}_{j}^{\ell}\}_{j=1}^{J}$ as given and focus on the endogenous determination of the amounts of deposits and loans in the equilibrium of this static model of multi-market oligopoly competition.\footnote{One can think of networks as resulting from a dynamic game of market entry-exit decisions with networks. Specifically, this dynamic game has the structure of an Ericson-Pakes model (\citeauthor{Ericson_Pakes_1995}, \citeyear{Ericson_Pakes_1995}). Every year, banks decide their respective deposit and loan networks for the following year (i.e., one-year time-to-build). Banks take as given their pre-determined networks and compete, statically, in prices for deposits and loans. In our earlier work \citeauthor{Aguirregabiria_Clark_2016} (\citeyear{Aguirregabiria_Clark_2016}), we developed a structural model of branch-network choice that employed a revealed preference approach and identified bank preferences towards geographic risk separately from the contribution of the costs of expansion, including economies of density and merger costs. That paper did not examine the expansion in banks' credit-provision networks, which are often different from their branch networks, such that banks may extend their credit-provision networks without expanding their branch networks.}

Each local market is populated by \textit{savers} who demand deposit products and \textit{investors} who demand loan products. Importantly, some savers are also investors and vice versa. These products are horizontally differentiated between banks due to different product characteristics and spatial differentiation within a local market. This view of banks' services as differentiated products is in the spirit of previous papers in the literature.\footnote{See \cite{Degryse_1996}, \cite{Schargrodsky_Sturzenegger_2000}, \citeauthor{Cohen_Mazzeo_2007} (\citeyear{Cohen_Mazzeo_2007} and \citeyear{Cohen_Mazzeo_2010}), \cite{Gowrisankaran_Krainer_2011}, or \cite{Egan_Hortacsu_2017}, among others.} A novel feature of our model, which is key for
our analysis, is that for traditional banks, it introduces endogenous links between deposit and
loan markets and between these markets at different geographic locations. 


For traditional bank $j$, the (variable) profit function  is equal to interest earnings from new loans (pre-existing loans and deposits are considered pre-determined fixed profits), minus payments to depositors, minus costs of managing  deposits and loans, and minus the costs (or returns) from the bank's activity in interbank wholesale markets: 
\begin{equation}
    \Pi_{j} \text{ } = \text{ }
    \sum_{m=1}^{M} p_{jm}^{\ell} \text{ } q_{jm}^{\ell}+
    p_{jm}^{d}\text{ }q_{jm}^{d} - 
    C_{jm}\left( q_{jm}^{\ell}, q_{jm}^{d}\right) -
    \left(r_{0}+c_{j0}\right) B_{j},  
    \label{variable profit function}
\end{equation}
where $p_{jm}^{\ell}$ and $p_{jm}^{d}$ are the prices for loans and deposits, respectively, for bank $j$ in market $m$, and $q_{jm}^{\ell}$ and $q_{jm}^{d}$ are the corresponding amounts (flows) of loans and deposits. Note that, typically, the price for loans will be positive ($p_{jm}^{\ell }>0$) because borrowers pay a positive interest rate to obtain a loan. In contrast, the price of deposits is typically negative ($p_{jm}^{d}<0$) because the bank pays savers to attract their deposits. Market $m=0$ represents the interbank wholesale market; $r_{0}$ is the interbank interest rate; $B_{j}$ is the net borrowing position of bank $j$ at the interbank market; and $c_{j0}$ is a bank-specific transaction cost associated with using the interbank market. The Federal Reserve determines the interbank interest rate, which is exogenous in this model.

The function $C_{jm}\left( q_{jm}^{\ell },q_{jm}^{d}\right) $ represents the cost of managing deposits and loans in market $m$. This cost includes the expected cost of loan default or pre-payment, as well as
the expected cost reduction associated with loan securitization. A bank's resources constraint implies that, $B_{j}=Q_{j}^{\ell }-Q_{j}^{d}$, where $Q_{j}^{\ell }\equiv \sum_{m=1}^{M}q_{jm}^{\ell }$ and $Q_{j}^{d}\equiv \sum_{m=1}^{M}q_{jm}^{d}$ are bank $j$'s total new loans and deposits, respectively.\footnote{We have that $B_{j} = S_{j}^{\ell} - S_{j}^{d} + Q_{j}^{\ell} -Q_{j}^{d}$, where $S_{j}^{\ell}$ and $S_{j}^{d}$ are stocks of live pre-existing loans and deposits, respectively. However, these stocks are pre-determined and do not affect the flow of variable profits.} Solving this restriction in the profit function, we have that $\Pi_{j}=\sum_{m=1}^{M}p_{jm}^{\ell }$ $q_{jm}^{\ell }+$ $p_{jm}^{d}$ $q_{jm}^{d}-$ $\widetilde{C}_{jm}\left( q_{jm}^{\ell },q_{jm}^{d}\right) $, with $\widetilde{C}_{jm}\left( q_{jm}^{\ell},q_{jm}^{d}\right) \equiv $ $C_{jm}\left( q_{jm}^{\ell },q_{jm}^{d}\right) +$ $\left( r_{0}+c_{j0}\right)$ $(q_{jm}^{\ell }-q_{jm}^{d})$. For the rest of the paper, we do not include the term $\left( r_{0}+c_{j0}\right) $ $(q_{jm}^{\ell}-q_{jm}^{d})$
explicitly in the variable cost function, but it should be understood that marginal costs include the component $r_{0}+c_{j0}$ with a positive sign for loans and a negative for deposits.

Given the interest rate in the interbank market, $r_0$, the equilibrium of our model determines the amounts of loans and deposits of every bank in every local market, and it also determines the net position of a bank in the interbank market, since $B_j = Q^{\ell}_j - Q^d_j$. Then, given the net positions of the private banks, the position of the Federal Reserve, represented by $B_0$, is also endogenously determined, such that the interbank market clears; that is, the equilibrium condition $\sum_{j=1}^{J}B_{j}+B_{0}=0$ is satisfied.

The profit function for shadow banks is similar in that they earn interest income collected directly or via servicing rights, but unlike traditional banks, they do not fund mortgages via deposits and do not pay interest as in equation  \eqref{variable profit function}. Rather, these institutions use an originate-to-distribute funding model to finance the loans they issue through securitization. We capture this funding model through our simple specification that includes securitization in $C_{jm}$.\footnote{Similar to the profit function for shadow banks specified in \cite{Buchak_etal_2024}.}

Section \ref{sec:model_demand} describes the demand system for deposits and loans. Section \ref{sec:model_cost} presents our specification of bank variable costs. In section \ref{sec:model_surplus}, we derive the expression for the social surplus implied by demand and cost functions. Section \ref{sec:model_equilibrium} describes the model's equilibrium.

\subsection{Demand for deposit and loan products \label{sec:model_demand}}

\noindent \textit{(a) Demand for deposit products.} There is a population of $H_{m}^{d}$ savers in market $m$. Each saver has a fixed wealth that we normalize to one unit.\footnote{See section \ref{sec:estimation} for a description of our measure of this `unit' and the number of consumers in the market, as well as our approach to deal with possible misspecification of these values.} A saver has to decide whether to deposit her savings unit in a bank and, if so, in which one. Due to transportation costs, savers consider only banks with branches in their local market. In other words, banks can only get deposits in markets with branches.\footnote{\cite{Honka_Hortacsu_2017} demonstrate the importance of branch presence for bank choice.} Banks provide differentiated deposit products. The (indirect) utility for a saver from depositing her wealth in bank $j$ in market $m$ is (omitting the individual-saver subindex in variables $u_{jm}^{d}$ and $\varepsilon _{jm}^{d}$):
\begin{equation}
    u_{jm}^{d} \text{ } = \text{ } 
    \mathbf{x}_{jm}^{d}\text{ }\beta^{d} - 
    \alpha^{d}\text{ } p_{jm}^{d} +
    \xi_{jm}^{d} + \varepsilon_{jm}^{d}.  
\label{depositors utility}
\end{equation}
$\mathbf{x}_{jm}^{d}$ is a vector of characteristics of bank $j$ (other than the deposit interest rate) and market $m$ that are valued by depositors and observable to the researcher, such as the number of branches of bank $j$ in the market, $n_{jm}$. The vector $\beta^{d}$ contains the marginal utilities of the characteristics $\mathbf{x}_{jm}^{d}$. Variable $p_{jm}^{d}$ is the price of deposit services (i.e., consumer fees minus the deposit interest rate), and $\alpha^{d}$ is the marginal utility of income. The term $\xi_{jm}^{d}$ represents other characteristics of bank $j$ in market $m$ that are observable and valuable to savers but unobservable for us as researchers. Variables $\varepsilon_{jm}^{d}$ represent savers' idiosyncratic preferences, and we assume that they are independently and identically distributed across banks with type 1 extreme value distribution. The utility from the outside alternative is
normalized to zero. Let $s_{jm}^{d}\equiv q_{jm}^{d}/H_{m}^{d}$ be the market share of bank $j$ in the market for deposits at location $m$. The model implies that:
\begin{equation}
    s_{jm}^{d} \text{ } = \text{ } 
    \frac{1\left\{m\in \mathcal{M}_{j}^{d}\right\} \text{ }
    \exp\left\{ 
        \mathbf{x}_{jm}^{d}\text{ }\beta^{d} -
        \alpha^{d}\text{ } p_{jm}^{d} +
        \xi_{jm}^{d}
    \right\} }
    {1+\sum\nolimits_{k=1}^{J}1
    \left\{ m \in \mathcal{M}_{k}^{d}\right\} \text{ }
    \exp \left\{ 
        \mathbf{x}_{km}^{d} \text{ }
        \beta^{d} - \alpha^{d}\text{ }p_{km}^{d} +
        \xi_{km}^{d}
    \right\} },
\label{deposits logit supply}
\end{equation}%
where $1\left\{ \text{.}\right\} $ is an indicator function such that $1\left\{ m\in \mathcal{M}_{j}^{d}\right\} $ is a dummy variable that indicates whether bank $j$ has branches in market $m$.

The vector of product characteristics $\mathbf{x}_{jm}^{d}$ includes three elements that are important for the implications of the model: (i) the number of branches, $n_{jm}$; (ii) the fraction of securitized loans, $sec_{jm}$;
(iii) the bank's loans in the local market, $\ln(1 + q_{jm}^{\ell})$; and (iv) the bank's total amount of deposits, $\ln(Q_{j}^{d})$. The number of branches captures the effects of consumer transportation costs and consumer awareness about the bank's presence. By including the bank's amount of loans in demand for deposits (and, as we will show below, the amount of deposits in the demand for loans), we try to capture, in a simple and parsimonious way, not only economies of scope and other synergies in demand for deposits and loans (i.e., one-stop banking) but also the two-sided-market nature of the banking business (see Section 4 of \cite{Vives_2016}).\footnote{To capture scope economies, we could consider a demand model for deposits and loans that endogenizes consumers' decisions to bundle deposits and mortgages in the same bank, as in \cite{Allen_etal_2019}. However, our dataset lacks information on bundling decisions, even in aggregate form. Consequently, our approach simplifies the way we capture this demand complementarity. Note that under scope economies in consumption, aggregating individual consumer decisions leads to the type of relationship between market shares represented in our model. This is our preferred interpretation of the underlying relationship in our model. Additionally, by including market share from the opposite side of the market, our specification can also account for other positive spillover effects between demand for loans and deposits at a market level. These effects could stem from two-sided markets or social interactions among consumers driven by factors such as concerns about financial stability.} The bank's total deposits capture consumers' concerns about the probability of default or bank run. Therefore, we have,
\begin{equation}
    \mathbf{x}_{jm}^{d}\beta^{d} =
    \beta^d_n \text{ } h(n_{jm}) + 
    \beta^d_{rs} \text{ } sec_{jm} +
    \beta_{\ell }^{d} \text{ } 
    \ln(1 + q_{jm}^{\ell}) +
    \beta_{Q}^{d} \text{ } \ln(Q_{j}^{d}).  
    \label{xbeta deposits}
\end{equation}
$h(.)$ is a monotonic function. We use the function $s_{jm}^{d}=d_{jm}(p_{jm}^{d},s_{jm}^{\ell },Q_{j}^{d})$ to represent the demand for deposits, where, for notational convenience, we include explicitly as arguments the endogenous variables $(p_{jm}^{d},s_{jm}^{\ell },Q_{j}^{d})$.

\medskip

\noindent \textit{(b) Demand for loan products.} Each local market is also populated by investors/borrowers. Let $H_{m}^{\ell }$ be the number of new borrowers in market $m$. Each (new) borrower is endowed with an investment project requiring a loan unit.\footnote{In our empirical application, this will be a real estate investment.} A borrower's set of possible choices is not limited to the banks with branches in the market. Some banks sell mortgages in the market but do not have physical branches (recall that $\mathcal{M}_{j}^{d}\subseteq \mathcal{M}_{j}^{\ell }$). However, borrowers may also value the geographic proximity of the bank as represented by the branches of the bank in the local market. Banks provide differentiated loan products. For a borrower located in market $m$, the
(indirect) utility of a loan from bank $j$ is:
\begin{equation}
    u_{jm}^{\ell} = 
    \mathbf{x}_{jm}^{\ell } \text{ } \beta^{\ell} -
    \alpha ^{\ell} \text{ } p_{jm}^{\ell} + 
    \xi_{jm}^{\ell} + \varepsilon_{jm}^{\ell}.
\label{utility borrowers}
\end{equation}
The variables and parameters in this utility function have a similar interpretation as in the utility for deposits presented above. Variable $p_{jm}^{\ell}$ represents a loan's interest rate from bank $j$ in market $m$. We also assume that the variables $\varepsilon_{jm}^{\ell}$
are identically distributed across banks with type 1 extreme value distribution, and the utility from the outside alternative is normalized to zero. Let $s_{jm}^{\ell }\equiv q_{jm}^{\ell }/H_{m}^{\ell}$
be the market share of bank $j$ in the market for loans at location $m$. According to the model, we have:
\begin{equation}
    s_{jm}^{\ell} = 
    \frac{1\left\{ m\in \mathcal{M}_{j}^{\ell }\right\} \text{ }
    \exp \left\{ \mathbf{x}_{jm}^{\ell }
    \text{ } \beta^{\ell} - \alpha^{\ell}
    \text{ } p_{jm}^{\ell} + \xi_{jm}^{\ell}\right\}}
    {1+ \sum\nolimits_{k=1}^{J}1
    \left\{ m\in \mathcal{M}_{k}^{\ell }\right\} 
    \text{ } 
    \exp \left\{ \mathbf{x}_{km}^{\ell}
    \text{ } \beta^{\ell} -\alpha^{\ell}
    \text{ } p_{km}^{\ell} + \xi_{km}^{\ell}\right\}}.
\label{loans logit equation}
\end{equation}

The vector of product characteristics $\mathbf{x}_{jm}^{\ell }$ includes: (i) the number of branches, $n_{jm}$; (ii) the fraction of securitized loans, $sec_{jm}$; (iii) the bank's amount of deposits in the local market, $\ln(1 + q_{jm}^{d})$; and (iv) the bank's total amount of deposits in all the markets, $\ln(Q_{j}^{d})$. As explained above for the demand for deposits, the number of branches captures consumer transportation cost and consumer awareness, and the amount of local deposits portrays synergies in the demand for deposits and loans if using the same bank. Consumers value a bank's total amount of deposits because it is related to the bank's risk of liquidity shortage and failure.\footnote{Borrowers are concerned with bank failure because of the risk the acquiring bank may not renew their loans.} Thus, we have,
\begin{equation}
    \mathbf{x}_{jm}^{\ell} \beta _{m}^{\ell} = 
    \beta^{\ell}_{n} \text{ } h(n_{jm}) +
    \beta^{\ell}_{rs} \text{ } sec_{jm} +
    \beta_{d}^{\ell} \text{ } \ln(1 + q_{jm}^{d}) + 
    \beta_{Q}^{\ell} \text{ } \ln(Q_{j}^{d}),  
    \label{xbeta loans}
\end{equation}
We use the function $s_{jm}^{\ell }=\ell _{jm}(p_{jm}^{\ell},s_{jm}^{d},Q_{j}^{d})$ to represent the demand for loans.

Naturally, there will be many instances where bank $j$'s share of loans in market $m$ is zero, and one might be concerned that these are primarily the result of a ``small sample" problem arising because of a small number of potential customers in a county (see \citeauthor{gandhi_lu_2022}, \citeyear{gandhi_lu_2022} for a discussion). However, this is not the reason for zeroes in our case.  Most bank-county-year observations in our dataset where loans are
zero occur because they are zero in the population. That is, there are many counties where a bank does not make any loans.\footnote{Suppose to the contrary that the observed zeroes were mostly the result of a small
number of potential customers in a county. When a bank has positive  loans  in a county the number of loan customers that the bank serves should be pretty small. This hypothesis is rejected in our data. For the bank-county-year observations with a positive amount of loans at year t and zero loans at year t-1 (5183 observations, 0.37\% of the sample), the sample mean for the number of loans is 112, and the median is 22. Similarly, for the observations with a positive amount of loans at year t and zero
loans at year t+1 (3263 observations, 0.23\% of the sample), the sample mean for the number of loans is 48, and the median is 10.} The observed zeroes in our market shares for loans (or deposits) are driven by banks' market entry decisions.

\medskip

\noindent \textit{(c) Demand system for deposits and loans.} The demand system can be represented by the equations $s_{jm}^{\ell} = \ell
_{jm}(p_{jm}^{\ell },s_{jm}^{d},Q_{j}^{d})$ and $%
s_{jm}^{d}=d_{jm}(p_{jm}^{d},s_{jm}^{\ell },Q_{j}^{d})$. This demand system establishes links between deposits and loans in the same local market and across different geographic markets. These links exist regardless of whether banks' pricing decisions internalize or not these spillover effects.

\subsection{Variable cost function \label{sec:model_cost}}

We consider the following specification for the variable cost function:
\begin{equation}
\begin{array}{ccl}
    \widetilde{C}_{jm}
    \left( q_{jm}^{\ell },q_{jm}^{d} \right) 
    & = & 
    \left( 
        \mathbf{x}_{jm}^{d} \text{ } \gamma^{d} +
        \omega_{jm}^{d}
    \right) \text{ } q_{jm}^{d} + 
    \left(
        \mathbf{x}_{jm}^{\ell } \text{ } 
        \gamma^{\ell} + \omega _{jm}^{\ell}
    \right) \text{ } q_{jm}^{\ell}.
\end{array}
\label{variable cost function}
\end{equation}
Therefore, the marginal costs for deposits and loans are $c_{jm}^{d} \equiv \mathbf{x}_{jm}^{d}$ $\gamma ^{d}+\omega _{jm}^{d}$ and $c_{jm}^{\ell
}\equiv \mathbf{x}_{jm}^{\ell }$ $\gamma ^{\ell }+\omega _{jm}^{\ell }$, respectively. Variables $\omega _{jm}^{\ell }$ and $\omega _{jm}^{d}$ are
unobservable to the researcher. The vector of observable variables $\mathbf{x}_{jm}$ includes the same variables as in the demand equations:
\begin{equation}
\begin{array}{rcl}
    \mathbf{x}_{jm}^{d} \text{ } \gamma^{d} 
    & = & 
    \gamma_{n}^{d} \text{  }h(n_{jm}) +
    \gamma_{sec}^{d} \text{ } sec_{jm} +
    \gamma_{\ell}^{d} \text{ } 
    \ln(1 + q_{jm}^{\ell}) +
    \gamma_{Q}^{d} \text{ } \ln(Q_{j}^{d}), \\ 
    &  &  \\ 
    \mathbf{x}_{jm}^{\ell} \text{ } \gamma^{\ell} 
    & = & 
    \gamma_{n}^{\ell} \text{ } h(n_{jm}) +
    \gamma_{sec}^{\ell} \text{ } sec_{jm} +
    \gamma_{d}^{\ell} \text{ } 
    \ln(1 + q_{jm}^{d}) +
    \gamma_{Q}^{\ell} \text{ } \ln(Q_{j}^{d}).
\end{array}
\label{x variables in costs}
\end{equation}
The terms $\gamma _{n}^{d}$ $h(n_{jm})$ and $\gamma _{n}^{\ell }$ $h(n_{jm})$ portray economies of scale and scope between branches of a bank in the same market. Some costs of providing deposits and loans are shared by multiple branches. The expressions $\gamma_{\ell}^{d}$ $\ln(s_{jm}^{\ell})$ and $\gamma_{d}^{\ell}$ $\ln(s_{jm}^{d})$ elucidate the economies of scope involved in managing deposits at the branch level. The component $\gamma_{Q}^{\ell}$ $\ln(Q_{j}^{d})$ delineates how the marginal cost of loans diminishes as the bank's aggregate deposit volume $Q_{j}^{d}$ increases. 

\subsection{Social surplus \label{sec:model_surplus}}

Let $ss_{jm}^{\ell}$ and $ss_{jm}^{d}$ be the unit 
\textit{social surpluses}, in dollar amounts, of the loan and deposit products of bank $j$ in market $m$. The unit social surplus is the (average) consumer willingness to pay minus the unit cost. Given our specification of demand and costs, we have the following expressions for the unit social surpluses:
\begin{equation}
\begin{array}{ccc}
    ss_{jm}^{\ell} & \equiv  & 
    \dfrac{1}{\alpha^{\ell}} 
    \left( 
        \mathbf{x}_{jm}^{\ell} \beta^{\ell} +
        \xi_{jm}^{\ell}
    \right) -
    \left(
        \mathbf{x}_{jm}^{\ell} \gamma^{\ell} +
        \omega_{jm}^{\ell}
    \right) =
    \dfrac{1}{\alpha^{\ell}} 
    \left(
        \mathbf{x}_{jm}^{\ell} \boldsymbol{\theta}^{\ell} +
        \eta_{jm}^{\ell}
    \right),  \\ 
    &  &  \\ 
    ss_{jm}^{d} & \equiv  & 
    \dfrac{1}{\alpha^{d}} 
    \left(
        \mathbf{x}_{jm}^{d} \beta^{d} +
        \xi_{jm}^{d}
    \right) -
    \left(
        \mathbf{x}_{jm}^{d} \gamma^{d} +
        \omega_{jm}^{d}
    \right) = 
    \dfrac{1}{\alpha^{d}}
    \left(
        \mathbf{x}_{jm}^{d} \boldsymbol{\theta}^{d} +
        \eta_{jm}^{d}
    \right),
\end{array}
\label{social_surpluses}
\end{equation}%
where $\boldsymbol{\theta}^{\ell} \equiv \beta^{\ell} - \alpha^{\ell}$ $\gamma^{\ell}$, $\boldsymbol{\theta}^{d} \equiv \beta^{d} - \alpha ^{d}$ $\gamma ^{d}$, $\eta_{jm}^{\ell } \equiv \xi_{jm}^{\ell}-\alpha^{\ell}$ $\omega_{jm}^{\ell}$, and $\eta_{jm}^{d} \equiv \xi_{jm}^{d} -\alpha^{d}$ $\omega_{jm}^{d}$.

\subsection{Bank competition and equilibrium \label{sec:model_equilibrium}}

A key feature of our model is the rich interactions between
deposit and loan markets and between these markets at different geographic locations. In our model, spillovers exist regardless of banks' pricing decisions; they are inherent to the specifications of market demand and costs described above. Any shock to local deposits changes local loan demand and marginal costs. Moreover, this change in the level of local deposits will affect bank $j$'s total deposits, $Q^{d}_{j}$, which in turn cascades into broader impacts on bank $j$'s operations across all counties in which it operates, influencing both deposits and loans.

For simplicity, we consider here a version of the model with separate Nash-Bertrand competition for deposits and loans at each local market.\footnote{Alternatively, one could also allow banks to internalize the spillover effects between loan and deposit markets, and between geographic markets when making pricing decisions. This generates an incentive to reduce price-cost margins in the two markets and positively affects market shares.  Without price data, the system of equilibrium equations for market shares is the same if banks internalize the spillover effects as if they set prices to maximize local profit in each market. We have also derived the best-response pricing equations when banks internalize local and global spillover effects. The derivation of these equations is available in a previous version of this paper, https://repec.cepr.org/repec/cpr/ceprdp/DP13741.pdf.} Each bank chooses its vectors of interest rates for deposits and loans, $\mathbf{p}_{j} \equiv 
\{p_{jm}^{d}:m\in \mathcal{M}_{j}^{d};$ $p_{jm}^{\ell
}:m\in \mathcal{M}_{j}^{\ell }\}$, to maximize its profit.\footnote{This assumption is supported by the empirical evidence in \cite{Amel_Stahl_2016} who show that banks' prices are tailored to local-market competition. They also find evidence of links in a bank's prices across markets, which is consistent with our network pricing model. However, even in the extreme scenario where a bank applies uniform pricing across all counties it operates in, the described spillover effects persist. See also \cite{Driscoll_Judson_2013} and \cite{HANNAN2006256} regarding non-uniformity of deposit pricing during our sample period, and \cite{Hendricks_etal_2024} for loan pricing.}  The marginal conditions of optimality imply the well-known pricing equations:
\begin{equation}
    p_{jm}^{d}-c_{jm}^{d} = 
    \dfrac{1}{\alpha ^{d}(1-s_{jm}^{d})}, 
    \qquad and \qquad
    p_{jm}^{\ell} - c_{jm}^{\ell } = 
    \dfrac{1}{\alpha ^{\ell }(1-s_{jm}^{\ell})},
\label{pricing equations}
\end{equation}
where $c_{jm}^{d} \equiv \dfrac{\partial \widetilde{C}_{jm}}{\partial q_{jm}^{d}}$ and 
$c_{jm}^{\ell} \equiv \dfrac{\partial \widetilde{C}_{jm}}{\partial q_{jm}^{\ell}}$ represent marginal costs.

For our empirical analysis, it is convenient to write the equilibrium conditions regarding the market shares as the only endogenous variables.
Let $s_{0m}^{d}$ and $s_{0m}^{\ell }$ be the market shares of the outside
alternative for deposits and loans in market $m$. The logit model implies
that $\ln (s_{jm}^{d}/s_{0m}^{d})=$ $\mathbf{x}_{jm}^{d}$ $\beta
_{m}^{d}-\alpha ^{d}$ $p_{jm}^{d}+\xi _{jm}^{d}$ and $\ln (s_{jm}^{\ell}/s_{0m}^{\ell})=$ $\mathbf{x}_{jm}^{\ell}$ $\beta
_{m}^{\ell}-\alpha ^{\ell}$ $p_{jm}^{\ell}+\xi _{jm}^{\ell}$.  Subbing the pricing
equations into these expressions, we obtain the following system of
equilibrium equations in terms of market shares:
\begin{equation}
\begin{array}{rl}
    \ln \left( \dfrac{s_{jm}^{d}}{s_{0m}^{d}}\right) +
    \dfrac{1}{1-s_{jm}^{d}}
    = & 
    \theta_{n}^{d} \text{  } h(n_{jm}) +
    \theta_{rs}^{d} \text{  } sec_{jm} +
    \theta_{\ell}^{d} \text{ } 
    \ln(1 + H_{m}^{\ell} s_{jm}^{\ell}) +
    \theta_{Q}^{d} \text{ } \ln(Q_{j}^{d}) + \eta_{jm}^{d}, 
    \\ &  \\ 
    \ln \left( \dfrac{s_{jm}^{\ell}}{s_{0m}^{\ell}}\right) +
    \dfrac{1}{1-s_{jm}^{\ell}}
    = & 
    \theta_{n}^{\ell} \text{  } h(n_{jm}) +
    \theta_{rs}^{\ell} \text{  } sec_{jm} +
    \theta_{d}^{\ell} \text{ } 
    \ln(1 + H_{m}^{d} s_{jm}^{d}) +
    \theta_{Q}^{\ell} \text{ } \ln(Q_{j}^{d}) + \eta_{jm}^{\ell},
\end{array}
\label{system compact}
\end{equation}
where the vectors of parameters $\boldsymbol{\theta}^{d}$ and $\boldsymbol{\theta}^{\ell}$, and the unobservables $\eta_{jm}^{d}$ and $\eta_{jm}^{\ell}$ are the same ones as in equation \eqref{social_surpluses} for social surpluses, i.e., $\boldsymbol{\theta}^{\ell} \equiv \beta^{\ell} - \alpha^{\ell}$ $\gamma^{\ell}$, $\boldsymbol{\theta}^{d} \equiv \beta^{d} - \alpha ^{d}$ $\gamma ^{d}$, $\eta_{jm}^{\ell} \equiv \xi_{jm}^{\ell}-\alpha^{\ell}$ $\omega_{jm}^{\ell}$, and $\eta_{jm}^{d} \equiv \xi_{jm}^{d} -\alpha^{d}$ $\omega_{jm}^{d}$.\footnote{ One might wonder why $\ln(Q^{\ell})$ is not also included in this system of equations, since it might naturally be thought to affect lending costs. We have considered a specification including $\ln(Q^{\ell})$. 
This has little impact on the main estimates; however, the sign on $\ln(Q^{d})$ becomes negative in the loan equation, when both it and $\ln(Q^{\ell})$ are included simultaneously. This is because  $\ln(Q^{\ell})$ and $\ln(Q^d)$ are highly correlated (correlation coefficient of 0.76). For these reasons, we omit $\ln(Q^{\ell})$ from our main analysis, but we report these results in Table \ref{tab:Ql_gmm_robust}, along with results from a specification in which the collinearity problem is addressed more symmetrically via the inclusion of the sum of $\ln(Q^{\ell})$ and $\ln(Q^{d})$.}

The vector of parameters $\boldsymbol{\theta}$, together with the exogenous variables of the model, contains all the information that we need to construct the equilibrium mapping of the model and obtain an equilibrium. Given this model
structure, we do not need to identify demand and cost parameters separately.
All our empirical results are based on the estimation of these parameters
and the implementation of counterfactual experiments using the equilibrium
mapping.  Furthermore, the indexes $\mathbf{x}_{jm}^{d} \boldsymbol{\theta}^{d} + \eta_{jm}^{d}$ and $\mathbf{x}_{jm}^{\ell} \boldsymbol{\theta}^{\ell} + \eta_{jm}^{\ell}$ are up-to-scale measures of the social surpluses of deposits and loans, respectively, of bank $j$ in market $m$. The identification of $\boldsymbol{\theta}^{d}$ and $\boldsymbol{\theta}^{\ell}$ (that we establish in Section \ref{sec:estimation}) implies the identification up to scale of the social surpluses in deposit and loan markets. Though these up-to-scale surpluses are not measured in dollar amounts, we can still use them to make welfare comparisons, e.g., percentage changes in the social surplus.

The term $1/(1-s_{jm})$ in equation \eqref{system compact} captures market power. Removing this term is equivalent to assuming that price is equal to marginal cost and it has two direct effects on market shares in the local market: first, it reduces the market share of the outside alternative and increases total quantity; second, it reduces the variance of market shares and the degree of market concentration.  In Appendix \ref{app:DSS} we provide evidence that this measure of market power is consistent with the stylized facts about margins and spreads presented in recent work on the banking industry (including by DSS), and in  Section 6.2, we simulate the equilibrium outcome that arises when this term is set to zero. Our findings suggest that markets become more competitive, implying that this term captures market power.

\section{Identification and estimation of the structural model\label{sec:estimation}}

The system of equations of the econometric model is:%
\begin{equation}
\begin{array}{ccc}
    y_{jmt}^{d} 
    & = & 
    \mathbf{z}_{mt}^{\prime} \theta_{0}^{d} +
    \sum\limits_{n=1}^{n_{\max}} \theta_{n}^{d}(n) \text{ }
    \mathbbm{1}(n_{jmt} \geq n) + 
    \theta_{rs}^{d} \text{ } sec_{jmt} + 
    \theta_{\ell}^{d} \text{ } 
    \ln(1 + q_{jm}^{\ell}) + 
    \theta_{Q}^{d} \text{ } \ln(Q_{jt}^{d}) + \eta_{jmt}^{d}, \\ 
    &  &  \\ 
    y_{jmt}^{\ell } 
    & = & 
    \mathbf{z}_{mt}^{\prime} \theta_{0}^{\ell}+
    \sum\limits_{n=1}^{n_{\max}} \theta_{n}^{\ell}(n) \text{ }
    \mathbbm{1}(n_{jmt} \geq n) + 
    \theta_{rs}^{\ell} \text{ } sec_{jmt} + 
    \theta_{d}^{\ell } \text{ } 
    \ln(1 + q_{jm}^{d}) + 
    \theta_{Q}^{\ell} \text{ } \ln(Q_{jt}^{d}) + \eta_{jmt}^{\ell},
\end{array}
\label{structural equations}
\end{equation}%
where $y_{jmt}^{d} \equiv \ln \left(s_{jm}^{d}/s_{0m}^{d}\right) + 1/(1-s_{jm}^{d})$, $y_{jmt}^{\ell} \equiv \ln \left(s_{jm}^{\ell}/s_{0m}^{\ell}\right) + 1/(1-s_{jm}^{\ell})$, and $\mathbbm{1}(n_{jmt} \geq n)$ is the binary variable indicating that the number
of branches $n_{jmt}$ is greater than or equal to $n$, and $\mathbf{z}_{mt}$ is a vector of
market characteristics capturing the relative value of the outside
alternative. More specifically, $\mathbf{z}_{mt}$ captures county-level time-varying exogenous variables, including a housing price
index, bankruptcy cases, income per capita, population, and
age distribution.

\medskip

\noindent \textit{(i)} {\bf Market size and market shares for deposits and loans}. To construct market shares we need first to construct market size variables $H_{mt}^{d}$ and $H_{mt}^{\ell }$. We postulate that market size is proportional to the total population in county $m$ at period $t$, represented by variable $POP_{mt}$: $H_{mt}^{d}=\lambda ^{d}$ $POP_{mt}$ and $H_{mt}^{\ell }=\lambda ^{\ell }$ $POP_{mt}$ where $\lambda^{d}$ and $\lambda^{\ell}$ are positive constants. The values of these coefficients are chosen such that the constructed market shares satisfy the model constraint that the sum of the market shares $\sum_j s_{jmt}^{d}=Q_{mt}^{d}/H_{mt}^{d}$ and $\sum_j s_{jmt}^{\ell }=Q_{mt}^{\ell }/H_{mt}^{\ell }$ are smaller than one for every county-year observation. More specifically, the values of these coefficients are $\lambda ^{d}=\max_{m,t}\left\{ \frac{Q_{mt}^{d}}{POP_{mt}}\right\} $ and \ $\lambda ^{\ell }=\max_{m,t}\left\{ \frac{Q_{mt}^{\ell }}{POP_{mt}}\right\} $, which in our data are $\lambda^{d}=382$ for deposit stocks and $\lambda^{\ell} = 84$ for new loans, measured in thousands of USD. Using $POP_{mt}$ as a measure of market size and assuming that $\lambda^{d}$ and $\lambda^{\ell}$ are constant across counties and over time may seem like strong restrictions. To control for measurement error, we include county $\times$ year fixed effects as explanatory variables in the model. Including these fixed effects makes our empirical results very robust to using alternative measures of market size, such as total county income.\footnote{One might wonder why we do not instead use the number of loan applications as our measure of market size in the mortgage markets. This is because, as explained in \cite{Agarwal_etal_2018}, many prospective borrowers apply multiple times for a loan before ultimately obtaining financing or abandoning their search altogether. According to their data on mortgages from a large government-sponsored entity in the US, the overall median number of applications per person is nine, and the median for those ultimately financed is two. Therefore, although we know the number of applications, since the HMDA data do not allow us to identify individual applicants, we cannot be sure of the number of applicants.}

\medskip

\noindent \textit{(ii)} {\bf Endogeneity.} The structural equations presented in \eqref{structural equations} incorporate exogenous variables $\mathbf{z}_{mt}$, $\mathbbm{1}(n_{jmt} \geq n)$, and $sec_{jmt}$ alongside endogenous variables $q_{jmt}^{\ell}$, $q_{jmt}^{d}$, and $Q_{jt}^{d}$. These endogenous variables are correlated with the error terms $\eta_{jmt}^{d}$ and $\eta_{jmt}^{\ell}$ due to simultaneity. Below, we outline our assumptions aimed at addressing endogeneity.

The identification and estimation of the model are based on four assumptions: (i) a rich fixed effects specification of the unobservables; the assumption that the remaining bank-county-year transitory shocks are (ii) not correlated with the observable exogenous county characteristics, and (iii) not serially correlated; and (iv) the bank-year effects are not correlated with observable exogenous county characteristics. Assumptions ID-1 to ID-4 provide a formal description of our identifying restrictions. For the rest of this section, to avoid repetition, we present the assumptions and moment conditions using the deposit equation, but it is important to note that these same assumptions and moment conditions apply to the loans equation.

\medskip

\noindent \textbf{Assumption ID-1 [Fixed Effects]:} The unobservables $\eta_{jmt}^{d}$ and $\eta_{jmt}^{\ell}$ have the following component structure:
\begin{equation}
    \eta _{jmt}^{d} = \eta _{jm}^{d(1)} +\eta_{t}^{d(2)} + \eta _{mt}^{d(3)} + \eta_{jt}^{d(4)} + \eta _{jmt}^{d(5)}.  \label{component structure of unobservables}
\end{equation}
$\eta_{jm}^{d(1)}$ denotes a bank-county fixed effect, $\eta_{t}^{d(2)}$ is a year fixed effect, $\eta_{mt}^{d(3)}$ represents a county-year fixed effect, $\eta_{jt}^{d(4)}$ indicates a bank-year fixed effect, and $\eta_{jmt}^{d(5)}$ accounts for shocks specific to each bank, county, and year combination. \qquad $\blacksquare $

\medskip

\noindent \textbf{Assumption ID-2:} Regressors $\mathbf{z}_{mt}$, $\mathbbm{1}(n_{jmt} \geq n)$, and $sec_{jmt}$ are strictly exogenous with respect to shocks $\eta_{jmt}^{d(5)}$ and $\eta _{jmt}^{\ell (5)}$. For any pair of markets $(m,m^{\prime})$ and any pair of years $(t,t^{\prime})$, we have that 
$\left( \mathbf{z}_{mt}, \text{ } \mathbbm{1}(n_{jmt} \geq n), \text{ } sec_{jmt} \right)$ are not correlated with $\left( \eta_{jm^{\prime} t^{\prime}}^{d(5)}, \text{ } \eta _{jm^{\prime}t^{\prime}}^{\ell (5)} \right)$. \qquad $\blacksquare$

\medskip

\noindent \textbf{Assumption ID-3: }Bank-county-year shocks $\eta _{jmt}^{d(5)}$ and $\eta _{jmt}^{\ell (5)}$ are not serially or spatially correlated. $\ \blacksquare $

\medskip

\noindent \textbf{Assumption ID-4:} County characteristics in $\mathbf{z}_{mt}$ are exogenous regressors with respect to the bank-year shocks $\eta_{jt}^{d(4)}$ and $\eta_{jt}^{\ell
(4)}$. For any market $m$ and bank $j$, $\mathbb{E}\left( \mathbf{z}_{mt} \text{ } \eta_{jt}^{d(4)} \right) =0$. \qquad $\blacksquare$

\medskip

Consider the following difference-in-difference (DiD) transformation of the structural equations in \eqref{structural equations}. First, a difference between the equations of two banks operating in the same county. This transformation eliminates the national-level effect $\eta_{t}^{d(2)}$ and the county-year effect $\eta_{mt}^{d(3)}$ from the error term. We use the $\sim$ symbol to represent this difference, e.g., $\widetilde{y}_{jmt}^{d}\equiv
y_{jmt}^{d}-y_{j_{m}^{\ast }mt}^{d}$, where $j_{m}^{\ast }$ is a baseline bank active at county $m$ that we select to make this transformation. Second, a time difference between the equations at two consecutive periods. This transformation eliminates the bank-county fixed effect $\eta _{jm}^{d(1)}$ from the error term. We use the symbol $\Delta$ to represent this time difference transformation, e.g., $\Delta \widetilde{y}_{jmt}^{d} \equiv 
\widetilde{y}_{jmt}^{d} - \widetilde{y}_{jm,t-1}^{d}$. The DiD transformation is:
\begin{equation}
\begin{array}{ccc}
    \Delta \widetilde{y}_{jmt}^{d} 
    & = & 
    \Delta \widetilde{\mathbf{x}}_{jmt}^{d}
    \text{ }\theta ^{d}+\Delta \widetilde{\eta}_{jt}^{d(4)} + \Delta \widetilde{\eta}_{jmt}^{d(5)},
\end{array}
\label{transformed DID equations}
\end{equation}
where $\mathbf{x}_{jmt}^{d}$ is the vector of regressors  defined in equation \eqref{xbeta deposits}.  

We can also apply a third difference to eliminate the bank-year component of the error term. Let the $\ast $ symbol represent the difference between two counties where the bank is active, e.g., $y_{jmt}^{\ast d} \equiv y_{jmt}^{d} - y_{jm_{j}^{\ast}t}^{d}$, where $m_{j}^{\ast}$ is a baseline
county in the network of bank $j$. Therefore, we have the
difference-in-difference-in-difference (DiDiD) transformation of the structural equations:
\begin{equation}
\begin{array}{ccc}
    \Delta \widetilde{y}_{jmt}^{\ast d} 
    & = & 
    \Delta \widetilde{\mathbf{x}}_{jmt}^{\ast d}\text{ }\theta ^{d}+\Delta \widetilde{\eta }_{jmt}^{\ast d(5)}.
\end{array}
\label{DiDiD transformed equations}
\end{equation}
Note this DiDiD transformation removes the bank's total deposits, $\ln Q_{jt}$, from the vector of explanatory variables. Therefore, this equation cannot be used to identify parameters $\theta_{Q}^{d}$ and $\theta_{Q}^{\ell}$. However, as shown below, these parameters can be identified from the DiD equation.

Assumptions ID-2, ID-3, and ID-4 imply moment conditions (or valid instrumental variables) in the transformed equations. First, assumptions ID-2 and ID-3 imply the following moment conditions in the DiDiD equations:
\begin{equation}
    \mathbb{E}\left( 
        \left[ 
            \mathbf{z}_{mt}, \text{ }
            \mathbbm{1}(n_{jmt} \geq n), \text{ }
            sec_{jmt}, \text{ }
            \mathbf{x}_{jm,t-s}, \text{ }
            \mathbf{x}_{jm't}
        \right] 
    \text{ } 
    \Delta \widetilde{\eta }_{jmt}^{\ast d(5)} 
    \text{ } \right) = 0, 
\label{DPD moment conditions}
\end{equation}
for any $s\geq 2$, and $m'$ represents any county that is a neighbor of a county contiguous to $m$. These moment conditions identify parameters $\theta_{n}^{d}(n)$, $\theta _{n}^{\ell }(n)$, $\theta _{\ell }^{d}$, and $\theta_{d}^{\ell }$, by combining dynamic panel models or Arellano-Bond moment conditions (\citeauthor{Arellano_Bond_1991}, \citeyear{Arellano_Bond_1991}) -- represented by $\mathbf{x}_{jm,t-s}$ -- with Hausman moment conditions (\cite{hausman_1996}) -- represented by $\mathbf{x}_{jm't}$. 

Importantly, assumption ID-3 of no serial and spatial correlation in $\eta_{jmt}^{d(5)}$ and $\eta_{jmt}^{\ell(5)}$ is testable using the residuals of the estimated equations. In our empirical results, we provide evidence supporting the assumption of no serial correlation in both the loan and deposit equations. As for the spatial correlation, for the loan equation, we reject the hypothesis that shocks are not correlated with those in neighboring counties or in neighbor-of-neighbor counties. Conversely, we cannot reject this latter hypothesis for the deposit equation.

Based on the results of these tests, we consider the following instruments in the DiDiD transformation of the model. In the estimation of the deposit equation, alongside the exogenous regressors, we instrument the endogenous regressor $\ln(1+q_{jmt}^{\ell})$ using $\ln(1+q_{jm,t-2}^{\ell})$, and the logarithm of loans issued by bank $j$ in all counties neighboring $m$'s nearest neighbors. In the estimation of the loan equation using the DiDiD transformation, we instrument the endogenous regressor $\ln(1 + q_{jmt}^{d})$ using $\ln(1+q_{jm,t-2}^{d})$ and the count of $j$'s branches in county $m$ lagged two periods,  and we do not use spatial instruments.

Second, assumptions ID-2 and ID-4 imply the following moment conditions in the DiD equations. For any $(m,m^{\prime },j)$:
\begin{equation}
    \mathbb{E}\left( 
        \mathbf{z}_{m^{\prime}t} \text{ } 
        \left[ 
            \Delta \widetilde{\eta}_{jt}^{d(4)} + 
            \Delta \widetilde{\eta}_{jmt}^{d(5)}
        \right] 
    \right) = 0.
\label{moment conditions second DID}
\end{equation}
These moment conditions identify  $\theta_{Q}^{d}$ and $\theta_{Q}^{\ell }$. Intuitively, they imply that, after controlling for bank $\times$ county and county $\times$ year fixed effects in the error terms, we can use the exogenous socioeconomic characteristics in markets other than $m$ where the bank is active in the deposit market, i.e., $\{\mathbf{z}_{m^{\prime} t}$ for $m^{\prime}\neq m$ with $m^{\prime} \in \mathcal{M}_{jt}^{d}\}$, to instrument the total amount of deposits $\ln Q_{jt}^{d}$. Socioeconomic characteristics in other markets do not have a direct effect on the
structural equation for market $m$, satisfying an exclusion
restriction. Furthermore, the model implies that these characteristics should affect the total volume of bank deposits; therefore, they are relevant instruments. This identification strategy is in the same spirit as the approaches in \cite{Gilje_etal_2016}, \cite{Cortes_Strahan_2017}, and \cite{Nguyen_2019}.\footnote{\cite{Gilje_etal_2016} use shale gas discoveries in a county as exogenous shocks and study how they generate increased lending in counties connected through branch networks.  \cite{Cortes_Strahan_2017} exploit exogenous variation provided by natural disasters, and \cite{Nguyen_2019} uses bank mergers. Our approach uses similar sources of local exogenous variation, but is more general since not limited to dramatic local shocks.}  We can apply this identification approach to every bank-county-year observation as long as the bank's network includes multiple counties and the county has more than one active bank. 

Specifically, when estimating the parameters $\theta_{Q}^{d}$ and $\theta_{Q}^{\ell}$ in the DiD transformation of deposit and loan equations, we instrument the endogenous regressor $\ln(Q^{d}_{jt})$ using the logarithm of the total per capita income from all other counties where bank $j$ operates branches in year $t$.

\medskip

\noindent \textit{(iii)} \textbf{Switching regressions in the loan equation.} Depository institutions face an important decision in loan markets: whether to operate with physical branches and local deposits or opt for a business model without such local infrastructure. Beyond the straightforward impact of branch numbers and deposit amounts on loan demand and costs, the rest of the parameters influencing the social surplus of loans may differ between these business models, e.g., the parameter associated with the securitization rate and the different fixed effects. Moreover, the choice between the two business models can be endogenous. Consequently, the loan equation can be framed as a switching regression model with two regimes as follows:
\begin{equation}
    y^{\ell}_{jmt} = 
    \begin{cases}
        \mathbf{x}^{\ell}_{jmt} \text{ } 
        \boldsymbol{\theta}^{\ell}(1) + 
        \eta^{\ell}_{jmt}(1) & \text{if } N_{jmt} = 1 \\
        \mathbf{x}^{\ell}_{jmt} \text{ } 
        \boldsymbol{\theta}^{\ell}(0) + 
        \eta^{\ell}_{jmt}(0) & \text{if } N_{jmt} = 0,
    \end{cases}
\label{eq:loan_switching_regression}
\end{equation}
\(N_{jmt} \in \{0,1\}\) denotes the binary decision regarding branch presence. $\boldsymbol{\theta}^{\ell}(0)$ and 
$\boldsymbol{\theta}^{\ell}(1)$ are two different vectors of parameters, with the elements of $\boldsymbol{\theta}^{\ell}(0)$ associated with a number of branches and amount of local deposits are constrained to be zero. We estimate the parameters of this model using a two-step control function approach {\it \`{a} la Heckman}, alongside GMM estimation. In the initial step, we estimate a linear probability model, with a branch presence indicator as the dependent variable, to estimate the propensity score. In the next step, we employ GMM estimation for the loan equation, integrating the propensity score as an additional regressor. This yields two sets of GMM estimates, one for each regime in the switching regression model. Identification hinges on an exclusion restriction where some regressors in the selection equation for the propensity score do not appear in the loan equation. These include dummies for branch presence in the preceding year in (i) the same county and (ii) in a neighboring county. The binary choice model incorporates all other exogenous variables and fixed effects from the loan equation as explanatory variables.

\medskip

\noindent \textit{(iv)} {\bf Treatment of shadow banks.} Shadow banks are not depository institutions and so are not included in the estimation of the deposit equation. Furthermore, even if shadow banks operate through multiple establishments, these offices play a very different role than branch networks in traditional banks. The decision-making for shadow banks with regard to lending activities does not involve the dilemma of operating in a county with or without branches. Consequently, for shadow banks, the switching regression model for the loans equation simplifies to a single regime without branches. This loan equation incorporates securitization (their main source of financing) along with the rich set of fixed effects. By integrating shadow banks into our model as `inside players', rather than merely considering them as part of the outside option, we can investigate the potential impacts of various simulated scenarios on these entities.

\begin{table}[htbp]
  \centering
\caption{Estimation of Structural Equation for Deposits \label{tab:estimation_results_deposits}}
    \begin{tabular}{p{15.5em}llll}
    \toprule
    \multicolumn{1}{r}{\textbf{Variable}} & \multicolumn{2}{c}{OLS\newline{}Fixed Effects} & \multicolumn{2}{c}{GMM\newline{}DiD \& DiDiD} \\
    \midrule
    \multicolumn{1}{l}{\textit{Number of branches}} &       &       &       &  \\
    \multicolumn{1}{r}{Second branch} & \multicolumn{1}{r}{0.5263***} & (0.0107) & \multicolumn{1}{r}{0.6730***} & (0.0102) \\
    \multicolumn{1}{r}{Third branch} & \multicolumn{1}{r}{0.2901***} & (0.0086) & \multicolumn{1}{r}{0.3764***} & (0.0081) \\
    \multicolumn{1}{r}{Fourth branch} & \multicolumn{1}{r}{0.2316***} & (0.0090) & \multicolumn{1}{r}{0.2863***} & (0.0084) \\
    \multicolumn{1}{r}{Fifth branch} & \multicolumn{1}{r}{0.2804***} & (0.0109) & \multicolumn{1}{r}{0.3552***} & (0.0105) \\
    \multicolumn{1}{r}{$>$ Fifth} & \multicolumn{1}{r}{0.0294***} & (0.0027) & \multicolumn{1}{r}{0.0315***} & (0.0024) \\
    \multicolumn{1}{r}{\% of loan being resold} & \multicolumn{1}{r}{-0.0568***} & (0.0062) & \multicolumn{1}{r}{-0.0408***} & (0.0148) \\
    \midrule
    \multicolumn{1}{l}{\textit{Econ. of scope and total deposits}} &       &       &       &  \\
    \multicolumn{1}{r}{log own loans in county} & \multicolumn{1}{r}{0.0445***} & (0.0019) & \multicolumn{1}{r}{0.0964***} & (0.0151) \\
    \multicolumn{1}{r}{log own total deposits} & \multicolumn{1}{r}{0.3117***} & (0.0095) & \multicolumn{1}{r}{0.0346***} & (0.0058) \\
    \midrule
    \multicolumn{1}{l}{\textit{Market characteristics}} &       &       &       &  \\
    \multicolumn{1}{r}{log county income} & \multicolumn{1}{r}{0.0507} & (0.0364) &       &  \\
    \multicolumn{1}{r}{log county population} & \multicolumn{1}{r}{-0.3039***} & (0.0546) &       &  \\
    \multicolumn{1}{r}{share population age $\leq$ 19} & \multicolumn{1}{r}{2.2449***} & (0.5778) &       &  \\
    \multicolumn{1}{r}{share population age $\geq$ 50} & \multicolumn{1}{r}{1.6438***} & (0.3618) &       &  \\
    \multicolumn{1}{r}{log housing price index} & \multicolumn{1}{r}{0.2537***} & (0.0212) &       &  \\
    \multicolumn{1}{r}{log number of bankruptcy filings} & \multicolumn{1}{r}{0.0121***} & (0.0044) &       &  \\
    \multicolumn{1}{r}{log number of loan applications} & \multicolumn{1}{r}{-0.0267***} & (0.0080) &       &  \\
    \midrule
    \multicolumn{1}{r}{Bank x County Fixed Effects} & \multicolumn{2}{c}{YES} & \multicolumn{2}{c}{YES (Implicit in DiD)} \\
    \multicolumn{1}{r}{Time Dummies} & \multicolumn{2}{c}{YES} & \multicolumn{2}{c}{NO} \\
    \multicolumn{1}{r}{County x Time Dummies} & \multicolumn{2}{c}{NO} & \multicolumn{2}{c}{YES (Implicit in DiD)} \\
    \multicolumn{1}{r}{Bank x Time Dummies} & \multicolumn{2}{c}{NO} & \multicolumn{2}{c}{YES (Implicit in DiDiD)} \\
    \multicolumn{1}{r}{Number of observations} & \multicolumn{2}{c}{236,498} & \multicolumn{2}{c}{241,911} \\
    \multicolumn{1}{r}{R-square} & \multicolumn{2}{c}{0.9550 } &       &  \\
    \multicolumn{1}{r}{S-H overid test: p-value} &       &       & \multicolumn{2}{c}{0.0000 } \\
    \multicolumn{1}{r}{No serial correlation-m2: p-value} &       &       & \multicolumn{2}{c}{0.2487 } \\
    \multicolumn{1}{r}{No spatial correlation: p-value} &       &       & \multicolumn{2}{c}{0.1769 } \\
    \midrule
    \multicolumn{5}{p{32.915em}}{Note: Sample Period is 1998-2010. Robust standard errors of serial correlation and heteroscedasticity are reported in parentheses. * means p-value < 0.05; ** means p-value < 0.01; *** means p-value < 0.001.} \\
    \end{tabular}%
\end{table}%

\begin{table}[htbp]
\centering
\caption{Estimation of Structural Equation for Loans \label{tab:estimation_results_loans}}
   {\scriptsize \begin{tabular}{p{13em}lllllllll}
    \midrule
   &  \multicolumn{6}{c}{\textbf{Depository Banks}} & \multicolumn{2}{c}{\textbf{Shadow Banks}} \\  \\
    &  \multicolumn{4}{c}{\textbf{With branches}} &   \multicolumn{2}{c}{\textbf{W/o branches}} &  & \\
    & \multicolumn{2}{c}{OLS} & \multicolumn{2}{c}{GMM} & \multicolumn{2}{c}{GMM} & \multicolumn{2}{c}{GMM} \\
    \multicolumn{1}{r}{\textbf{Variable}} & \multicolumn{2}{c}{Fixed Effects} & \multicolumn{2}{c}{DiD \& DiDiD} & \multicolumn{2}{c}{DiD \& DiDiD} & \multicolumn{2}{c}{DiDiD } \\
    \midrule
    \multicolumn{1}{l}{\textit{Number of branches}}  & \multicolumn{8}{c}{} \\
    \multicolumn{1}{r}{Second branch} & \multicolumn{1}{r}{0.130$^{***}$} & (0.016) & \multicolumn{1}{r}{0.1338$^{***}$} & (0.0240) &   &  &  &  \\
     \multicolumn{1}{r}{Third branch} & \multicolumn{1}{r}{0.088$^{***}$} & (0.016) & \multicolumn{1}{r}{0.0826$^{***}$} & (0.0165) &   &    & & \\
    \multicolumn{1}{r}{Fourth branch} & \multicolumn{1}{r}{0.087$^{***}$} & (0.017) & \multicolumn{1}{r}{0.0828$^{***}$} & (0.0155) &   &    & & \\
    \multicolumn{1}{r}{Fifth branch} & \multicolumn{1}{r}{0.138$^{***}$} & (0.020) & \multicolumn{1}{r}{0.1084$^{***}$} & (0.0181) &  &   &  &  \\
    \multicolumn{1}{r}{$>$ Fifth} & \multicolumn{1}{r}{0.020$^{***}$} & (0.002) & \multicolumn{1}{r}{0.0101$^{***}$} & (0.0020) &  &  &   &  \\
    \midrule
    \multicolumn{1}{l}{\textit{Securitization}}
    & \multicolumn{8}{c}{} \\
    \multicolumn{1}{r}{\% of loans resold} & \multicolumn{1}{r}{1.016$^{***}$} & (0.018) & \multicolumn{1}{r}{0.6632$^{***}$} & (0.0306) & \multicolumn{1}{r}{0.6978$^{***}$} & (0.0057) & \multicolumn{1}{r}{0.0613$^{***}$}  & (0.0036) \\
    \midrule
    \multicolumn{1}{l}{\textit{Econ of scope and $Q^d$}} &       &       &       &       &       & & & \\
    \multicolumn{1}{r}{log own local deposits} & \multicolumn{1}{r}{0.190$^{***}$} & (0.009) & \multicolumn{1}{r}{0.2963$^{***}$} & (0.0334) &  &  &   &  \\
    \multicolumn{1}{r}{log own total deposits} & \multicolumn{1}{r}{0.291$^{***}$} & (0.015) & \multicolumn{1}{r}{0.1727$^{***}$} & (0.0157) & 
     \multicolumn{1}{r}{0.4111$^{***}$} & (0.0104) & & \\
    \midrule
    \multicolumn{1}{l}{\textit{Market characteristics}} 
    & \multicolumn{8}{c}{} \\
    \multicolumn{1}{r}{log county income} & \multicolumn{1}{r}{0.319$^{***}$} & (0.071) &       &   &    &    &   &  \\
    \multicolumn{1}{r}{log county population} & \multicolumn{1}{r}{-1.222$^{***}$} & (0.096) &       &   &    &    &   &  \\
    \multicolumn{1}{r}{share pop. age $\leq$ \$ 19} & \multicolumn{1}{r}{-3.592$^{***}$} & (0.929) &       &       &   &    & & \\
    \multicolumn{1}{r}{share pop. age $\geq$ 50} & \multicolumn{1}{l}{-0.734$^{}$} & (0.667) &    &   &    &   &       &  \\
    \multicolumn{1}{r}{log house price index} & \multicolumn{1}{r}{0.347$^{***}$} & (0.042) &       &   &    &    &   &  \\
    \multicolumn{1}{r}{log nbr bankruptcy } & \multicolumn{1}{r}{-0.050$^{***}$} & (0.010) &       &  &     &   &    &  \\
   \multicolumn{1}{r}{log nbr loan applications} & \multicolumn{1}{r}{0.670$^{***}$} & (0.017) &    &   &       &   &    &  \\
   \midrule
    \multicolumn{1}{l}{\textit{Selection -- Control function}} 
    & \multicolumn{8}{c}{} \\
    \multicolumn{1}{r}{Propensity score} & \multicolumn{1}{r}{0.053} & (0.374) & \multicolumn{1}{r}{-0.6455$^{*}$} & (0.3409) & \multicolumn{1}{r}{2.2301$^{***}$} & (0.0563) & &\\
    \multicolumn{1}{r}{Propensity score square} & \multicolumn{1}{r}{-0.047} & (0.364) & \multicolumn{1}{r}{0.9942$^{***}$} & (0.3392) & \multicolumn{1}{r}{-5.8633$^{***}$} & (0.5022)& &\\
    \multicolumn{1}{r}{Propensity score cubic} & \multicolumn{1}{r}{-0.041} & (0.116) & \multicolumn{1}{r}{-0.3515$^{***}$} & (0.1052) & \multicolumn{1}{r}{7.9259$^{***}$} & (0.8210)& &\\
    \midrule
    \multicolumn{1}{l}{\textit{Fixed effects}} 
    & \multicolumn{8}{c}{} \\
    \multicolumn{1}{r}{Bank x County} & \multicolumn{2}{c}{YES} & \multicolumn{2}{c}{YES} & \multicolumn{2}{c}{YES} & \multicolumn{2}{c}{YES} \\
    \multicolumn{1}{r}{Time} & \multicolumn{2}{c}{YES} & \multicolumn{2}{c}{NO } & \multicolumn{2}{c}{NO}&\multicolumn{2}{c}{NO} \\
    \multicolumn{1}{r}{County x Time} & \multicolumn{2}{c}{NO} & \multicolumn{2}{c}{YES } & \multicolumn{2}{c}{YES}&\multicolumn{2}{c}{YES} \\
    \multicolumn{1}{r}{Bank x Time} & \multicolumn{2}{c}{NO} & \multicolumn{2}{c}{YES } & \multicolumn{2}{c}{YES}&\multicolumn{2}{c}{YES} \\
    \midrule
    \multicolumn{1}{l}{Number of observations} & \multicolumn{2}{c}{194,267} & \multicolumn{2}{c}{196,090} & \multicolumn{2}{c}{797,927}&\multicolumn{2}{c}{2,351,846} \\
    \multicolumn{1}{l}{R-square} & \multicolumn{2}{c}{0.872} &       &       & \multicolumn{2}{c}{}& &\\
    \multicolumn{1}{l}{Joint sign. Selection: pvalue} & \multicolumn{2}{c}{0.1008 } & \multicolumn{2}{c}{0.0000 } & \multicolumn{2}{c}{0.0000 } &&\\
    \multicolumn{1}{l}{S-H overid test: pvalue} &   
    \multicolumn{2}{c}{} & 
    \multicolumn{2}{c}{0.0360 } &  \multicolumn{2}{c}{}  & \multicolumn{2}{c}{} \\
    \multicolumn{1}{l}{No serial correlation-m2: pvalue} &
    \multicolumn{2}{c}{} & \multicolumn{2}{c}{0.2659 } &
    \multicolumn{2}{c}{} & \multicolumn{2}{c}{} \\
    \multicolumn{1}{l}{No spatial corrrelation: pvalue} &
    \multicolumn{2}{c}{} & \multicolumn{2}{c}{0.0000 } &
    \multicolumn{2}{c}{} & \multicolumn{2}{c}{} \\
    \midrule

    \end{tabular}}
\begin{minipage}{0.90\textwidth}
    \footnotesize{Note: Sample Period is 1998-2010.  Robust standard errors of serial correlation and heteroscedasticity are reported in parentheses. $^{*}$ means p-value $<$ 0.05; $^{**}$ means p-value $<$ 0.01; $^{***}$ means p-value $<$ 0.001.}
\end{minipage}
\end{table}

\section{Estimation results\label{sec:results}}

Tables \ref{tab:estimation_results_deposits} and \ref{tab:estimation_results_loans} present estimation results of the structural equation for deposits and loans, respectively. As mentioned above, the right-hand side of the equilibrium equations expressed in equation \eqref{structural equations} represents social surplus. This surplus and the $\theta$ parameters are not measured in monetary units (dollars) but in utils. However, the dependent variables in these regressions are close to the logarithm of market shares. This enables us to effectively compare the parameters of the two equations, interpreting them as elasticities (provided the explanatory variable is also in logarithmic form) or semi-elasticities.

We report OLS Fixed-Effects (without instrumenting) and GMM (DiD and DiDiD) estimates. Compared to GMM, OLS underestimates the effect of the number of branches and the magnitude of local spillovers between deposits and loans. The main difference between the two sets of estimates is in the effect of total deposits on local loans and deposits. Statistical tests support the validity of our moment conditions/instruments.\footnote{Our GMM estimates successfully pass the Arellano-Bond (m2) serial correlation test, supporting the validity of the dynamic instruments. We also show in Appendix \ref{app:additional_estimation_results} that our results remain economically consistent when using different instruments sets, although the chosen specifications do not pass over-identifying restriction tests.} For the rest of the paper, we focus on the GMM estimates.\footnote{County characteristic effects can be estimated by OLS, with all coefficients of expected sign.}

Looking first at Table  \ref{tab:estimation_results_deposits} we see that the number of branches a bank has in a county has a substantial effect on the social surplus for a deposit product. The marginal effect of an additional branch declines with the number of branches: a second branch increases the social surplus by $67\%$; a third branch by $38\%$; a fourth branch by $29\%$; a fifth branch $36\%$;
and subsequent branches by (on average) $3\%$.  Turning to our main parameters of interest, we identify moderate economies of scope between deposits and loans: the elasticity of deposits with respect to loans is $0.10$. Finally, a bank's total amount of deposits at the national level has a small effect on its social surplus at every local market where it operates.

Loan equation results are presented in Table \ref{tab:estimation_results_loans}. As discussed in Section \ref{sec:estimation} estimation of the loan equation involves overcoming a selection problem  by using a  control function approach a la Heckman. In the first step, both the number of branches in the last period and the number of branches in the neighboring counties in the last period,  are strong predictors of having a branch in the local county in the current period. Their coefficients are 0.012 and 0.003, with standard errors being 0.0013 and 0.00021, respectively. We also control for all the other exogenous variables and fixed effects used in the loan equation. The goodness-of-fit in this first step is very high with an R-squared of 0.94.  The control function consists of a third order polynomial in the propensity score, which is the fitted value obtained from the first step. The estimates of the control function coefficients are significant, both individually and jointly. Columns (2) and (3) present estimates from observations where depository institutions make loans and receive deposits. Columns (4) and (5) present results for observations where depository institutions make loans but do not receive deposits, while columns (5) and (6) presents results for shadow banks.

The effect of the number of branches on the social surplus of a loan product (semi-elasticity) is important but smaller than for deposits: a second branch increases social surplus by $13\%$; a third by $8\%$; a fourth by $8\%$;  a fifth by $11\%$; and subsequent branches by $1\%$.\footnote{The estimated parameters for the impact of the fifth branch appear anomalous as they disrupt the expected monotonicity in the regression function with respect to the number of branches. This apparent deviation from monotonicity arises from the linear specification for branches exceeding a certain threshold, denoted as $n_{max}$. It is worth noting that observations with more than 5 branches are scarce.} The positive and significant coefficient associated with securitization underscores the pivotal role of this funding mechanism in facilitating lending activities for traditional banks. This impact is large for these depository institutions, regardless of whether they have branches in the county or not.

We find that doubling a bank's deposits within a county implies a $30\%$ increase in the social surplus of the bank’s loans. Furthermore, a bank’s total amount of deposits at the national level substantially affects the social surplus in all local markets where it operates. Specifically, a $100\%$ surge in a bank's nationwide deposits leads to an increase in the social surplus of loans of $17\%$ if the bank has local branches and of $41\%$ if it does not. This asymmetric effect appears intuitive since, in the absence of local deposits generated by the presence of local branches, a bank must heavily rely on the liquidity generated in other parts of its branch network to finance its local loans. These findings support the notion that a bank's internal liquidity is pivotal in facilitating lending.

It is noteworthy that the elasticity of the social surplus of loans in relation to the securitization rate differs significantly between shadow banks and depository banks. The elasticity stands at $0.06$ for shadow banks, contrasting with $0.66$ and $0.70$ for depository banks. The substantial gap between the average securitization rates of traditional banks ($21\%$) and shadow banks ($77\%$) is likely a significant factor in explaining the difference in the marginal effects of securitization between these two groups of banks. In particular, there should be a considerable difference in the risk levels of the marginal securitized loans between these two types of institutions.

\section{Counterfactual experiments \label{sec:cfs}}

In this section we use the estimates from our model to learn more about the factors that influence the geographic flow of funds and the provision of credit, and to perform a number of policy-related experiments. We split our analysis in two parts. In Section \ref{sec:decomposition} we present a decomposition exercise in which we alternatively shut down (i) liquidity flows from branch networks, (ii) local spillovers, (iii) local market power, (iv) and shadow banks. Results are presented in Table \ref{tab:decomposition}. Then, in Section \ref{sec:policy_CFs} we first evaluate the impact of the Riegle-Neal Act on the flow of funds. Second, we study the consequences from the introduction of a deposit tax that approximates the impact of inflation taxing away the real value of nominal deposits. Results are presented in Table \ref{tab:policy_CFs}. Additional results related to these experiments are presented in Appendix \ref{sec:CF_results_app}.

For all the experiments, we use the GMM estimates for the structural parameters $\boldsymbol{\theta}$, obtain the model residuals, and then apply OLS to estimate the five different groups of fixed effects in the error terms $\eta _{jmt}^{d}$
and $\eta _{jmt}^{\ell }$. With the exception of the branch-network experiments, in conducting our analysis, we maintain the number of branches of a bank within each county at the observed value in the dataset. It seems intuitive that branch networks will evolve over time in response to the changes under consideration. Therefore, these experiments should be viewed as short-run responses to unanticipated changes in policy.

The system of equilibrium equations in \eqref{system compact} can be succinctly expressed as $\Psi(\mathbf{s}) = 0$, with $\mathbf{s}$ denoting the vector with the market shares of deposits and loans across all bank-counties within the networks of banks, namely $\mathcal{M}_{j}^{d}$ and $\mathcal{M}_{j}^{\ell}$. A solution to this system is an equilibrium of the model. By Brower's Theorem, an equilibrium exists, as $\Psi(.)$ is a continuous mapping, and the space of $\mathbf{s}$ is compact. To deal with multiplicity of equilibria in implementing counterfactual experiments, we apply a simple algorithm. In the first step, we compute the unique equilibrium for the vector of loan shares ($s_{jm}^{\ell}$) in every county  taking as given $Q_{j}^{d}$  and the $s_{jm}^{d}$  for every $j$. We then compute the unique equilibrium of deposit shares ($s_{jm}^{d}$) taking as given $Q_{j}^{d}$  and the $s_{jm}^{\ell}$  for every $j$  computed in the first step, and then we aggregate over counties to get $Q_{j}^{d} \text{ for every } j$. Finally, we repeat each of these steps such that each endogenous variable has been updated twice.\footnote{We present results for two rounds of updating, but our results are similar if we update until $Q_{j}^{d}$ converges.} A nice feature of this approach is that the local equilibrium at each step is unique and can be easily computed using a Bisection algorithm.\footnote{Note that we can also use Newton's method to solve for equilibrium.}  We provide a detailed description of the algorithm in Appendix \ref{appendix_equilibrium_algorithm}.

We measure the effects of these counterfactual experiments by looking at the following statistics or outcome variables: (i) National imbalance index,\footnote{In Tables \ref{tab:decomposition} and \ref{tab:policy_CFs} we present the mean national imbalance index over the 13 years of our sample. In Figure \ref{fig:II_DF_vs_DS} of Appendix \ref{sec:CF_results_app}, we present the full evolution across the years.}  (ii) Median bank-level imbalance index (for depository institutions), (iii) Value of loans at the national level, (iv) Total social surplus in loan markets, (v) Value of loans broken down by depository and shadow banks, (vi) Value of loans across the top 100 counties and across the bottom 2500 counties ranked by loan amounts / income per capita / \% urban population, (vii) Value of deposits at the national level, and (viii) Total social surplus in deposit markets.

The statistics that we present here  capture a key trade-off in the geographic distribution of credit. A higher imbalance index implies that a larger share of bank funds is moved across counties such that credit can be used in those locations with more social surplus for loans. However, this movement of bank funds can generate not only winners, but also losers. Some counties may end up with very limited amounts of credit.

\begin{table}[h!]
  \centering
  \caption{Decomposition Experiments} \label{tab:decomposition}
    \begin{tabular}{llllll}
    \toprule
    \toprule
          & Data  & Exp. 1 & Exp. 2 & Exp. 3 & Exp. 4 \\
    Outcomes (13 yr avr) &       & \multicolumn{1}{p{6.165em}}{No branch \newline{}network} & \multicolumn{1}{p{6.165em}}{No EOS} & \multicolumn{1}{p{6.335em}}{No market \newline{}power} & \multicolumn{1}{p{5.54em}}{No \newline{}Shadowbanks} \\
    \midrule
    \textbf{Imbalanced index} &       &       &       &       &  \\
    National level & 0.2984 & 0.3123(4.7\%) & 0.3475(16.5\%) & 0.2489(-16.6\%) & 0.2926(-1.9\%) \\
    Bank level (Median) & 0.2701 & 0.0364(-86.5\%) & 0.8397(210.9\%) & 0.2606(-3.5\%) & 0.2709(0.3\%) \\
          & \textcolor[rgb]{ .459,  .443,  .443}{} & \textcolor[rgb]{ .459,  .443,  .443}{} & \textcolor[rgb]{ .459,  .443,  .443}{} & \textcolor[rgb]{ .459,  .443,  .443}{} & \textcolor[rgb]{ .459,  .443,  .443}{} \\
    \textbf{Loans (in B\$)} &       &       &       &       &  \\
    Tot. value & 1935  & 1400(-27.6\%) & 1366(-29.4\%) & 5134(165.3\%) & 944(-51.2\%) \\
    Tot. social surplus & 2083  & 1482(-28.9\%) & 1444(-30.7\%) & 6165(196.0\%) & 979(-53.0\%) \\
    By bank types &       &       &       &       &  \\
        \quad Depository banks & 876   & 299(-65.9\%) & 263(-70.0\%) & 2896(230.6\%) & 944(7.8\%) \\
        \quad Shadow banks & 1058  & 1100(4.0\%) & 1103(4.3\%) & 2239(111.6\%) & 0(-100.0\%) \\
    By county loans &       &       &       &       &    \\
        \quad Top 100 & 1103  & 827(-25.0\%) & 789(-28.5\%) & 2755(149.8\%) & 528(-52.1\%) \\
        \quad Bottom 2500 & 165   & 104(-37.0\%) & 107(-35.2\%) & 524(217.6\%) & 92(-44.2\%) \\
    By county income/cap &       &       &       &       &    \\
        \quad Top 100 & 539   & 393(-27.1\%) & 382(-29.1\%) & 1295(140.3\%) & 278(-48.4\%) \\
        \quad Bottom 2500 & 454   & 313(-31.1\%) & 317(-30.2\%) & 1328(192.5\%) & 220(-51.5\%) \\
    By county urban\% &       &       &       &       &  \\
        \quad Top 100 & 765   & 575(-24.8\%) & 539(-29.5\%) & 1949(154.8\%) & 369(-51.8\%) \\
        \quad Bottom 2500 & 268   & 173(-35.4\%) & 181(-32.5\%) & 799(198.1\%) & 144(-46.3\%) \\
          &       &       &       &       &  \\
    \textbf{Deposit (in B\$)} &       &       &       &       &  \\
    Tot. value & 4542  & 4039(-11.1\%) & 1737(-61.8\%) & 12599(177.4\%) & 4562(0.4\%) \\
    Tot. social surplus & 4826  & 4291(-11.1\%) & 1794(-62.8\%) & 14281(195.9\%) & 4849(0.5\%) \\
    \bottomrule
    \end{tabular}%
  \label{tab:addlabel}%
\end{table}%

\subsection{Decomposition experiments}\label{sec:decomposition}

\subsubsection{Experiment 1: Eliminating liquidity flows through branch networks \label{sec_branch_networks}}

First, we investigate the importance of branch networks for generating funding by considering the counterfactual equilibrium that would arise if each county in which a bank operated relied only on deposits collected from that county and not from its entire branch network. To operationalize this experiment, we set to zero all liquidity flowing into  market (county) $m$ from other markets and examine the impact on local lending and the other outcome variables. This experiment has a significant impact on banks whose operations feature counties where the amount of credit offered exceeds the locally generated deposits since, in these counties, they rely largely on liquidity from elsewhere. At the extreme, many banks make loans in counties where they have no branch presence at all, so under this experiment, access to credit would fall sharply therein.\footnote{It is therefore important to note that we hold fixed the amount of securitized loans, when naturally it is likely that a bank without access to funding from its deposit base would exhibit a substantially higher local securitization rate than what we observe in the data.} Therefore, eliminating liquidity flows from throughout the branch networks should make local loans more closely related to local deposits, such that bank-level imbalance scores fall. 

Our findings are consistent with this. From Table \ref{tab:decomposition}, we can see that the median bank imbalance index declines substantially from $0.270$ to $0.036$ -- that is, the proportion of funds that banks transfer between counties in which they operate declines by roughly 87 percent. Overall, there is a substantial reduction in the total volume of loans, from about \$1.9 to \$1.4 trillion. Note that this decrease is driven entirely by depository institutions, who see their lending activity fall by over 65 percent. The decrease in lending is more pronounced in counties that are (i) smaller in initial lending activity, (ii) poorer, and (iii) more rural.

Note, that the large effect on lending activity is the result not only of the elimination of  total liquidity, but also the subsequent negative synergies elimination entails. The total value of deposits decreases by over 11\% because of economies of scope, and this decrease  in turn causes lending to fall further. We also find that shadow banks actually benefit indirectly under this experiment. They experience a slight increase in lending activity because depository institutions become less attractive.

Although bank-level imbalance scores fall, the national imbalance index actually increases. This is somewhat counterintuitive, but can be explained by the fact that, at the county level, there is a positive correlation between the total volume of deposits and the presence of banks making loans without branches. Counties with high levels of deposits are the large/rich counties to which it is profitable to extend credit even without any branches. As explained above, counties with many banks making loans without branches are affected most in this experiment, suffering larger loan volume losses. The result is that in these counties, the correlation between deposits and loans falls, causing the national imbalance index to increase.

{\bf Takeaway:} The results from this experiment confirm the importance of branch networks for spreading liquidity across the different markets in which banks operate, especially in socially disadvantaged regions.

\subsubsection{Experiment 2: Eliminating local synergies \label{sec_local_synergies}}

In this experiment, we study the effects of eliminating 
local synergies (economies of scope)  between deposits and loans. We implement this experiment by setting the parameters $\theta_{\ell }^{d}$ and $\theta _{d}^{\ell}$ to zero and computing the new equilibrium of the model. 

Economies of scope contribute to the home bias in credit distribution, resulting in lower imbalance scores. Therefore, intuitively, eliminating economies of scope should push the bank-level imbalance index higher. Indeed, our results suggest that the  median bank-level imbalance index increases by 211\%. The national imbalance index also increases by 16.5\%. The overall level of credit provision falls by 29.4\%, with almost all of the decrease coming from traditional banks. Since they do not have any deposit funding, shadow banks are only affected in this experiment because traditional banks become relatively less attractive. Eliminating synergies implies that the decrease in lending activity has a effect on deposit activity,  such that the total value of deposits falls by 61.8\%.

Our findings also suggest that eliminating synergies has a stronger effect on counties at the bottom of the tail of the loan distribution,  that are poorer, and that are more rural. These counties benefit most from synergies, keeping funds for use in the markets where they are generated.

{\bf Takeaway}: The results from this experiment demonstrate that economies of scope are important for generating a home bias and preventing funds from flowing away from where they are generated. This is especially true in socially disadvantaged regions.

\subsubsection{Experiment 3: Removing local market power \label{sec_market_power}}

In this experiment we obtain the equilibrium of the model under the
condition that banks behave competitively in all local markets banks, charging
prices equal to marginal cost.  Recall from our system of equations in \eqref{structural equations} that the left-hand-sides are defined as $y_{jmt}^{d} \equiv \ln \left(s_{jm}^{d}/s_{0m}^{d}\right) + 1/(1-s_{jm}^{d})$ and $y_{jmt}^{\ell} \equiv \ln \left(s_{jm}^{\ell}/s_{0m}^{\ell}\right) + 1/(1-s_{jm}^{\ell})$, where $1/(1-s_{jm}^{d})$ and  $1/(1-s_{jm}^{\ell })$ are the price - cost margins of firm $j$ in market $m$, measured in utils. The experiment involves eliminating the two margin terms.

Our findings suggest that the median bank-level imbalance index falls by 3.5\% and that the national imbalance index decreases by 16.6\%. The total volume of both loans and deposits increases substantially, implying that market power has a strong negative effect on social surplus. The effect is largely driven by depository institutions, but shadow banks also engage in significantly more lending as a result of the removal of local market power. 

Eliminating market power has a stronger effect in smaller/poorer counties, where loan volumes roughly triple. Market power is higher in these counties such that banks operating in them reduce the supply of loans relative to the level that would arise under perfect competition, and they move these funds to richer more competitive counties where they may not have branches/deposits. As a result smaller/poorer counties typically have lower loan-to-deposit ratios. Removing local market power will therefore increase the supply of loans from banks in these smaller/poorer counties, making loans closer to deposits, reducing the II.

Finally, it is worth pointing out that the fact that all of the results from this experiment are consistent with our expectations of the impact of shutting market power down helps to validate our structural approach and confirm that the absence of price data does not prevent us from developing a meaningful measure of market power.

{\bf Takeaway}: Local market power has a strong negative effect on lending and social surplus. The effect is felt most strongly in socially disadvantaged counties.

\subsubsection{Experiment 4: Removing shadow banks \label{sec_shadow _banks}}

Our final decomposition experiment sheds light on the role of shadow banks. We operationalize this experiment by eliminating shadow banks altogether. Doing so reduces lending by 51.2\%, with the decrease being felt more strongly in high-volume and urban markets. Note from column 1 of Table \ref{tab:decomposition} that shadow banks actually account for 54.7\% of loan volume, but when they are removed from the market, depository institutions pick up some of the slack, such that their lending activity increases by 7.8\%. Because of economies of scope, deposits increase by 0.4\%. 

The National II falls by 2\%. This is because shadow banks are more likely to operate (make loans) in counties where depository institutions also lend heavily, such that they have relatively high loan-to-deposit ratios. Therefore, when shadow banks are removed, the distribution of loans approaches the distribution of deposits at the county level, decreasing the II. In Appendix \ref{sec:CF_results_app}, panel (D) of Figure \ref{fig:II_DF_vs_DS} displays the counterfactual evolution of the national II over the 13 years of our sample. While the average decrease resulting from this experiment is 2\%, that actually masks important heterogeneity. In the early years of the sample, the counterfactual index is well below the factual, with a maximum decrease of 6\% in 2001. As we approach the Great Financial Crisis, the decrease becomes smaller and actually becomes positive in 2009 and 2010.

{\bf Takeaway}: Shadow banks play an important role in credit provision, especially in large and urban markets. The ability of depository banks to substitute for shadow bank credit provision is limited.

\subsection{Policy Counterfactuals}\label{sec:policy_CFs}

\subsubsection{Experiment 5: Splitting up multi-state banks (i.e., undoing Riegle-Neal) \label{sec_RN}}

This experiment is related to Experiment 1, where we eliminated the flow of liquidity throughout the entire branch network so that in any given local market, banks have access to only their local deposits. Here, we use our model to understand the impact on the flow of funds of the Riegle-Neal Act of 1994, which allowed banks to operate branch networks in multiple states. To operationalize this experiment we divide every multi-state bank in our sample into different independent banks, one for each state. Only about $10\%$ of depository institutions in our sample have branches in multiple states, but over $50\%$ make loans in multiple states.\footnote{For details, see Table \ref{tab:multistate_banks} in Appendix \ref{sec:appendix_multistate}. Note also that a significantly larger proportion of shadow banks extend their operations across state borders, with nearly $75\%$ engaging in lending activities across multiple states.}

Multi-state branch networks help credit flow across regions and make local lending less reliant on local deposits. A direct effect of breaking up a traditional bank's branch network  is to shrink its total deposit volume  since it no longer has access to any deposits generated out-of-state. This reduction in total deposits contributes to reducing the social surpluses of loans and deposits, by 19.7\% and 4.5\%, respectively. Furthermore, at the local level, loans become more closely related to deposits, such that the bank-level imbalance index falls by about 66\%. For the same reasons as in Experiment 1, the national imbalance index increases. Finally, shadow banks actually benefit slightly under this scenario, since depository institutions become relatively less attractive.\footnote{In Appendix \ref{app:RN_test} we present results from a test of this counterfactual against pre-merger data that confirm the ability of this experiment to capture the impact of the policy.}

{\bf Takeaway}: Riegle-Neal impacted lending activity and the flow of funds, but the effect was moderated by the fact that cross-border lending activity without branches is somewhat limited.

\begin{table}[h!]
  \centering
  \caption{Policy Counterfactuals} \label{tab:policy_CFs}%
    \begin{tabular}{lll}
    \toprule
    \toprule
          & Exp. 5 & Exp. 6 \\
    Outcomes (13 yr avr) & \multicolumn{1}{p{5.665em}}{Single state\newline{}networks} & \multicolumn{1}{p{5.415em}}{Deposit\newline{}tax} \\
    \midrule
    \textbf{Imbalanced index} &       &  \\
    National level & 0.3044(2.0\%) & 0.3055(2.4\%) \\
    Bank level (Median) & 0.0924(-65.8\%) & 0.2718(0.6\%) \\
          & \textcolor[rgb]{ .459,  .443,  .443}{} & \textcolor[rgb]{ .459,  .443,  .443}{} \\
    \textbf{Loans (in B\$)} &       &  \\
    Tot. value & 1570(-18.9\%) & 1866(-3.6\%) \\
    Tot. social surplus & 1672(-19.7\%) & 2005(-3.7\%) \\
    By bank types &       &  \\
       \quad Depository banks & 484(-44.7\%) & 803(-8.3\%) \\
       \quad Shadow banks & 1086(2.6\%) & 1063(0.5\%) \\
    By county loans &       &  \\
       \quad Top 100 & 908(-17.7\%) & 1067(-3.3\%) \\
       \quad Bottom 2500 & 131(-20.6\%) & 157(-4.8\%) \\
    By county income/cap &       &    \\
       \quad Top 100 & 436(-19.1\%) & 521(-3.3\%) \\
       \quad Bottom 2500 & 369(-18.7\%) & 436(-4.0\%) \\
    By county urban\% &       &  \\
       \quad Top 100 & 626(-18.2\%) & 739(-3.4\%) \\
       \quad Bottom 2500 & 212(-20.9\%) & 256(-4.5\%) \\
          &       &  \\
    \textbf{Deposit (in B\$)} &       &  \\
    Tot. value & 4343(-4.4\%) & 3642(-19.8\%) \\
    Tot. social surplus & 4610(-4.5\%) & 3845(-20.3\%) \\
    \bottomrule
    \end{tabular}%
  \label{tab:addlabel}%
\end{table}%

\subsubsection{Experiment 6: Tax on deposits}\label{sec:tax_on_deposits}

We evaluate how a counterfactual tax on deposits would affect the provision of credit and its geographic distribution. This experiment offers valuable insight into how exogenous changes in deposit rates affect the geographic distribution of deposits and loans. It can be viewed through the lens of a monetary policy experiment, given that inflation ``taxes" away the real value of nominal deposits. We implement this experiment via an ad-valorem tax such that the marginal cost of deposits in every market increases by the same magnitude of $\tau$, and social surplus in the deposit market declines by $\alpha^{d}\tau $. The value of $\tau$ is chosen such that social surplus in the median county -- according to the empirical distribution of
social surplus of deposits -- declines by $20\%$.

Our findings suggest that the targeted reduction in deposits results in an average deposit volume decline of $19.8\%$ for depository institutions, but a smaller decrease in lending activity, felt more strongly in smaller, poorer, and more rural counties. This implies that for traditional banks the elasticity of credit relative to deposits is roughly $0.5$. These banks experience a reduction in their lending activity of 8.3\% in response to a 19.8\% decrease in their deposits. Note that shadow bank lending is barely affected, since they do not collect deposits, and shadow banks make up roughly half of total lending. As a result,  overall, total lending activity only falls by 3.6\%. This effect is felt more strongly in smaller, poorer, and more rural counties.

These changes result in only a small impact on the imbalance indices. The  median bank-level imbalance index increases slightly (by 0.6\%), and the national imbalance index grows by 2.4\%. To understand this result note that taxing deposits affects a bank's lending activity through two channels: local synergies and total deposits. Counties in which the bank collects and makes loans are affected through both channels, while counties where it only makes loans are influenced just by total deposits, such that former experience a larger decrease in lending activity. Supposing that the deposit decrease hits all of the bank's counties equally, then the fact that these two types of counties are differentially affected implies that, across the bank's network, deposits and loans will be less correlated, thereby increasing its imbalance index. The same forces explain why the national imbalance index increases.

{\bf Takeaway}: Imposition of a deposit tax has a moderate effect on lending activity. The effect is felt more strongly in socially disadvantaged markets.

\section{Conclusions\label{sec:conclusions}}

In this paper we use data from the SOD and HMDA data sets for the period 1998-2010 to study the extent to
which deposits and loans are segregated, and to investigate the factors that
contribute to this imbalance. We make two main contributions. First, we adapt techniques developed in sociology and labor to measure the
degree of segregation of deposits and loans. Our \textit{imbalance indexes}
provide information on the transfer of funds within branch networks of US
banks, and across counties. Our results reveal that the majority of banks
exhibit a strong home bias and some regions have limited access to credit
relative to their share of deposits. Second, we develop and estimate a structural model of bank oligopoly
competition that allows for rich interconnections across geographic
locations and between deposit and loan markets such that local shocks in
demand for deposits or loans can affect endogenously the volume of loans and
deposits in every local market. The estimated model reveals that a bank's
total deposits have a  significant effect on the bank's shares in
loan markets. We also find evidence that is consistent with significant
economies of scope between deposits and loans at the local level.

An important advantage of our structural approach is that we can study counterfactual scenarios in which we adjust parameters or impose relevant policy-related restrictions. 
Our  experiments show that multi-state branch networks contribute significantly to the geographic flow of credit, but benefit especially larger/richer counties. Scope economies are important for generating a home bias and preventing funds from flowing away from where they are generated. Local market power, on the other hand, has a substantial negative effect on the geographic flow of credit, with its adverse impact most pronounced in economically disadvantaged counties. Furthermore, our findings underscore the considerable contribution of shadow banks in credit provision, particularly within large and urbanized markets. The capacity of traditional depository institutions to supplant the credit function provided by shadow banking entities proves to be constrained. In terms of policy experiments, we find that Riegle-Neal impacted lending activity and the flow of funds, albeit moderately due to the inherent limitations on cross-border lending activity. Additionally, introducing a deposit tax has a moderate negative effect on lending activity, with its effects disproportionately felt in socially disadvantaged markets.

\newpage
\clearpage

\appendix

\setcounter{table}{0}
\renewcommand{\thetable}{A\arabic{table}}
\setcounter{figure}{0}
\renewcommand\thefigure{A\arabic{figure}}   

\section{Data and Descriptive Evidence}\label{sec:data_appendix}

\subsection{Datasets and observation level}

We rely on four main sources for the construction of our dataset. The first is the summary of deposits (SOD) collected by the Federal Deposit Insurance Corporation (FDIC), which provides county-level information on deposits outstanding (the “stock” measure) for each FDIC-insured bank. 

The second is the Reports of Condition and Income (CALL) report collected by the FDIC. The CALL reports contain important bank-level characteristics, such as assets, total deposits (stock), total (mortgage) loans (stock), etc. It also provides loan turnover rate that we use to construct the “adjusted deposit flow” measures in deposits and loans described in detail in Section \ref{appendix_construction_flows}. 

Third, we make use of Home Mortgage Disclosure Act (HMDA) data  collected by the Consumer Financial Protection Bureau, which provide county-level total mortgage loans issued by each reported institutions in a given year (a “flow” measure of loans).

Lastly, we obtain credit union characteristics from the CALL reports (CALL-CU afterwards) collected by National Credit Union Administration (NCUA). The NCUA CALL reports are generally organized in a similar fashion as the FDIC CALL reports. However, one important feature worth mentioning is that they contain not only the mortgage loan outstanding for each credit union, but also the amount of mortgage loans newly issued in the corresponding calendar year. Similar information on loan flow is not available among the depository institutions in the FDIC CALL reports. 

Table \ref{tab:data_sources} summarizes the above information. 
The SOD and HMDA data are available annually during our study periods (1998-2010). The CALL reports are reported quarterly. In practice, we found that the June (the second quarter) CALL reports deliver the highest matching rate with the SOD and HMDA datasets each year. We therefore use the second quarter CALL reports for all the FDIC insured banks. For credit unions, we must use the December CALL data because only this quarter can give us the mortgage loans issued in the whole calendar year (the loan “flow” measure).

\begin{table}[htbp]
  \centering
  \caption{Data sources}\label{tab:data_sources}
    \begin{tabular}{p{9em}p{8.735em}p{9.3em}p{13.8em}}
    \toprule
    \toprule
    Dataset & Collector & Detailed Level & Variables \\
    \midrule
    Summary of Deposits\newline{}(SOD) & Federal Deposit Insurance Corporation\newline{}(FDIC) & Institution-county level\newline{}(Depository banks) & Amount of deposits (stock)\newline{}Number of branches \\
    \midrule
    Home Mortgage Disclosure Act Data\newline{}(HDMA) & Consumer Financial Protection Bureau & Institution-county level\newline{}(Depository banks \& \newline{}Shadow banks) & Amount of loans issued (flow)\newline{}Securitization rate \\
    \midrule
    Reports of Condition and Income \newline{}(CALL report) & Federal Deposit Insurance Corporation\newline{}(FDIC) & Institution level \newline{}(Depository banks) & Amount of deposits (stock)\newline{}Amount of loans (stock)\newline{}Turn over rate of loans \\
    \midrule
    Reports of Condition and Income \newline{}for  Credit Union\newline{}(CU-CALL report) & National Credit Union Administration \newline{}(NCUA) & Institution level\newline{}(Credit Unions) & Amount of deposits (stock)\newline{}Amount of loans (stock and flow)\newline{}Turn over rate of loans \\
    \bottomrule
    \end{tabular}%
\end{table}%

\subsection{Construction of the working sample}\label{sec:working_sample}
Table \ref{tab:types_banks} describes how all of the datasets just described are matched. The last three columns report the number of institutions, the share of deposits, and the share of loans of each case at the national level in year 2000.

\begin{table}[h!]
\caption{Classification of Banks \& Matching Cases Between SOD, HMDA, and CALL \label{tab:types_banks}}
\centering
\resizebox{\textwidth}{!}{\begin{tabular}{llccccccc}
\hline\hline
\multicolumn{2}{c}{} & 
\multicolumn{4}{c}{Availability in Datasets} & 
\multicolumn{3}{c}{Year 2000} \\ 
\hline
 
\footnotesize{} & \footnotesize{} & \footnotesize{} & \footnotesize{} & 
\footnotesize{} & \footnotesize{} & \footnotesize{Nbr}
& \footnotesize{Share} & \footnotesize{Share}\\
\footnotesize{Case} & \footnotesize{Inst. Category \& Matching Case} & \footnotesize{HMDA} & \footnotesize{SOD} & 
\footnotesize{CALL} & \footnotesize{CU CALL} & \footnotesize{Banks} & \footnotesize{Deps} & \footnotesize{Loan}\\
\hline
&  &  &  &  &  &  &  & \\ 
\multicolumn{1}{c}{\small{1}} & {\footnotesize FDIC Core banks matched with HMDA} & {\footnotesize X} & {\footnotesize X} & {\footnotesize X} & {\footnotesize } 
& {\footnotesize 3971} & {\footnotesize 68.7\%} & {\footnotesize 23.5\%} \\ 
\multicolumn{1}{c}{\footnotesize{2}} & {\footnotesize FDIC Core banks unmatched with HMDA} & {\footnotesize } & {\footnotesize X} & {\footnotesize X} & {\footnotesize } & {\footnotesize 5038} & {\footnotesize 10.9\%} & {\footnotesize 2.7\%} \\ 
\multicolumn{1}{c}{\footnotesize{3}} & {\footnotesize FDIC Savings Asso. matched with HMDA} & {\footnotesize X} & {\footnotesize X} & {\footnotesize } & {\footnotesize } 
& {\footnotesize 696} & {\footnotesize 10.9\%} & {\footnotesize 15.7\%} \\ 
 \multicolumn{1}{c}{\footnotesize{4}} & {\footnotesize FDIC Savings Asso. unmatched with HMDA} & {\footnotesize } & {\footnotesize X} & {\footnotesize } & {\footnotesize } 
& {\footnotesize 393} & {\footnotesize 0.87\%} & {\footnotesize 1.8\%} \\ 
\multicolumn{1}{c}{\footnotesize{5}} & {\footnotesize Foreign banks matched with HMDA} & {\footnotesize X} & {\footnotesize } & {\footnotesize X} & {\footnotesize } 
& {\footnotesize 0} & {\footnotesize -} & {\footnotesize -} \\ 
\multicolumn{1}{c}{\footnotesize{6}} & {\footnotesize Foreign banks unmatched with HMDA} & {\footnotesize } & {\footnotesize } & {\footnotesize X} & {\footnotesize } 
& {\footnotesize 408} & {\footnotesize -} & {\footnotesize -} \\ 
\multicolumn{1}{c}{\footnotesize{7}} & {\footnotesize Shadow banks in HMDA} & {\footnotesize X} & {\footnotesize } & {\footnotesize } & {\footnotesize } 
& {\footnotesize 1047} & {\footnotesize 0.0\%} & {\footnotesize 54.4\%} \\ 
\multicolumn{1}{c}{\footnotesize{8}} & {\footnotesize Credit Unions in HMDA \& CU CALL} & {\footnotesize X} & {\footnotesize } & {\footnotesize } & {\footnotesize X} 
& {\footnotesize 1674} & {\footnotesize 6.5\%} & {\footnotesize 1.9\%} \\ 
\multicolumn{1}{c}{\footnotesize{9}} & {\footnotesize Credit Unions in HMDA but no CU CALL} & {\footnotesize X} & {\footnotesize } & {\footnotesize } & {\footnotesize } 
& {\footnotesize 11} & {\footnotesize -} & {\footnotesize -} \\ 
\multicolumn{1}{c}{\footnotesize{10}} & {\footnotesize Credit Unions in CU CALL but no HMDA} & {\footnotesize  } & {\footnotesize } & {\footnotesize } & {\footnotesize X} 
& {\footnotesize 8932} & {\footnotesize 2.1\%} & {\footnotesize 0.3\%} \\ 
&  &  &  &  &  &  &  & \\ \hline\hline
\end{tabular}}
\end{table}

\textbf{Institutions in both SOD and HMDA:}

The depository institutions in our working sample are those banks (case 1) and saving associations (case 3) that can be matched between SOD and HMDA. In year 2000, there are 3971 case 1 banks,  accounting for 68.7\% of the national deposits and 23.5\% of the national loans. Meanwhile, there are 696 case 3 saving associations, accounting for 10.9\% of the national deposits and 15.7\% of the national loans. 

Notice that case 1 banks can also be matched with the CALL report, while case 3 cannot. Matching with CALL reports is important because they contain the institutional level of loan turnover rate, which is used in constructing our “adjusted flow” measure of deposits. For case 3, we impute their turnover rate by the average turnover rate of the case 1 banks in the same county. 

\textbf{Institutions in HMDA but not in SOD:}

If an institution appears in HMDA but not in SOD, and if they cannot be further matched with the CALL reports or the Credit Union Call Report, then they are labelled a shadow bank (case 7). These are mainly mortgage companies that finance their loans from sources other than deposits. In 2000, there are 1047 shadow banks, accounting for 54.4\% of national loans.

	Institutions that are not in SOD but are in the CALL reports are foreign banks (case 5). They are small in numbers, and also account for a tiny proportion of national deposits and loans. Therefore, they are excluded from our sample. 
 
	Credit unions (cases 8, 9) are large in number, but they are excluded from the sample for two reasons. First, their deposits and loans are only available at the institution level, not at the institution-county level. Second, they only account for a small proportion of national deposits and loans (8.6\% and 2.2\%, respectively, in 2000).

\textbf{Institutions in SOD but not in HMDA:}

Mortgage reporting for institutions smaller than a certain threshold is not compulsory. This is why most of the banks in SOD but not in HMDA (case 2 and 4 banks) tend to be small banks in assets. Although they are large in numbers (5431 institutions in year 2000), but they only account for 4.5\% of total loans and 11.8\% of total deposits. 

\textbf{Institutions not in SOD, not in HMDA:}

Some foreign banks (case 6) and some credit unions (case 10) do not appear in either SOD or HMDA, but these groups are very small in terms of both deposits and loans.

\textbf{Final sample:} Our final working sample includes Case 1, 3, and 7 institutions. These are the institutions for which we know both their deposit distribution (Cases 1 and 3) and their loan distribution (Cases 1, 3, and 7) at the county level. 

\subsection{Multi-state banks}\label{sec:appendix_multistate}

\begin{table}[!h]
    \centering
    \caption{Multi-state banks}\label{tab:multistate_banks}
    \begin{tabular}{c|c|c|c|c|c|c|c|c}
    \hline
        & All & \multicolumn{4}{c}{Depository Banks} & \multicolumn{2}{c}{Shadow banks} \\ \hline
        year & \# & \# & ms\_branch & ms\_loan & br OR loan & \# & ms\_loan \\ \hline
        1998 & 6123 & 4740 & 188 & 1861 & 1861 & 1383 & 936 \\ 
        1999 & 6014 & 4667 & 235 & 1979 & 1979 & 1347 & 934 \\ 
        2000 & 5892 & 4649 & 263 & 2026 & 2026 & 1243 & 912 \\ 
        2001 & 5774 & 4592 & 306 & 2135 & 2135 & 1182 & 878 \\ 
        2002 & 5856 & 4622 & 334 & 2294 & 2294 & 1234 & 923 \\ 
        2003 & 6080 & 4624 & 358 & 2451 & 2451 & 1456 & 1124 \\ 
        2004 & 6619 & 4906 & 376 & 2632 & 2632 & 1713 & 1250 \\ 
        2005 & 6615 & 4830 & 422 & 2662 & 2662 & 1785 & 1310 \\ 
        2006 & 6659 & 4793 & 441 & 2723 & 2723 & 1866 & 1371 \\ 
        2007 & 6404 & 4767 & 466 & 2721 & 2721 & 1637 & 1202 \\ 
        2008 & 6166 & 4789 & 486 & 2810 & 2810 & 1377 & 1008 \\ 
        2009 & 5913 & 4723 & 492 & 2657 & 2657 & 1190 & 890 \\ 
        2010 & 5715 & 4576 & 495 & 2473 & 2473 & 1139 & 860 \\ \hline
    \end{tabular}
\end{table}

\clearpage
\newpage

\subsection{Alternative Measures of the Imbalance Index -- Adjusted Deposit Flows and Loan Stocks  \label{appendix_construction_flows}}

As mentioned in Section \ref{sec:data_features}, the loan information in HMDA refers to the mortgage loans issued in a year, and should therefore be considered a \textit{flow} measure. On the deposit side, the SOD provides information on the deposits available on a day of the year (June 30th), and it is a \textit{stock} variable. One problem with using stocks is that some past deposit inflows,  which make up today's stock of deposits, have already been used to fund prior mortgages. To address this, deposit flows can be constructed from the SOD dataset as  the net change in deposit stocks at the bank-county level by first-differencing by year. However, this represents a net change -- newly attracted deposits minus withdrawn deposits. 

 There are two problems with using the net change. First, it can be negative, which makes it difficult to calculate the deposit share at the county level.  Second, it is also the case that this net change in deposits would underestimate the funds available to banks to create new loans, since a fraction of existing loans are repaid (come due) every year. 
 
 We therefore construct what we call an {\it adjusted flow} measure, that includes net deposit changes along with those deposits that are freed up today as a fraction of previously funded loans are paid off. The Call Reports provide information for each bank on the fraction of loans coming due each year and we use this to construct our measure. 

 In our main analysis we continue to focus on stocks of deposits for the reasons mentioned in Section \ref{sec:data_features}. However, we have performed all of our descriptive analysis using new loans and adjusted deposit flows. Our results are robust to this alternative measure of deposits -- The Imbalance Index for adjusted deposit flows vs loan flows follows a similar pattern to stocks vs flows. 

We also repeat the analysis with deposit stocks and replacing loan flows with loan stocks in the  construction of the Imbalance Index. Again, our findings suggest that there is little change to the Imbalance Index from such a modification. 
 
 In what follows we provide a detailed explanation of the implementation of these two robustness checks, along with a discussion of the consequences for our analysis. 
 
\medskip

{\bf I. Adjusted deposit flows vs loan flows:}

\medskip

We start by introducing the following notation and variable definitions. For bank $j$ in county $m$ and year $t$, define:
\begin{itemize}
    \item[$\square$] $q^d_{jmt}$:  amount of deposits (stock) held by $j$ in $m$ as of $t$,
    \item[$\square$] $L_{jmt}$: proportion of $j$'s deposits in $m$ as of $t$ used to fund mortgages throughout $j$'s network,
    \item[$\square$] $R_{jmt}$: loan turnover rate for $j$ in $m$ in $t$ (i.e. the proportion of loans that will mature within a year),
    \item[$\square$] $DF_{jmt}$: amount of deposits newly available for $j$ in $m$ in $t$ (i.e. the adjusted flow of deposits).
\end{itemize}

\medskip

$DF_{jmt}$ is constructed as follows:
\begin{eqnarray*}
DF_{jmt}=q^d_{jmt}-q^d_{jmt-1}+R_{jmt-1}{\times}L_{jmt-1}.
\end{eqnarray*}
This adjusted flow measure is made up of two components:
\begin{itemize}
    \item[i.] $q^d_{jmt}-q^d_{jmt-1}$: net change in the stock of deposits, reflecting the savings and withdrawal decisions of the local consumers.
    \item[ii.] $R_{jmt-1}{\times}L_{jmt-1}$: loans maturing and turning over within a year -- funds that should be available for the purpose of credit.
\end{itemize}

\medskip

There are two simplifying assumptions we must make when constructing the adjusted deposit flow measure:
\begin{itemize}
    \item[$\square$] We do not have the loan turnover rate at the bank-market level. To deal with this, we use the bank level turnover rate to proxy for the bank-market level measure (i.e., we use $R_{jt-1}$ in place of $R_{jmt-1}$).
\item[$\square$] We do not observe the  amount of deposits issued  as  loans at the bank-county level. Instead, we infer this  using observables in the dataset: the total deposit stock ($Q^d_{jt}$), the total loan stock ($L_{jt}$), and the bank-county level deposit stock ($q^d_{jmt}$). The first two come from the CALL reports, while the last comes from the SOD dataset. Specifically, we assume that  at the bank-market level the share of deposit stock used to fund loans throughout $j$'s network is the same as the bank-level share of deposit stock used to fund loans (i.e., $L_{jmt}/q^d_{jmt}=L_{jt}/Q^d_{jt}$). Therefore, 
\begin{eqnarray*}
L_{jmt-1}&=&L_{jt-1}/Q^d_{jt-1}{\times}q^d_{jmt-1}\\
&=&\rho_{jt-1}{\times}q^d_{jmt-1},
\end{eqnarray*}
where $\rho_{jt-1}=L_{jt-1}/Q^d_{jt-1}$ is the ratio between the loan (stock) and deposit (stock) at the bank level. 
\end{itemize}
\bigskip
With the above adjustments, the adjusted deposit flow measure can be calculated as:
\begin{eqnarray}\label{eq:deposit_flow}
DF_{jmt} = q^d_{jmt} -q^d_{jmt-1} + R_{jt-1} {\times} \rho_{jt-1}{\times}q^d_{jmt-1}.              
\end{eqnarray}
Data on deposits at the bank-county level come from the SOD dataset, while variables at the bank level ($R$ and $\rho$) come from the CALL reports.

\medskip

An additional complication arises in constructing our adjusted deposit flow measure because of mergers and the  entry and exit of banks in our sample period. The construction of our adjusted flow measure requires bank-level and bank-county level information from two continuous years. Over time, there are banks that  enter into or exit from the local market. Entry leads to missing values in year $t-1$, while exit leads to missing values in year $t$. The more complicated cases are entry or exit due to mergers, where a bank’s ID should be replaced by another one.

\begin{table}[!ht]
 \caption{Entry and Exit Scenarios}\label{tab:entry_exit}
\begin{center}
{\tiny    \begin{tabular}{llllllllll}
    \hline
        &&&&Year $t-1$ & Year $t$ & Flow calc  & Flow calc w/ & Nbr obs & \% \\ 
         &  &  &  &  &  & w/o adjust. & merger adjust. & in 2000 & share \\ \hline
        Scenario 0 &  &  &  & A & A & Standard formula & Standard formula eq \eqref{eq:deposit_flow} & 22.434 & 87 \\ \hline
        Scenario 1 & Exit &  &  & A &  & Missing value in A & Missing value in A & 322 & 1.2 \\ \hline
         &  & 2.1 & B takes over & A  & B  &Missing value in B  & \color{red} Replace A's ID w/ B's in &  &  \\ 
         &  &  & A's branches &  & &  & \color{red} $t-1$, apply st form & 1,541 & 6 \\ 
         & Exit &  &  &  &  &  &  & &  \\
        Scenario 2 & through & 2.2 & Within & A,B & B & Part of B's $t-1$ & \color{red} Replace A's ID w/ B's & 315 & 1.2 \\
         & merger &  & market M\&A & & & not counted & \color{red} $t-1$, apply st formula &  &  \\ 
         &  &  &  &  &  &  &  &  &  \\ \hline
        Scenario 3 & B denovo& &  & & B & Missing value in B & \color{red} B's deposits in $t$ & 1,148 & 4.5 \\ 
          &  enter &  &  &  &  &  &  &  &  \\ \hline
    \end{tabular}}
    \end{center}
   {\tiny Note: A and B are banks, with B being the surviving bank. Year $t-1$ and Year $t$ columns indicate which of A and B have branches in the given county in the given year.}
\end{table}

Table \ref{tab:entry_exit} lists all the possible scenarios when we compare observations at the bank-county level between year $t-1$ and $t$. The scenarios listed in red  are those in need of modification using the merger information, otherwise we will have missing values in the deposit flow for these cases. 

\begin{itemize}
    \item[$\square$] 	In 87\% of the cases, a bank exists in both years, so that the standard formula of equation (1) can be used. This is scenario 0. 
    \item[$\square$] Scenario 1 describes simple exit from a county. This is infrequent and represents just 1.2\% of observations.
\item[$\square$]	Most banks exit from a market because of a merger (scenario 2). Their ID in year $t-1$ needs to be replaced by the acquiring bank before applying the standard formula to the acquiring banks in year $t$. There are three subcases depending on whether surviving bank B's acquisition of bank A involves taking over A's branches in counties in which it wasn't active (2.1), and in which it was (2.2). 
\item[$\square$] 	Scenario 3 describes denovo entry (about 4.5\%). We assign the deposit stock of this bank in year $t$ as its adjusted deposit flow in year $t$.  
\end{itemize}

\medskip

{\bf Summary statistics and results:}

\begin{itemize}
\item[$\square$]  {\bf Loan turnover rate, $R_{jt}$:} The CALL reports provide information  at the bank level on “loans and leases with a remaining maturity of one year or less.”  The ratio between this variable and the “total loans and leases”  gives us the bank-level turnover rate for loans.  The median (also the mean) turnover rate is around 30\%.

\item[$\square$] {\bf Loan to Deposit ratio, $\rho_{jt}$:} The CALL reports provide information on both the loan stock and deposit stock variable. The ratio between these two can be calculated accordingly. 
In the year 2000, the median value of the loan/deposit ratio is 0.90. We truncate the ratio at the 5\% and 95\% percentile, which means that the final value of the ratio will range between 0.49 and 1.28.\footnote{All values below 0.49 or above 1.28 will be replaced by 0.49 or 1.28, respectively.}

\item[$\square$] {\bf The adjustment factor, $R_{jt} {\times} \rho_{jt}$:} Figure  \ref{fig:adjustment_factor} shows the distribution of $R_{jt} {\times} \rho_{jt}$ over time. The median is around 0.3, which is comforting because the resulting  adjusted deposit flows are unlikely to be negative.

\begin{figure}[!h]
\caption{Adjustment factor, $R_{jt} {\times} \rho_{jt}$}\label{fig:adjustment_factor}
\centering
\centering
\includegraphics[width=0.6\linewidth]{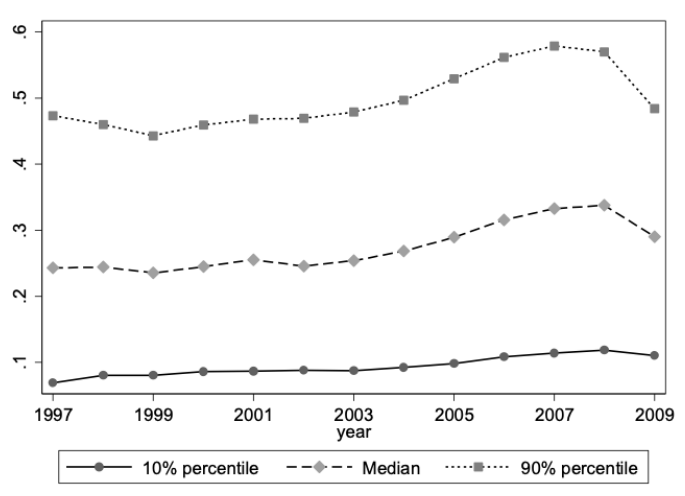}

\end{figure}

\end{itemize}

Equation \eqref{eq:deposit_flow} does not guarantee our adjusted deposit flow measure will be positive.  Fortunately, observations with negative values only account for a small proportion of the sample. The share of negative observations at the bank-county level is around 3.5\%, and the share of negative observations at the bank level  is around 1.3\%. 

\begin{table}[!h]
    \caption{Summary Statistics}\label{tbl:depo_flows_sum_stats}
    \centering
    \begin{tabular}{lccccc}
   
      \multicolumn{6}{c}{ Panel A: Bank Level Statistics: Depository Banks (61,475 bank-year obs.)}   \\ \hline
        Variable & Mean & S.D. & Pctile 5 & Median & Pctile 95 \\ \hline
        Total deposits (stock) & 1018 & 11676 & 38 & 154 & 1799 \\ 
        Total deposits flow & 342 & 4265 & 7 & 47 & 591 \\ 
        Total loan flow & 188 & 3182 & 1 & 13 & 259 \\ \hline
         &  &  &  &  &  \\ 
         \multicolumn{6}{c}{Panel B: County Level Statistics: Depository Banks (40,733 county-year obs.)} \\ \hline
        Variable & Mean & S.D. & Pctile 5 & Median & Pctile 95 \\ \hline
        HHI of deposits (stock) & 4321 & 2891 & 1175 & 3392 & 10000 \\ 
        HHI of deposits flow & 4515 & 2902 & 1232 & 3595 & 10000 \\ 
        HHI of loan flow & 1803 & 1346 & 626 & 1399 & 4357 \\ \hline
    \end{tabular}
\end{table}

Table \ref{tbl:depo_flows_sum_stats} reports the summary statistics of the adjusted deposit flow, and compares it with the deposit stock and the loan flow (Total new loans). In Panel A, all variables are aggregated at the bank level. In Panel B, all variables are aggregated at the county level. The sample we use is the SOD-HMDA matched sample of depository institutions (cases 1 and 3). 

From Panel A we can see that bank-level adjusted deposit flows are about a third of deposits, on average. However, despite their differences in magnitude, the deposit flow (DF) and deposit stock (DS) measures are highly correlated with each other. The correlation coefficient between the two is 0.86. Moreover, as can be seen from Panel B, the adjusted flow and stock measures imply very similar levels of concentration, on average.

\medskip

{\bf Implications of the adjusted deposit flow for the Imbalance Index}

\medskip

\begin{figure}[!h]
\caption{Bank-level Imbalance Index -- Adjusted Deposit flow and Loan flow Distributions}\label{fig:bank_II_DF}
\centering
\begin{minipage}{\linewidth}
\centering
\includegraphics[width=0.6\linewidth]{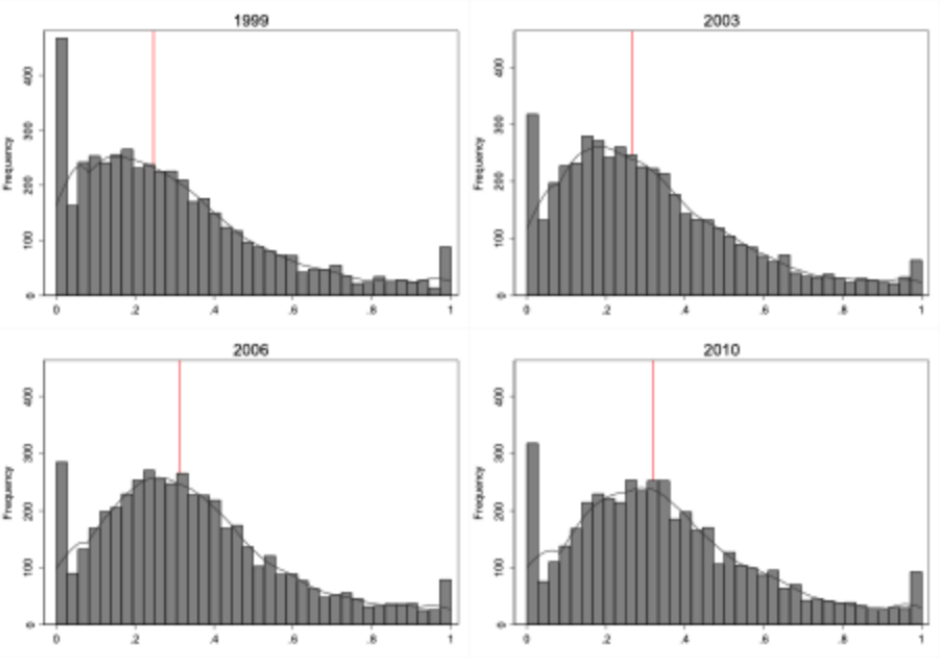}
\end{minipage}
\end{figure}

{\bf Bank level:} Using our sample of matched depository institutions, and  dropping all observations with negative adjusted deposit flows, we can calculate the bank-level Imbalance Index  between the distributions of adjusted deposit flows (DF) and loan flows (LF). Figure \ref{fig:bank_II_DF} of this response reports the distribution of the Imbalance Index changes over time. This figure can be directly compared to  Figure 3 of the revised version of the paper. Doing so reveals that both in terms of magnitudes and  evolution over time, findings are similar whether using adjusted deposit flows or deposit stocks. 

	Comparison of the bank-level Imbalance Index constructed using adjusted deposit flows (DF) with the Imbalance Index using deposit stocks (DS) reveals that they are, surprisingly, very similar. For example, in year 2000, among the 4668 matched depository banks, the mean value of Imbalance Index using adjusted deposit flows is 0.304, while the mean value of Imbalance Index using deposit stocks is 0.301.  Figure \ref{fig:II_DF_vs_DS} plots $II\_DF$ against $II\_DS$, and most of the dots are tightly located around the 45 degree line. The correlation coefficient between $II\_DF$ and $II\_DS$ is 0.944.
	
	\begin{figure}[!h]
\caption{Correlation: Imbalance Index using Adjusted Deposit Flow and Deposit Stock}\label{fig:II_DF_vs_DS}
\centering
\begin{minipage}{\linewidth}
\centering
\includegraphics[width=0.6\linewidth]{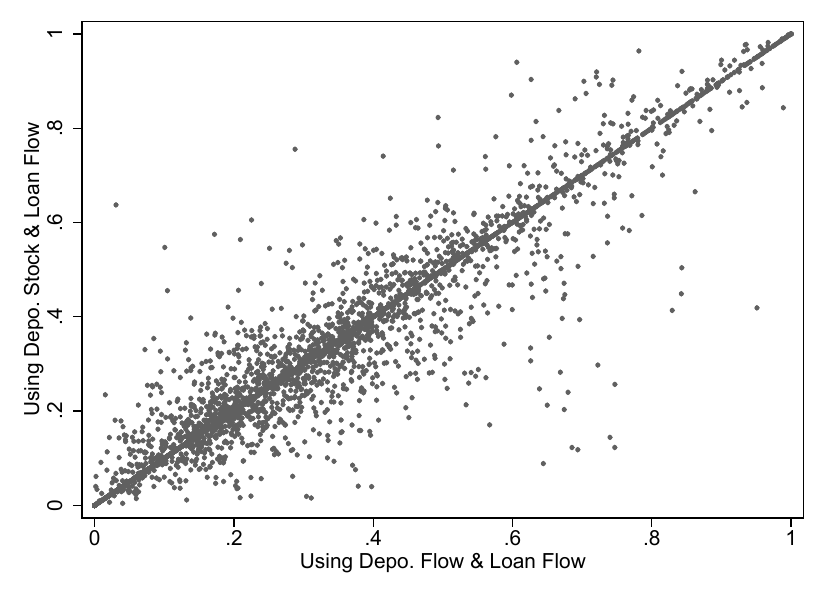}

\end{minipage}
\end{figure}
	
Table \ref{tbl:depo_flows_sum_stats} reports the percentile of the Imbalance Index at the bank level using adjusted flows (panel A) and deposit stocks (panel B). Again, they all look very similar. Their trends are almost identical.

	\begin{table}[!ht]
    \caption{Percentile of Imbalance Index at the Bank Level}\label{tbl:depo_flows_sum_stats}
    \centering
{\small    \begin{tabular}{cccccc|cccccc}
\multicolumn{6}{c}{Panel A: II using DF}&\multicolumn{6}{c}{Panel B: II using DS}\\ \hline
year	&	nbk	&	mean	&	10\%	&	23\%	&	90\%	&	year	&	nbk	&	mean	&	10\%	&	50\%	&	90\%	\\	\hline
1998	&	4745	&	0.28	&	0.02	&	0.23	&	0.63	&	1998	&	4751	&	0.28	&	0.02	&	0.23	&	0.63	\\	
1999	&	4678	&	0.3	&	0.03	&	0.25	&	0.64	&	1999	&	4682	&	0.3	&	0.03	&	0.25	&	0.64	\\	
2000	&	4654	&	0.3	&	0.03	&	0.26	&	0.64	&	2000	&	4668	&	0.3	&	0.04	&	0.26	&	0.63	\\	
2001	&	4615	&	0.3	&	0.04	&	0.25	&	0.62	&	2001	&	4620	&	0.3	&	0.04	&	0.26	&	0.62	\\	
2002	&	4631	&	0.3	&	0.04	&	0.26	&	0.62	&	2002	&	4635	&	0.31	&	0.05	&	0.27	&	0.63	\\	
2003	&	4635	&	0.31	&	0.06	&	0.27	&	0.64	&	2003	&	4637	&	0.32	&	0.06	&	0.27	&	0.63	\\	
2004	&	4904	&	0.33	&	0.06	&	0.29	&	0.65	&	2004	&	4912	&	0.33	&	0.08	&	0.29	&	0.64	\\	
2005	&	4836	&	0.34	&	0.07	&	0.31	&	0.67	&	2005	&	4839	&	0.34	&	0.08	&	0.31	&	0.66	\\	
2006	&	4792	&	0.35	&	0.08	&	0.31	&	0.68	&	2006	&	4804	&	0.35	&	0.09	&	0.31	&	0.67	\\	
2007	&	4775	&	0.36	&	0.08	&	0.32	&	0.69	&	2007	&	4782	&	0.36	&	0.09	&	0.32	&	0.69	\\	
2008	&	4794	&	0.36	&	0.09	&	0.33	&	0.68	&	2008	&	4801	&	0.36	&	0.1	&	0.33	&	0.67	\\	
2009	&	4745	&	0.35	&	0.08	&	0.32	&	0.68	&	2009	&	4747	&	0.35	&	0.09	&	0.31	&	0.68	\\	
2010	&	4591	&	0.35	&	0.07	&	0.32	&	0.69	&	2010	&	4597	&	0.36	&	0.08	&	0.32	&	0.69	\\	\hline

    \end{tabular}}
\end{table}

\begin{figure}[!h]
    \caption{Imbalance Index: Deposit stocks vs Deposit Flows}
    \label{fig:DSLF_DFLF}
    \centering
    \subfloat[Matched depository \& shadow banks]
    {\includegraphics[width=0.48\linewidth]{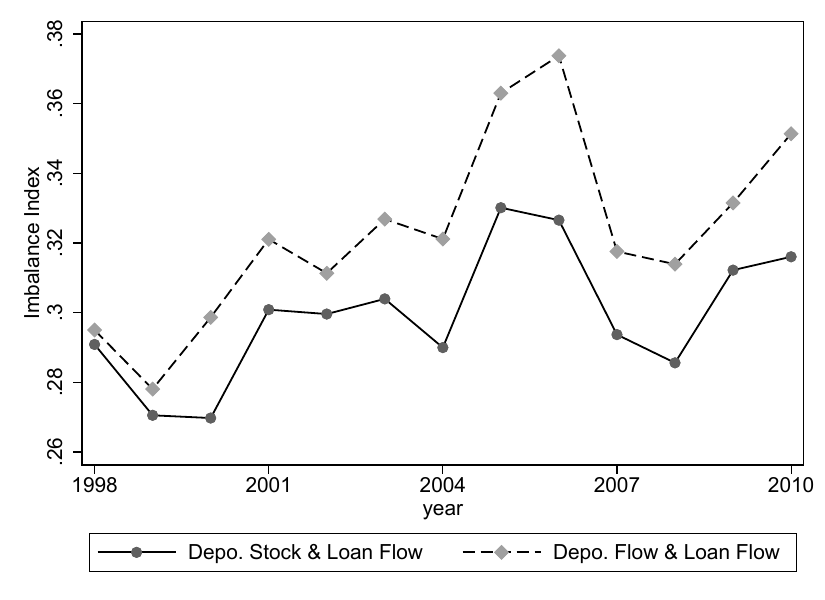}
    \label{fig:DSLF_DFLF_Case137}}
    \subfloat[Matched depository banks]
    {\includegraphics[width=0.48\linewidth]{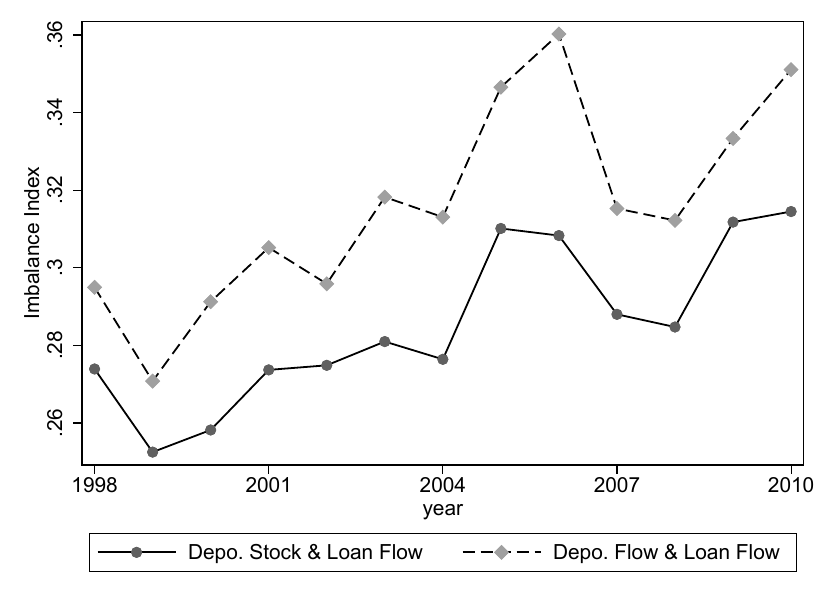}
    \label{fig:DSLF_DFLF_Case13}}
\end{figure}

	{\bf National-level Imbalance Index:} Figure  \ref{fig:DSLF_DFLF} duplicates Figure \ref{fig:national_segregation}, which plots the National-level Imbalance Index. Panel (a) shows results for matched depository \& shadow banks (cases 1, 3, and 7). The loan variables are from HMDA (i.e., loan flow measure). The dashed curve uses the deposit stock, while the solid curve uses the adjusted deposit flow measure. The Imbalance Index constructed using adjusted flows rather than stocks is higher than for stocks in every year, but the two curves follow very similar patterns. Panel (b) shows the same information, but this time just for depository institutions. Again, the Imbalance Index based on adjusted flows is everywhere greater than for stocks, but the two curves are quite similar. Based on this, we think it is reasonable to stick with our stock measure for our analysis.
	
	\medskip
	
{\bf II. Deposit stocks vs loan stocks:}

\medskip

We also construct the National-level Imbalance Index using loan stocks, so that we can compare results to those derived using our main measure of loan flow (constructed from HMDA data). Mortgage-loan stocks are constructed using the CALL reports, under the assumption that stocks have the same bank-county distribution as flows from HMDA.  We also construct a version based on total loans (i.e., not just mortgages), again using the CALL reports. 

	\begin{figure}[!h]
\caption{Imbalance Index: Loan Flow vs Loan Stock}\label{fig:DSLSmgg}
\centering
\begin{minipage}{\linewidth}
\centering
\includegraphics[width=0.6\linewidth]{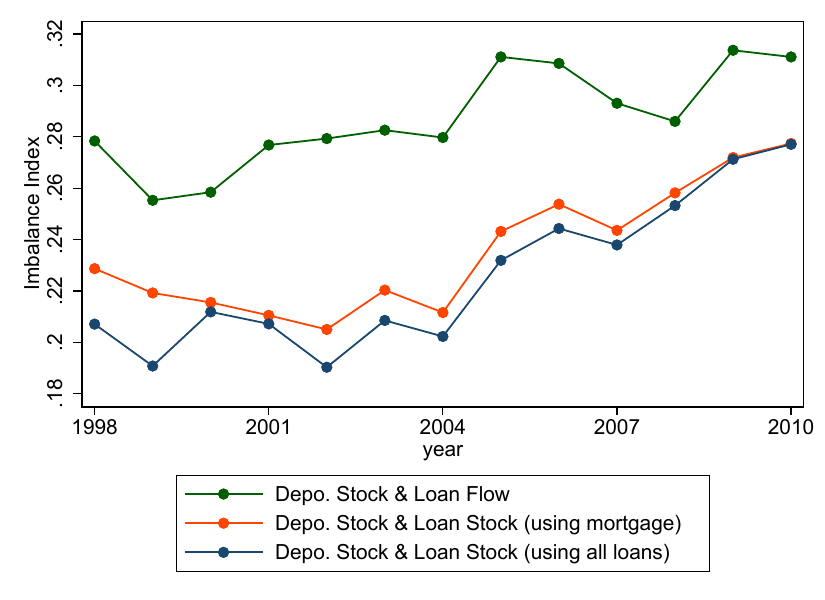}

\end{minipage}
\end{figure}

 Note that this comparison focuses only on depository institutions, since we do not have CALL information for shadow banks.
 
 Our findings suggest that the Imbalance Index that uses the loan flow measure is slightly higher than the Imbalance Index constructed with loan stocks. But the overall trends of the two curves are very similar.

\clearpage
\newpage

\subsection{Bank heterogeneity in Imbalance Index \label{sec:appendix_heterogeneity_bank_ii}}

\begin{table}[htbp]
  \centering
  \footnotesize
    \caption{Evolution of the Imbalance Index of the top 10 banks (by assets)}\label{tab:II_top_10}
    \begin{tabular}{lrr}
    \toprule
    \multicolumn{1}{p{3.5em}}{Rank} & \multicolumn{1}{l}{Bank Name} & \multicolumn{1}{c}{II} \\
    \midrule
          & \multicolumn{1}{c}{1999} &  \\
    \multicolumn{1}{c}{1} & \multicolumn{1}{l}{CITIBANK, N. A.} & 0.37  \\
    \multicolumn{1}{c}{2} & \multicolumn{1}{l}{CHASE MANHATTAN BANK} & 0.47  \\
    \multicolumn{1}{c}{3} & \multicolumn{1}{l}{BANK OF AMERICA NA} & 0.50  \\
    \multicolumn{1}{c}{4} & \multicolumn{1}{l}{FIRST UNION NATIONAL BANK} & 0.36  \\
    \multicolumn{1}{c}{5} & \multicolumn{1}{l}{WASHINGTON MUTUAL BANK, FA} & 0.37  \\
    \multicolumn{1}{c}{6} & \multicolumn{1}{l}{WELLS FARGO BANK, N. A.} & 0.33  \\
    \multicolumn{1}{c}{7} & \multicolumn{1}{l}{FLEET NATIONAL BANK} & 0.45  \\
    \multicolumn{1}{c}{8} & \multicolumn{1}{l}{FIRST NATIONAL BANK OF CHICAGO} & 0.17  \\
    \multicolumn{1}{c}{9} & \multicolumn{1}{l}{BANKERS TRUST COMPANY} & 0.97  \\
    \multicolumn{1}{c}{10} & \multicolumn{1}{l}{KEYBANK NATIONAL ASSN} & 0.37  \\
    Average &       & 0.44  \\
    \midrule
          & \multicolumn{1}{c}{2004} &  \\
    \multicolumn{1}{c}{1} & \multicolumn{1}{l}{BANK OF AMERICA NA} & 0.35  \\
    \multicolumn{1}{c}{2} & \multicolumn{1}{l}{JPMORGAN CHASE BANK} & 0.41  \\
    \multicolumn{1}{c}{3} & \multicolumn{1}{l}{CITIBANK NATIONAL ASSN} & 0.38  \\
    \multicolumn{1}{c}{4} & \multicolumn{1}{l}{WACHOVIA BANK NATIONAL ASSN} & 0.41  \\
    \multicolumn{1}{c}{5} & \multicolumn{1}{l}{WELLS FARGO BANK NA} & 0.51  \\
    \multicolumn{1}{c}{6} & \multicolumn{1}{l}{WASHINGTON MUTUAL BANK FA} & 0.39  \\
    \multicolumn{1}{c}{7} & \multicolumn{1}{l}{BANK ONE NATIONAL ASSN} & 0.70  \\
    \multicolumn{1}{c}{8} & \multicolumn{1}{l}{FLEET NATIONAL BANK} & 0.37  \\
    \multicolumn{1}{c}{9} & \multicolumn{1}{l}{U S BANK NATIONAL ASSN} & 0.45  \\
    \multicolumn{1}{c}{10} & \multicolumn{1}{l}{SUNTRUST BANK} & 0.43  \\
    Average &       & 0.44  \\
    \midrule
          & \multicolumn{1}{c}{2009} &  \\
    \multicolumn{1}{c}{1} & \multicolumn{1}{l}{JPMORGAN CHASE BANK NA} & 0.52  \\
    \multicolumn{1}{c}{2} & \multicolumn{1}{l}{BANK OF AMERICA NA} & 0.43  \\
    \multicolumn{1}{c}{3} & \multicolumn{1}{l}{CITIBANK NATIONAL ASSN} & 0.75  \\
    \multicolumn{1}{c}{4} & \multicolumn{1}{l}{WACHOVIA BANK NATIONAL ASSN} & 0.54  \\
    \multicolumn{1}{c}{5} & \multicolumn{1}{l}{WELLS FARGO BANK NA} & 0.51  \\
    \multicolumn{1}{c}{6} & \multicolumn{1}{l}{U S BANK NATIONAL ASSN} & 0.49  \\
    \multicolumn{1}{c}{7} & \multicolumn{1}{l}{SUNTRUST BANK} & 0.61  \\
    \multicolumn{1}{c}{8} & \multicolumn{1}{l}{HSBC BANK USA NATIONAL ASSN} & 0.66  \\
    \multicolumn{1}{c}{9} & \multicolumn{1}{l}{BRANCH BANKING\&TRUST CO} & 0.33  \\
    \multicolumn{1}{c}{10} & \multicolumn{1}{l}{PNC BANK NATIONAL ASSN} & 0.77  \\
    Average &       & 0.56  \\
    \bottomrule
    \end{tabular}%
\end{table}%

\footnotesize NOTE: Ranking based on banks' assets in each year.

\clearpage\newpage

\begin{figure}[!h]
\caption{Breakdown of imbalance index by bank size -- size measured by counties}\label{fig:breakdown_counties}
\centering
\begin{minipage}{\linewidth}
\centering
\subfloat{\includegraphics[width=0.96\linewidth]{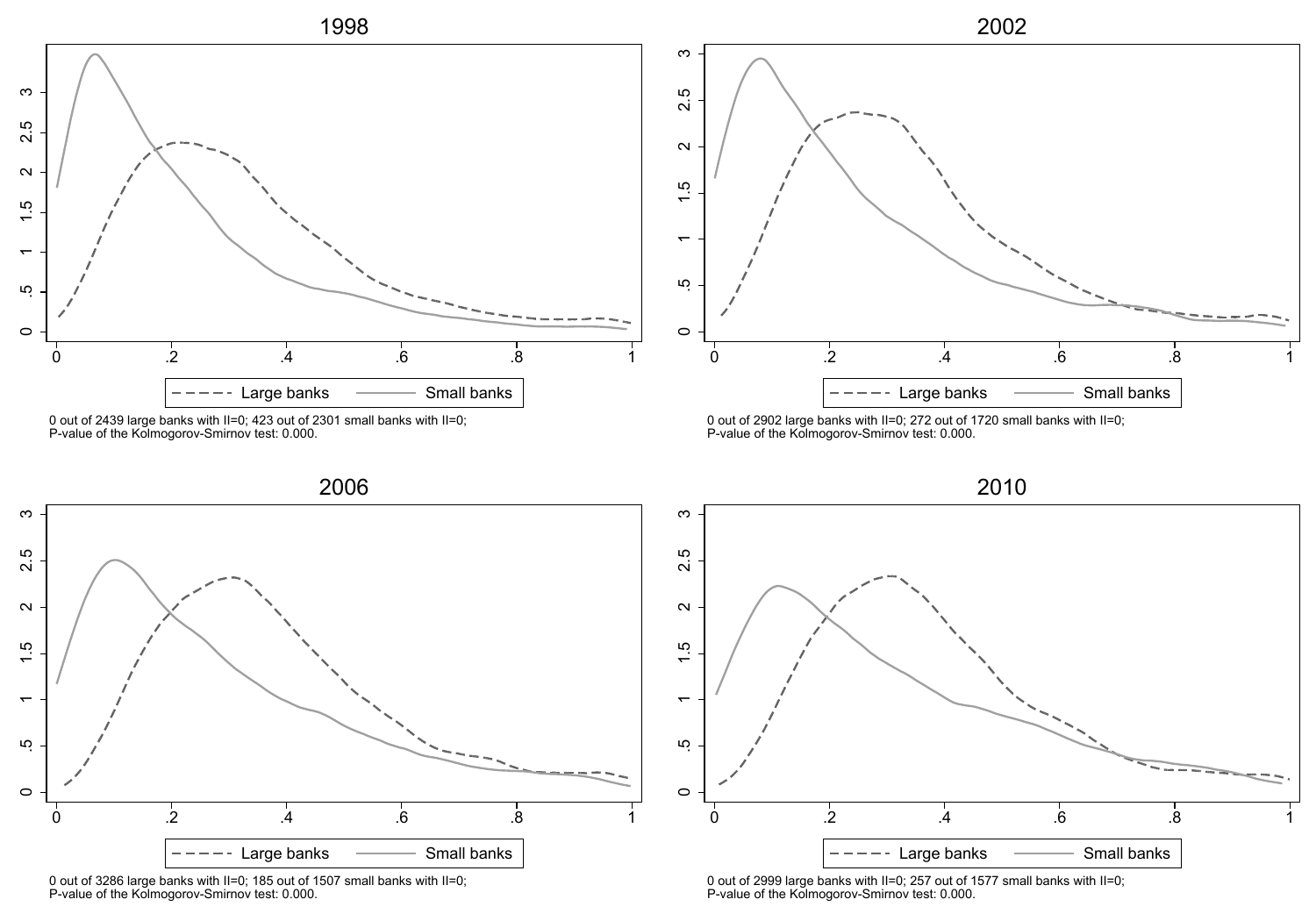}}\\
\bigskip
\bigskip
\subfloat{\includegraphics[width=0.96\linewidth]{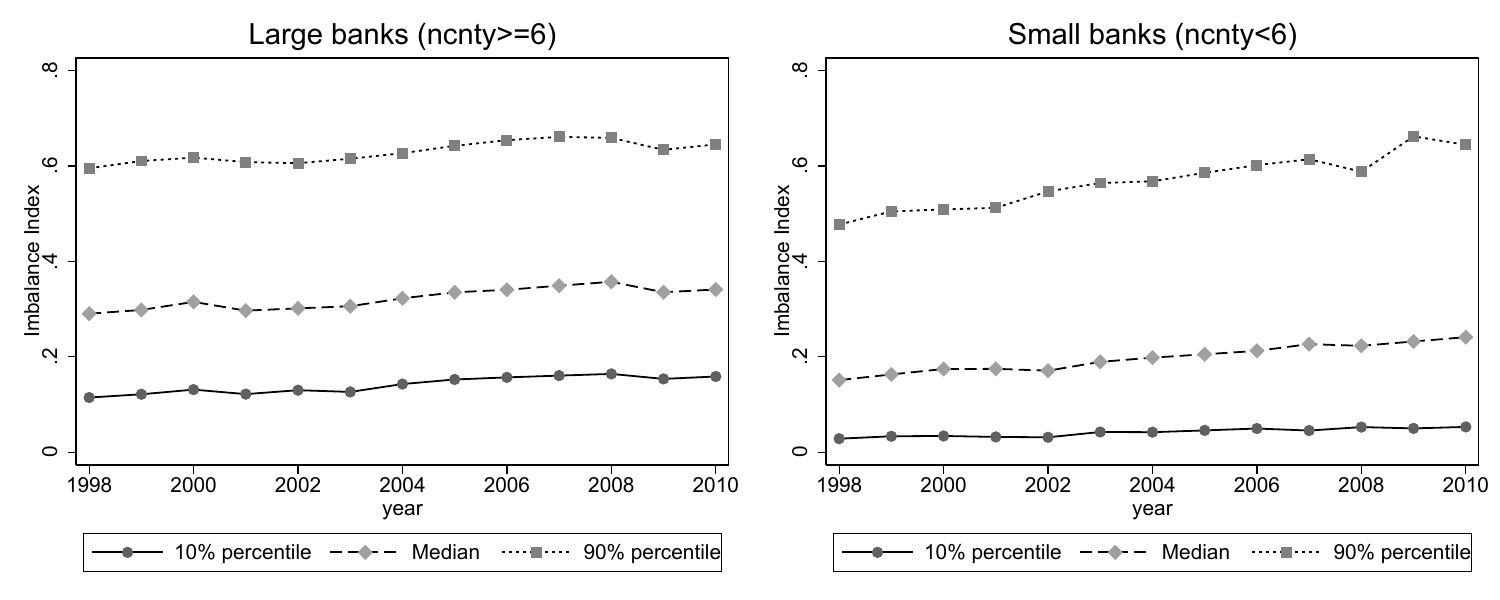}}

\end{minipage}
\end{figure}

	\begin{figure}[!h]
\caption{Breakdown of imbalance index by bank size -- size measured by assets}\label{fig:breakdown_assets}
\centering
\begin{minipage}{\linewidth}
\centering
\subfloat{\includegraphics[width=0.96\linewidth]{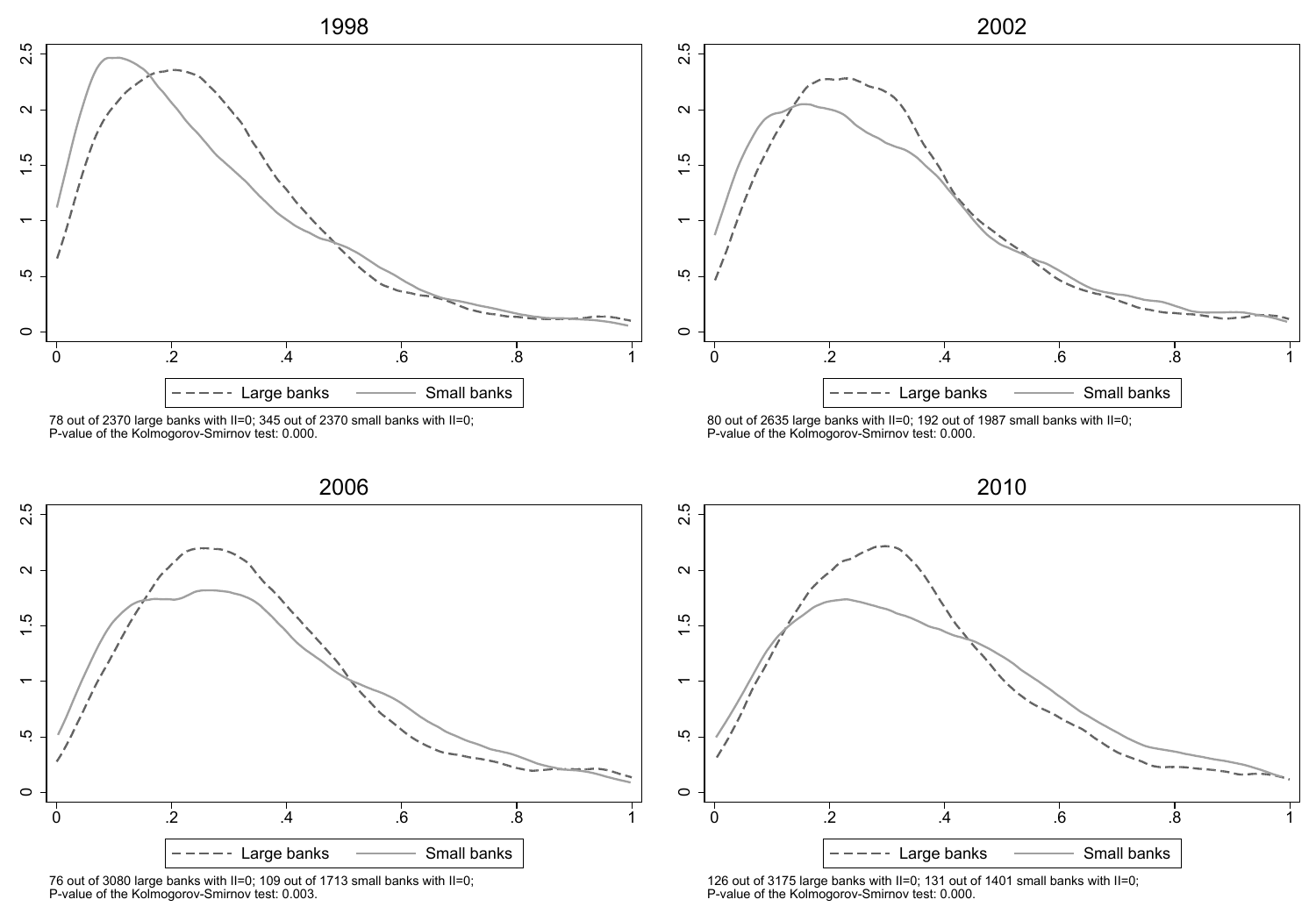}}\\
\bigskip
\bigskip
\subfloat{\includegraphics[width=0.96\linewidth]{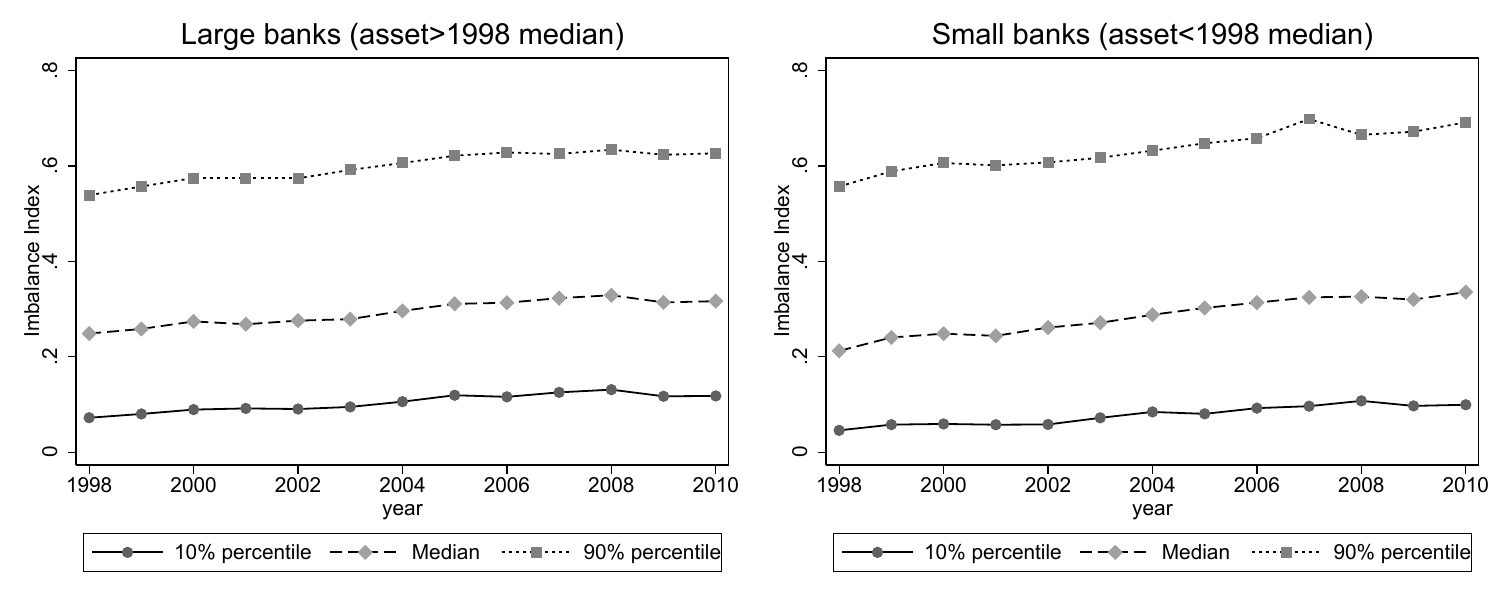}}

\end{minipage}
\end{figure}

\begin{figure}[!h]
\caption{Breakdown of imbalance index by bank size -- size measured by deposits}\label{fig:breakdown_deposits}
\centering
\begin{minipage}{\linewidth}
\centering
\subfloat{\includegraphics[width=0.96\linewidth]{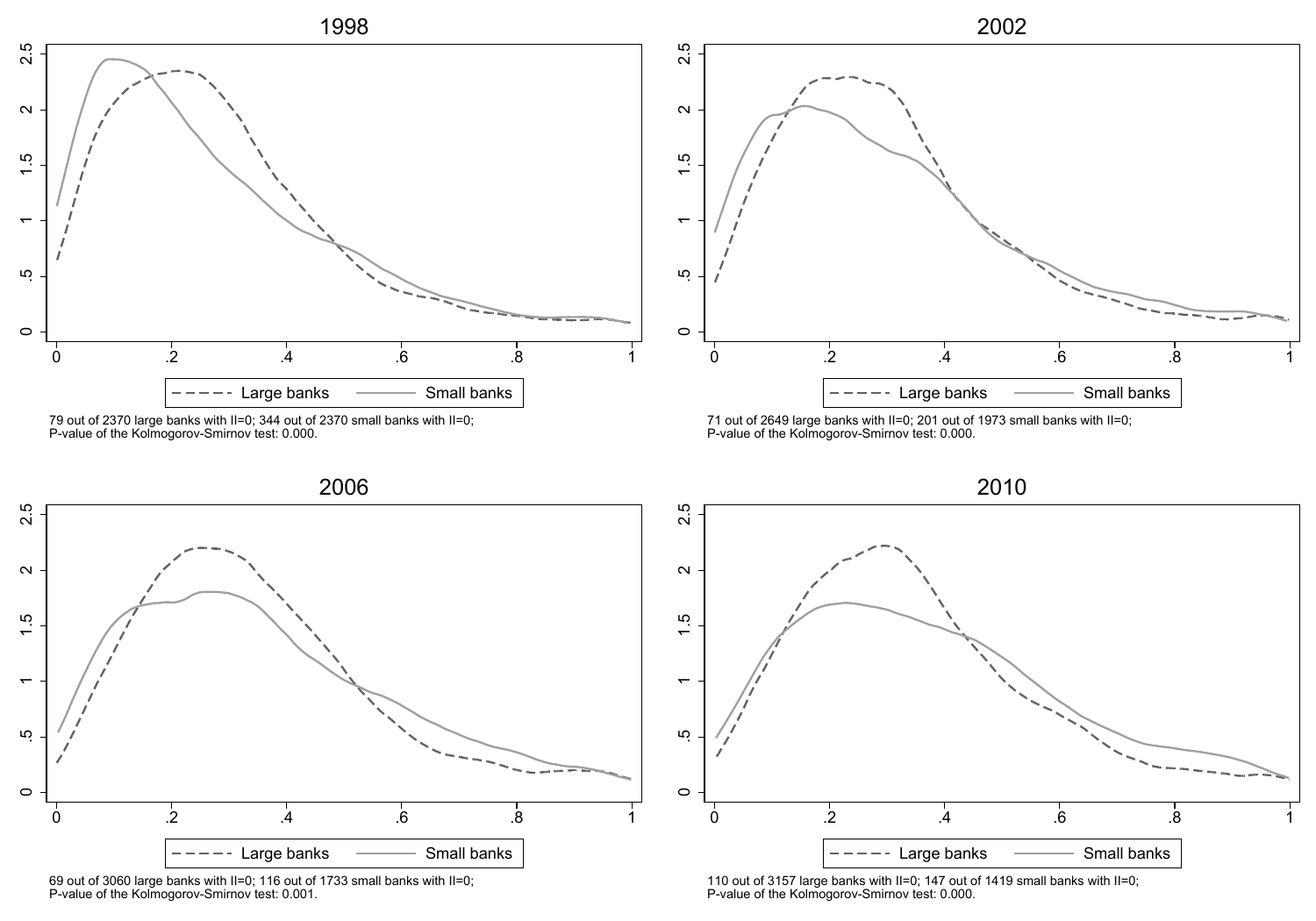}}\\
\bigskip
\bigskip
\subfloat{\includegraphics[width=0.96\linewidth]{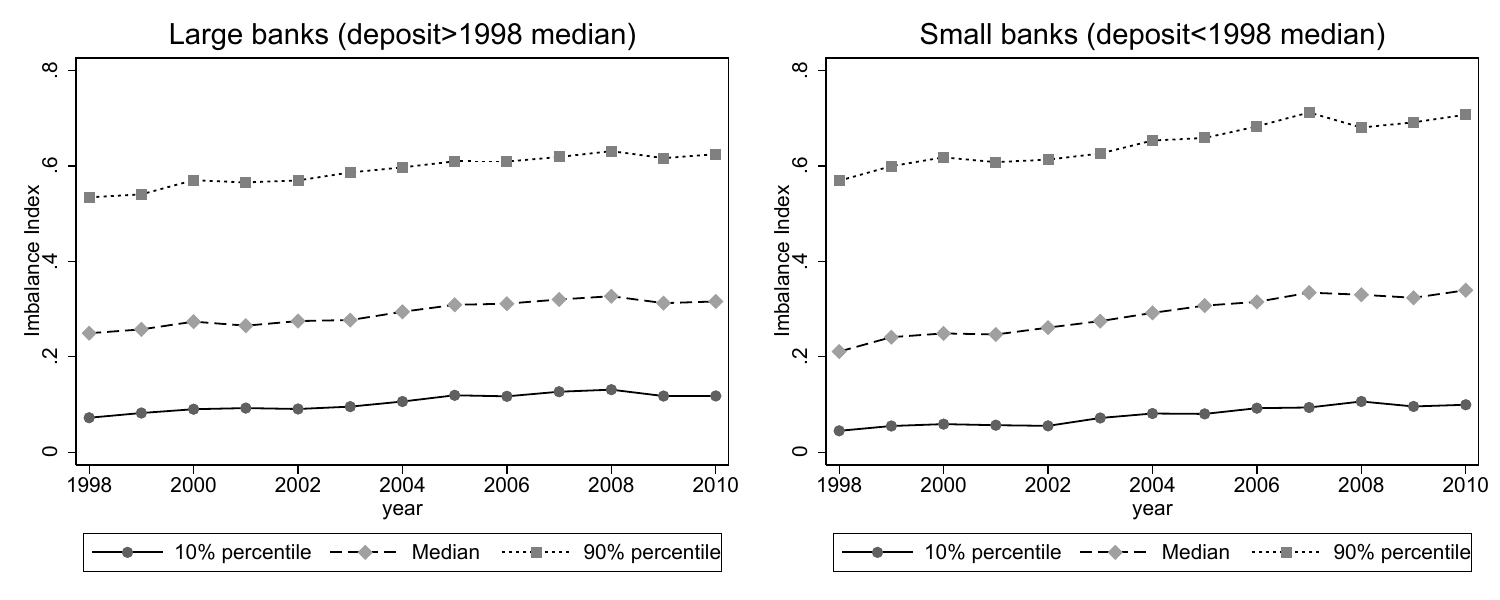}}

\end{minipage}
\end{figure}

\begin{figure}[!h]
\caption{Breakdown of imbalance index by location -- urban vs rural}\label{fig:breakdown_geo}
\centering
\begin{minipage}{\linewidth}
\centering
\subfloat{\includegraphics[width=0.96\linewidth]{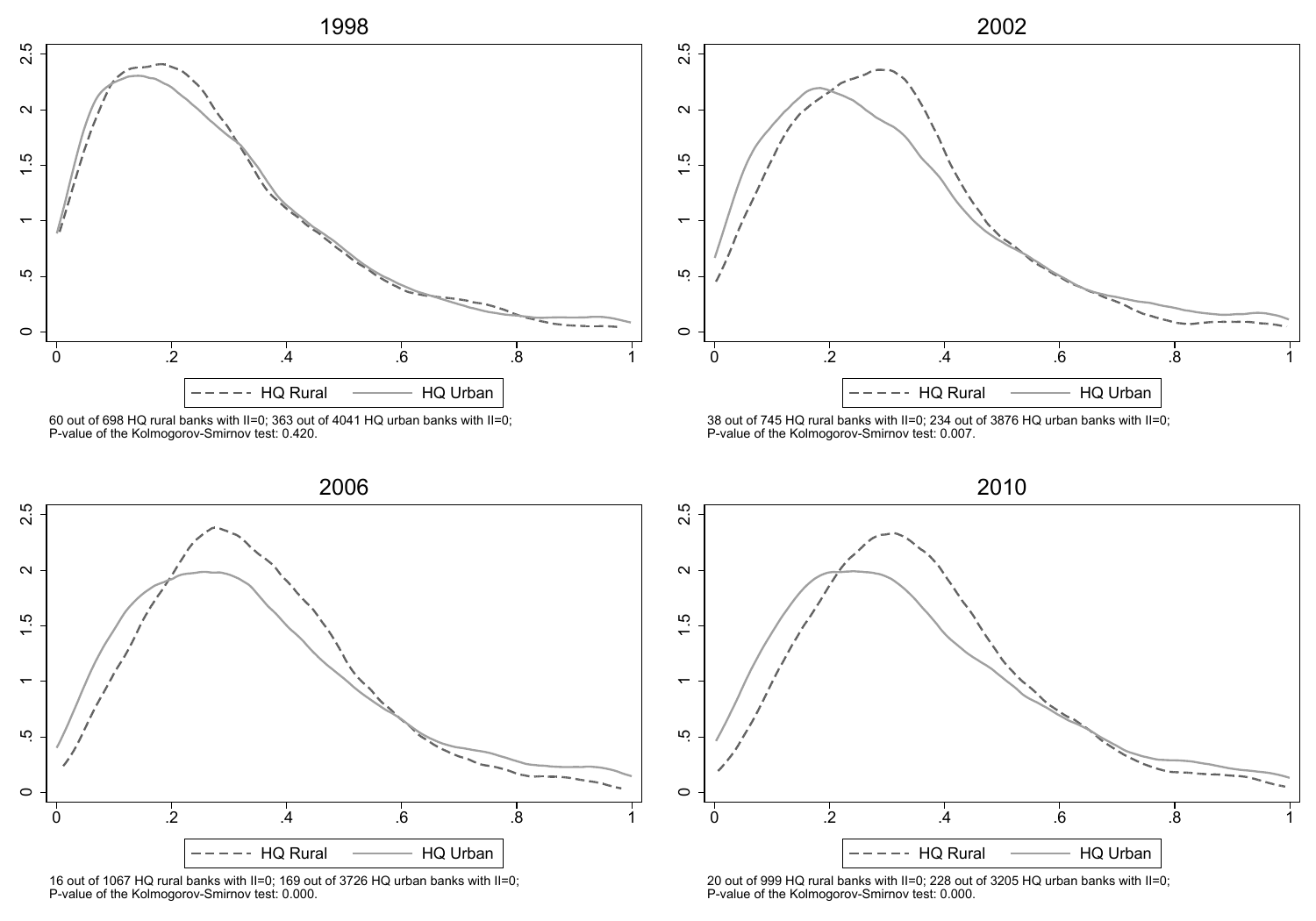}}\\
\bigskip
\bigskip
\subfloat{\includegraphics[width=0.96\linewidth]{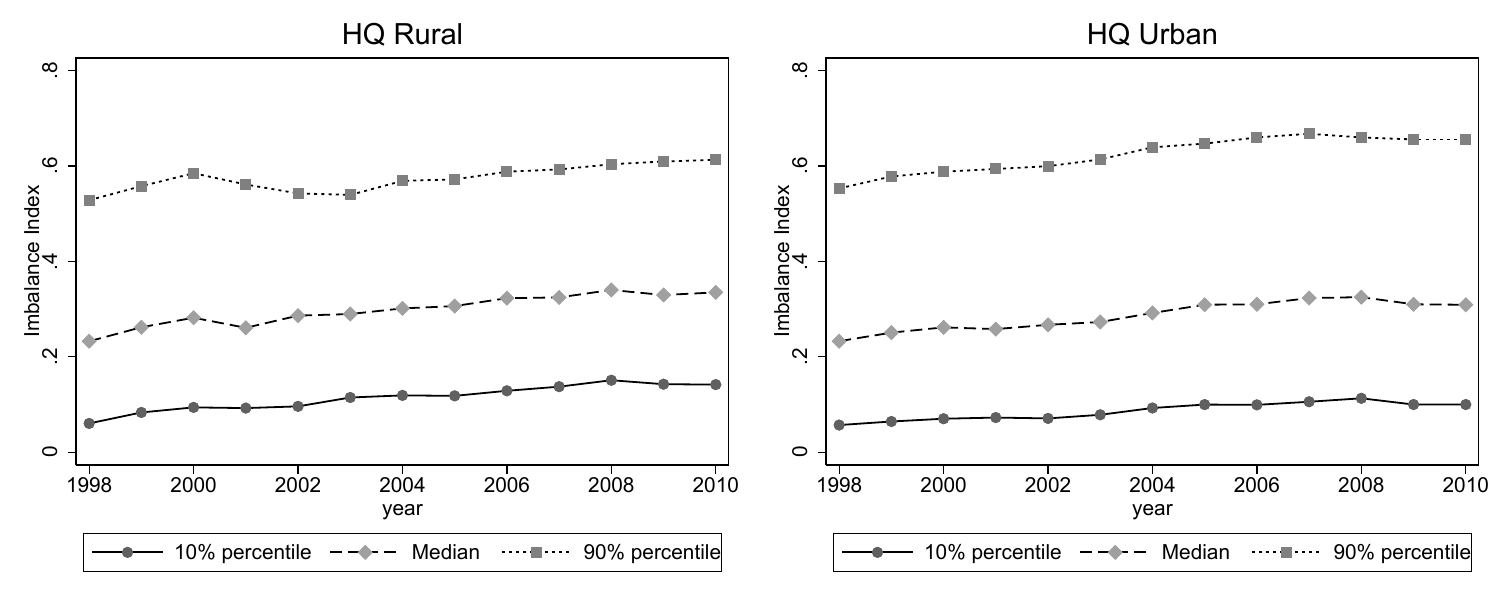}}
\footnotetext{A “rural county” is defined as rural population share is larger than 50\%, based on population census in 2010. Reference: 
https://www.census.gov/programs-surveys/geography/guidance/geo-areas/urban-rural/2010-urban-rural.html
}
\end{minipage}
\end{figure}

\begin{figure}[!h]
\caption{Breakdown of imbalance index by location -- North, South, East, West}\label{fig:breakdown_region}
\centering
\begin{minipage}{\linewidth}
\centering
\subfloat{\includegraphics[width=0.96\linewidth]{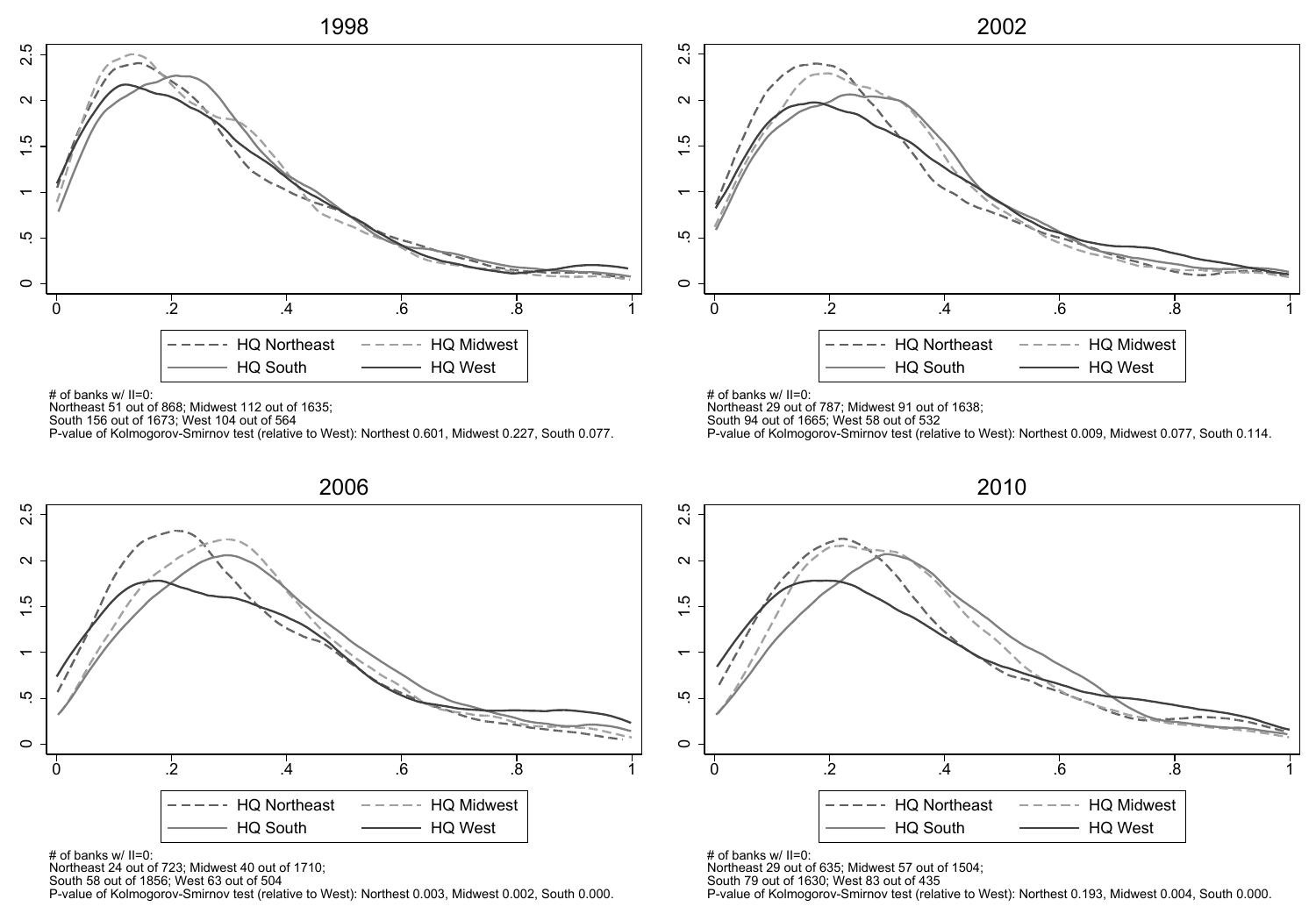}}\\
\bigskip
\bigskip
\subfloat{\includegraphics[width=0.96\linewidth]{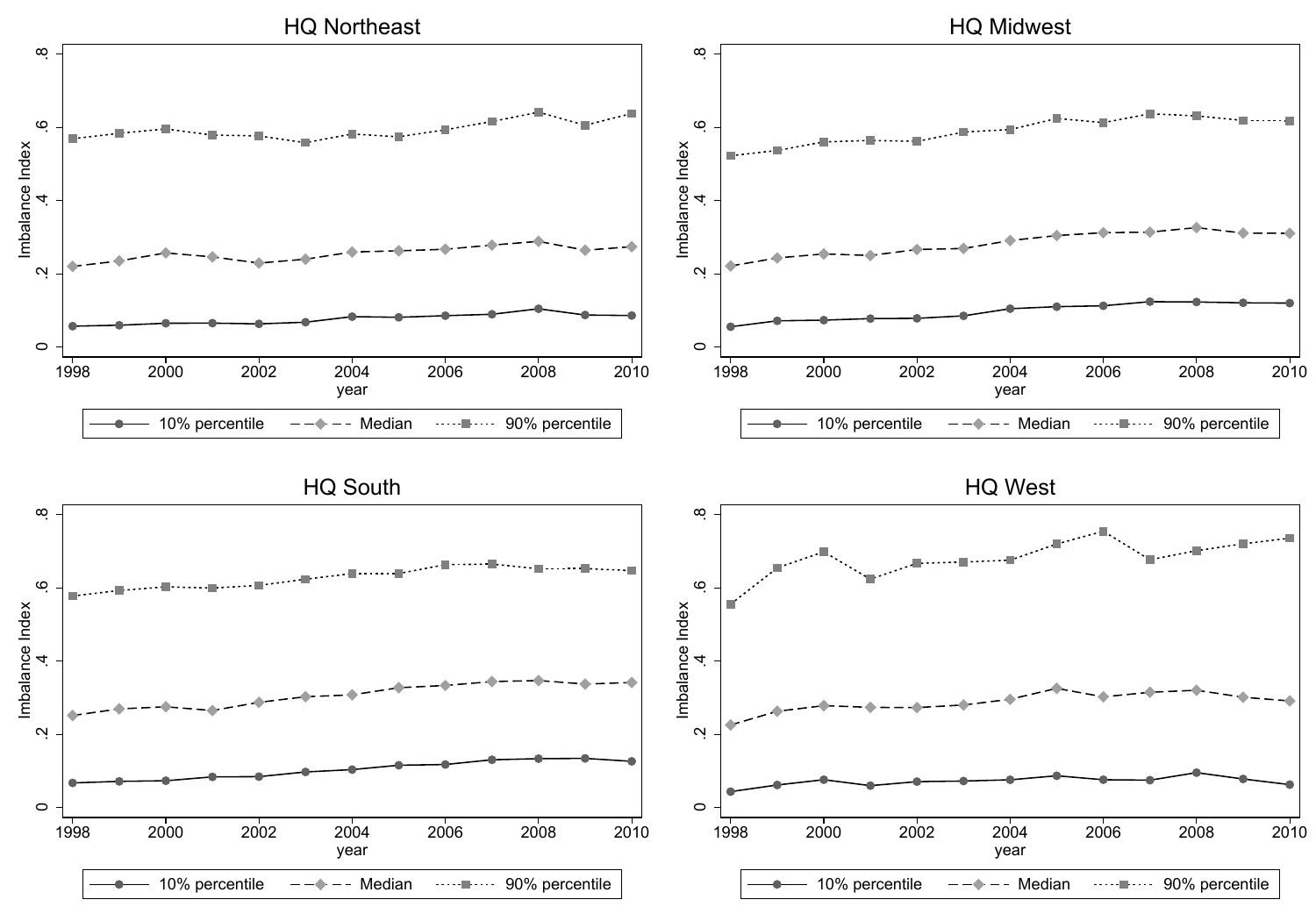}}

\end{minipage}
\end{figure}

\begin{figure}[!h]
\caption{Breakdown of imbalance index by HQ-county income level (cutoff = 1998 median income)}\label{fig:breakdown_wealth}
\centering
\begin{minipage}{\linewidth}
\centering
\subfloat{\includegraphics[width=0.96\linewidth]{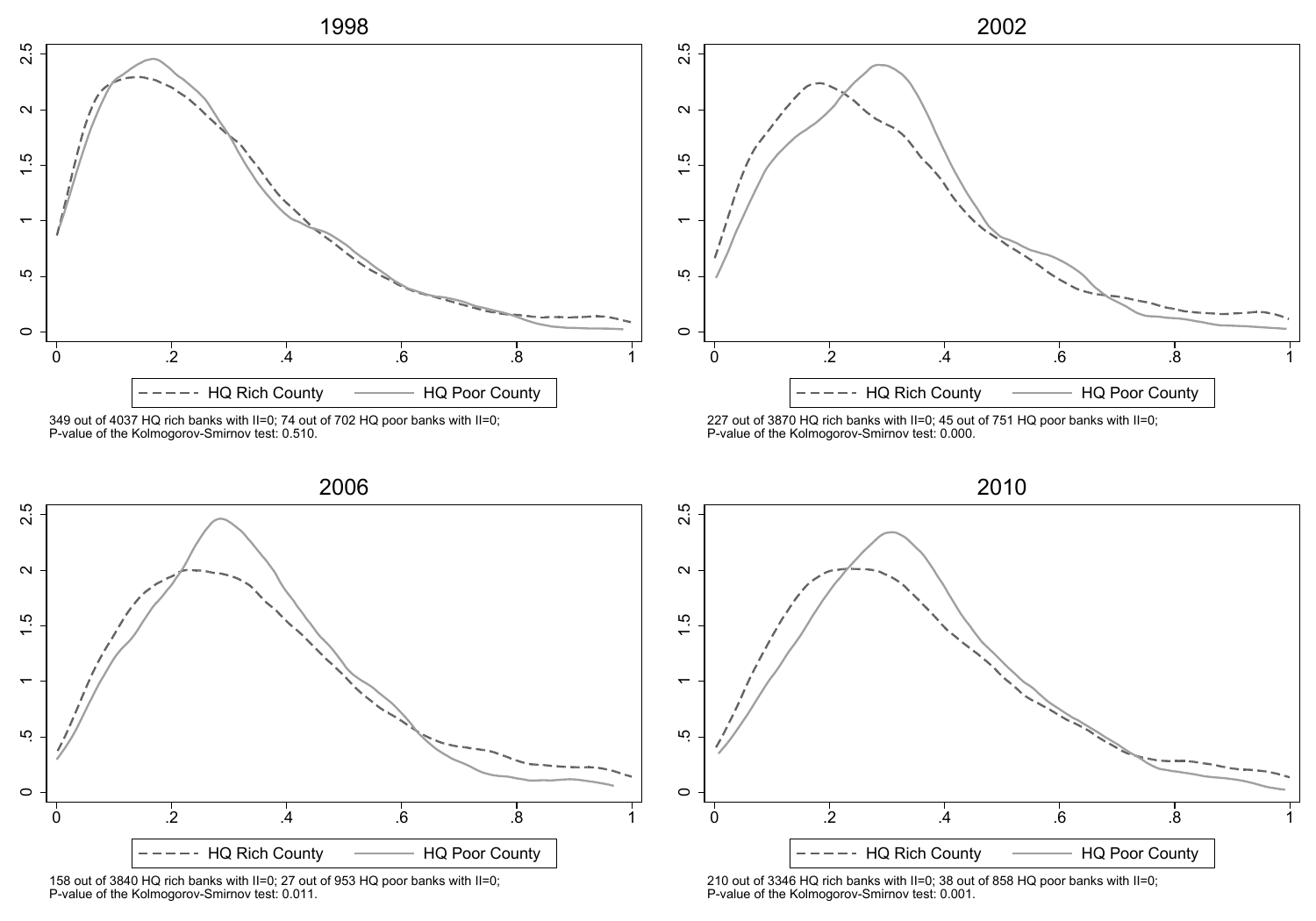}}\\
\bigskip
\bigskip
\subfloat{\includegraphics[width=0.96\linewidth]{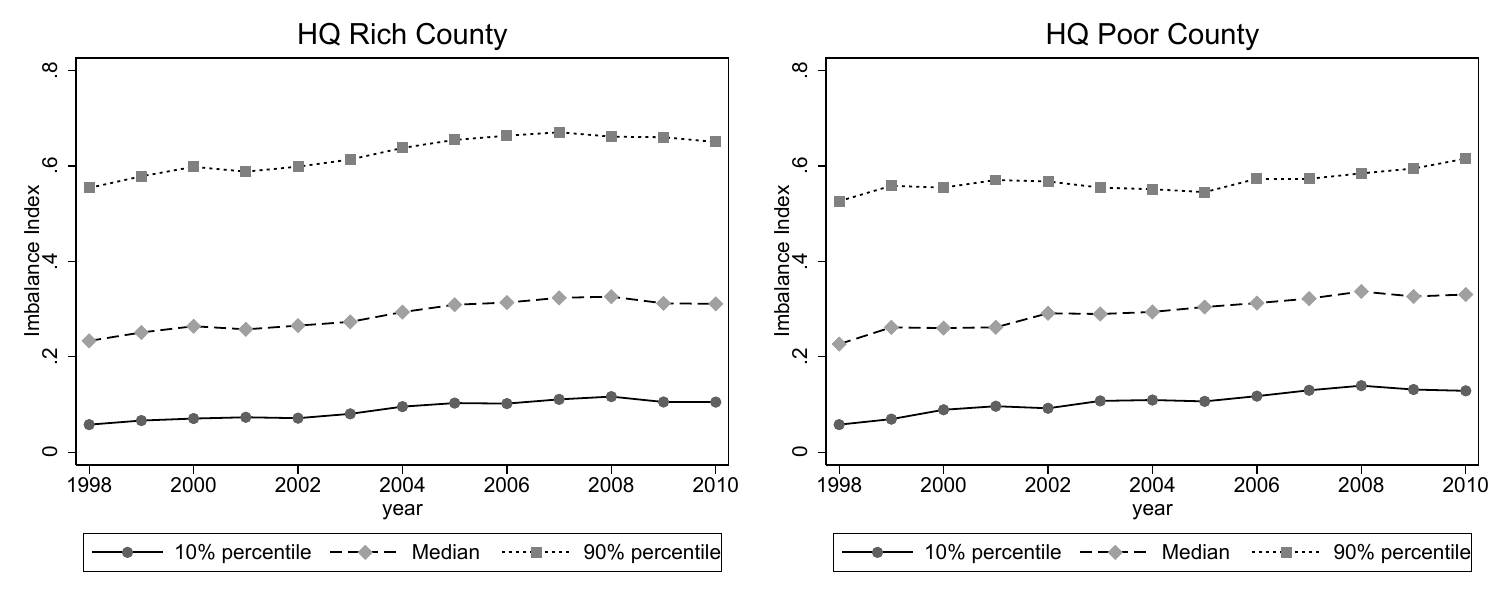}}

\end{minipage}
\end{figure}

\begin{figure}[!h]
\caption{Breakdown of imbalance index by HQ-county white or non-white (cutoff = 1998 median white\%)}\label{fig:breakdown_race}
\centering
\begin{minipage}{\linewidth}
\centering
\subfloat{\includegraphics[width=0.96\linewidth]{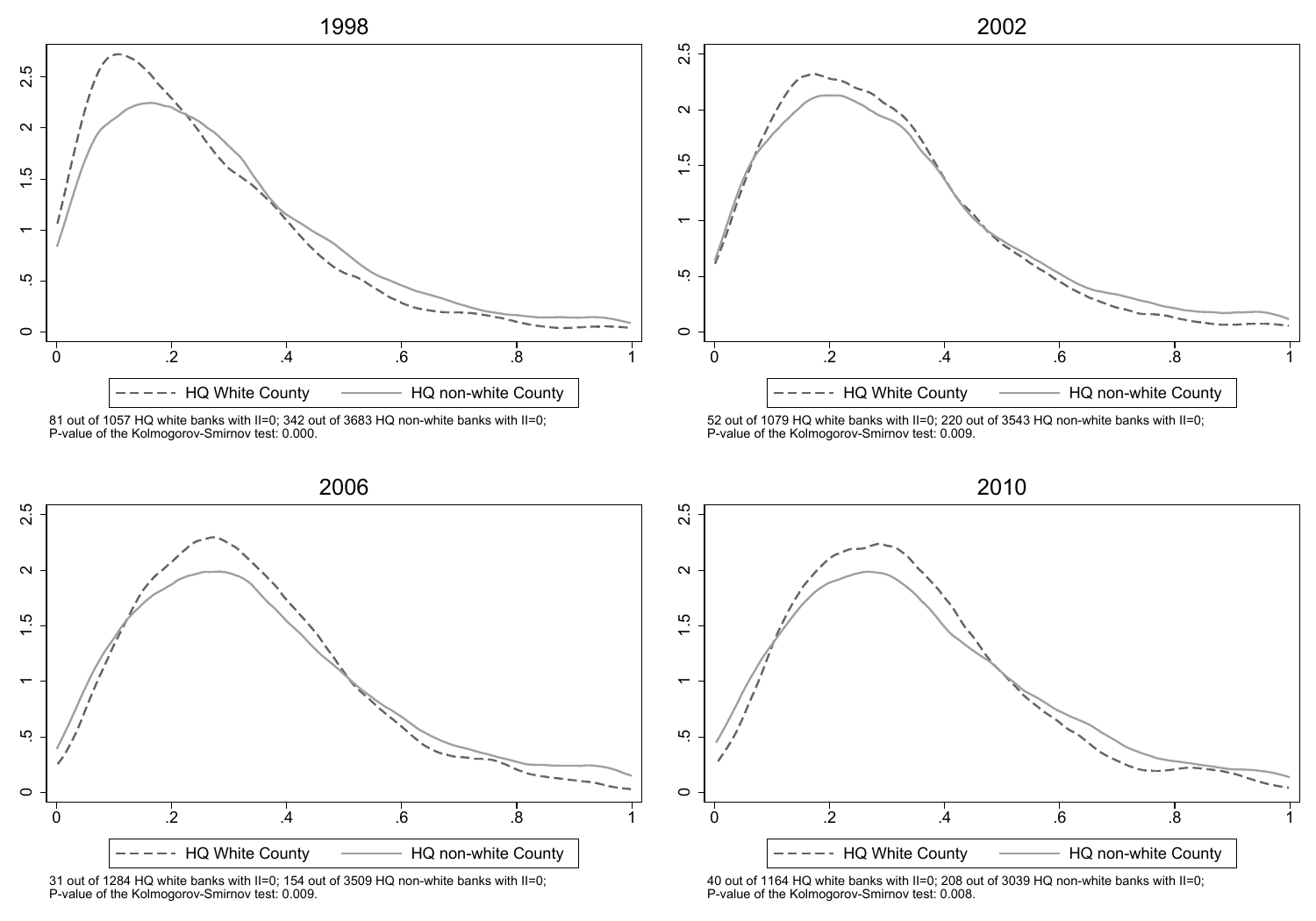}}\\
\bigskip
\bigskip
\subfloat{\includegraphics[width=0.96\linewidth]{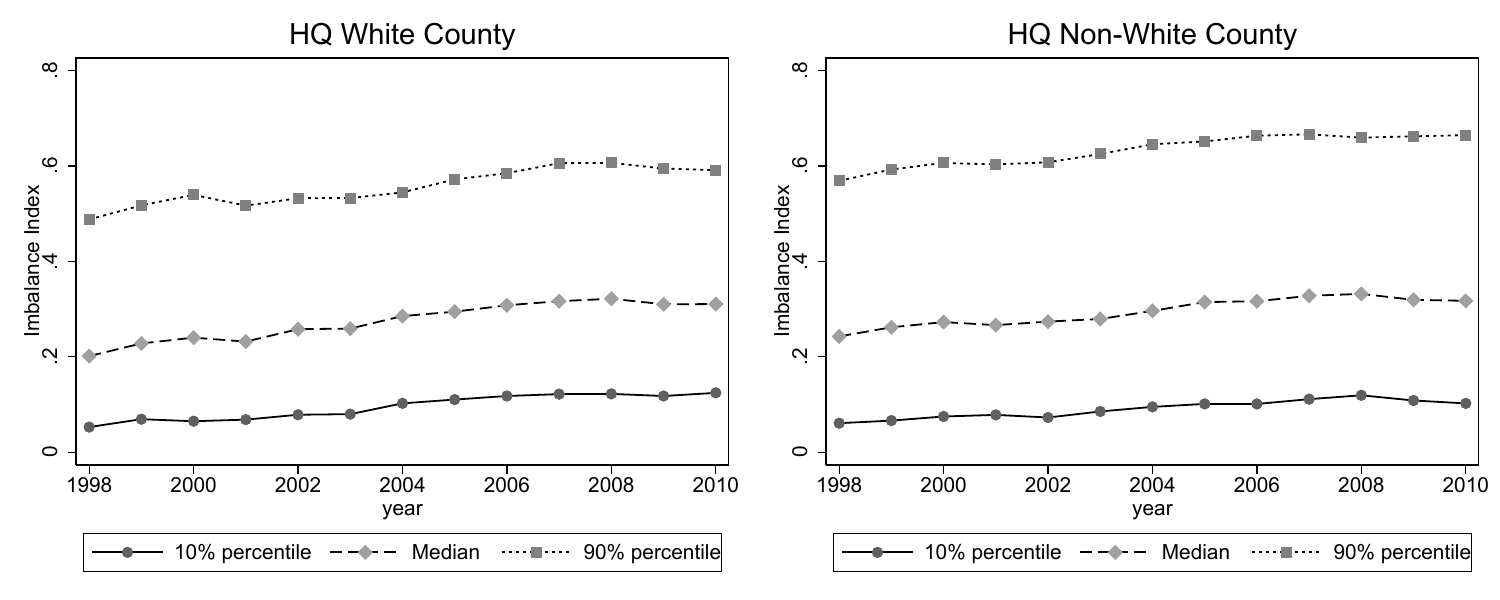}}

\end{minipage}
\end{figure}

\clearpage
\newpage

\section{Markup predictions and stylized facts about margins}\label{app:DSS}
\small

The predictions of our model are consistent with the stylized facts about margins and spreads presented in recent work by  \citeauthor{Drechsler_Savov_2017}  (DSS) in their 2017 QJE paper.

The main finding in DSS is that when the Fed funds rate rises,  deposit spreads rise and the amount of deposit declines, and that these effects are stronger in more concentrated markets. Using  our structural model we can measure market power in the local deposit market (i.e.,  the deposit spread (margin) at the bank-county-year level) by $1/(1-s_{jmt})$. Then, using this measure we can investigate how it changes when the Fed funds rate  rises, and how this varies depending on market concentration. 

Our findings are presented below in Table \ref{tab:DSS} and confirm these results. Market power in local deposit markets is positively correlated with the Fed funds rate (as measured by T-bills), and this correlations is stronger if the HHI of deposits in the local market is higher. 

\begin{table}[htbp]
\centering
\begin{threeparttable}
\footnotesize
    
    \caption{Market power in local deposit markets}\label{tab:DSS}
    \begin{tabular}{lllll}

    \hline
        DV: $1/(1-s_j)$ & (1) & (2) & (3) & (4) \\ \hline
        T-bill (log) & 0.002*** & 0.003*** & 0.016*** & ~ \\ 
         & (0.001) & (0.000) & (0.002) & ~ \\
        HHI deposit (log) & ~ & 0.318*** & 0.313*** & 0.197*** \\ 
        & ~ & (0.001) & (0.001) & (0.004) \\ 
        T-bill x HHI (log) & ~ & ~ & 0.008*** & 0.014*** \\ 
         & ~ & ~ & (0.001) & (0.001) \\ 
        Constant & 1.195*** & 1.714*** & 1.707*** & 1.538*** \\ 
        ~ & (0.001) & (0.002) & (0.002) &(0.006) \\\hline
        County FE & ~ & ~ & ~ & Y \\
        Bank-year FE & ~ & ~ & ~ & Y \\ 
        Observations & 225,513 & 225,513 & 225,513 & 197,620 \\ 
        R-squared & 0.000 & 0.249 & 0.249 & 0.522 \\ \hline
    \end{tabular}
    \begin{tablenotes}
        \scriptsize
        \item Note: Robust standard errors of serial correlation and heteroscedasticity are reported in parentheses. * means p-value < 0.05; ** means p-value < 0.01; *** means p-value < 0.001.
    \end{tablenotes}
  \label{tab:addlabel}%
  \end{threeparttable}
\end{table}%

\clearpage
\newpage

\section{Additional estimation results}\label{app:additional_estimation_results}

\begin{table}[htbp]
\scriptsize
  \begin{threeparttable}
  \centering
  \caption{Alternative specifications of the loan equation}\label{tab:Ql_gmm_robust}
    \begin{tabular}{lcccccccc}
    \toprule
    \multicolumn{1}{r}{\textbf{Variables}} & \multicolumn{4}{c}{OLS Fixed Effects} & \multicolumn{4}{c}{GMM DiD \& DiDiD} \\
    \midrule
    \textit{Number of branches} &       &       &       &       &       &       &       &  \\
    \multicolumn{1}{r}{Second branch} & \multicolumn{1}{r}{0.117***} & \multicolumn{1}{l}{(0.012)} & \multicolumn{1}{r}{0.123***} & \multicolumn{1}{l}{(0.013)} & \multicolumn{1}{l}{0.1036***} & \multicolumn{1}{l}{(0.0231)} & \multicolumn{1}{l}{0.1586***} & \multicolumn{1}{l}{(0.0231)} \\
    \multicolumn{1}{r}{Third branch} & \multicolumn{1}{r}{0.085***} & \multicolumn{1}{l}{(0.011)} & \multicolumn{1}{r}{0.078***} & \multicolumn{1}{l}{(0.013)} & \multicolumn{1}{l}{0.0675***} & \multicolumn{1}{l}{(0.0158)} & \multicolumn{1}{l}{0.0946***} & \multicolumn{1}{l}{(0.0161)} \\
    \multicolumn{1}{r}{Fourth branch} & \multicolumn{1}{r}{0.080***} & \multicolumn{1}{l}{(0.012)} & \multicolumn{1}{r}{0.062***} & \multicolumn{1}{l}{(0.014)} & \multicolumn{1}{l}{0.0666***} & \multicolumn{1}{l}{(0.0144)} & \multicolumn{1}{l}{0.0793***} & \multicolumn{1}{l}{(0.0152)} \\
    \multicolumn{1}{r}{Fifth branch} & \multicolumn{1}{r}{0.093***} & \multicolumn{1}{l}{(0.013)} & \multicolumn{1}{r}{0.092***} & \multicolumn{1}{l}{(0.015)} & \multicolumn{1}{l}{0.0785***} & \multicolumn{1}{l}{(0.0168)} & \multicolumn{1}{l}{0.0999***} & \multicolumn{1}{l}{(0.0174)} \\
    \multicolumn{1}{r}{> Fifth} & \multicolumn{1}{r}{0.008***} & \multicolumn{1}{l}{(0.002)} & \multicolumn{1}{r}{0.009***} & \multicolumn{1}{l}{(0.002)} & \multicolumn{1}{l}{0.0043**} & \multicolumn{1}{l}{(0.0020)} & \multicolumn{1}{l}{0.0079***} & \multicolumn{1}{l}{(0.0023)} \\
    \midrule
    \textit{Securitization} &       &       &       &       &       &       &       &  \\
    \multicolumn{1}{r}{\% of loan resold} & \multicolumn{1}{r}{0.158***} & \multicolumn{1}{l}{(0.011)} & \multicolumn{1}{r}{0.352***} & \multicolumn{1}{l}{(0.013)} & \multicolumn{1}{l}{0.2207***} & \multicolumn{1}{l}{(0.0244)} & \multicolumn{1}{l}{0.3480***} & \multicolumn{1}{l}{(0.0281)} \\
    \midrule
    \textit{Econ of scope, Qd, and Ql} &       &       &       &       &       &       &       &  \\
    \multicolumn{1}{r}{log own local deposit} & \multicolumn{1}{r}{0.210***} & \multicolumn{1}{l}{(0.008)} & \multicolumn{1}{r}{0.098***} & \multicolumn{1}{l}{(0.007)} & \multicolumn{1}{l}{0.3599***} & \multicolumn{1}{l}{(0.0319)} & \multicolumn{1}{l}{0.2094***} & \multicolumn{1}{l}{(0.0320)} \\
    \multicolumn{1}{r}{log own total deposit} & \multicolumn{1}{r}{-0.432***} & \multicolumn{1}{l}{(0.011)} & \multicolumn{1}{r}{} & \multicolumn{1}{l}{} & \multicolumn{1}{l}{-0.6726***} & \multicolumn{1}{l}{(0.0209)} &       &  \\
    \multicolumn{1}{r}{log own total loans} & \multicolumn{1}{r}{0.871***} & \multicolumn{1}{l}{(0.006)} & \multicolumn{1}{r}{} & \multicolumn{1}{l}{} & \multicolumn{1}{l}{0.7639***} & \multicolumn{1}{l}{(0.0159)} &       &  \\
    \multicolumn{1}{r}{log(total depo + total loan)} & \multicolumn{1}{r}{} & \multicolumn{1}{l}{} & \multicolumn{1}{r}{0.640***} & \multicolumn{1}{l}{(0.005)} &       &       & \multicolumn{1}{l}{0.2799***} & \multicolumn{1}{l}{(0.0092)} \\
    \midrule
    \textit{Market characteristics} & \multicolumn{1}{r}{} & \multicolumn{1}{l}{} & \multicolumn{1}{r}{} & \multicolumn{1}{l}{} &       &       &       &  \\
    \multicolumn{1}{r}{log county income} & \multicolumn{1}{r}{-0.025} & \multicolumn{1}{l}{(0.052)} & \multicolumn{1}{r}{0.036} & \multicolumn{1}{l}{(0.058)} &       &       &       &  \\
    \multicolumn{1}{r}{log county population} & \multicolumn{1}{r}{-0.778***} & \multicolumn{1}{l}{(0.068)} & \multicolumn{1}{r}{-0.936***} & \multicolumn{1}{l}{(0.077)} &       &       &       &  \\
    \multicolumn{1}{r}{share pop. age< 19} & \multicolumn{1}{r}{-4.381***} & \multicolumn{1}{l}{(0.700)} & \multicolumn{1}{r}{-3.519***} & \multicolumn{1}{l}{(0.788)} &       &       &       &  \\
    \multicolumn{1}{r}{share pop. age >50} & \multicolumn{1}{r}{-1.767***} & \multicolumn{1}{l}{(0.477)} & \multicolumn{1}{r}{-1.377***} & \multicolumn{1}{l}{(0.531)} &       &       &       &  \\
    \multicolumn{1}{r}{log house price index} & \multicolumn{1}{r}{0.350***} & \multicolumn{1}{l}{(0.031)} & \multicolumn{1}{r}{0.184***} & \multicolumn{1}{l}{(0.034)} &       &       &       &  \\
    \multicolumn{1}{r}{log nbr bankuptcy} & \multicolumn{1}{r}{-0.021***} & \multicolumn{1}{l}{(0.008)} & \multicolumn{1}{r}{-0.022***} & \multicolumn{1}{l}{(0.008)} &       &       &       &  \\
    \multicolumn{1}{r}{log nbr loan applications} & \multicolumn{1}{r}{0.448***} & \multicolumn{1}{l}{(0.013)} & \multicolumn{1}{r}{0.499***} & \multicolumn{1}{l}{(0.014)} &       &       &       &  \\
    \midrule
    \textit{Selection - Control function} &       &       &       &       &       &       &       &  \\
    \multicolumn{1}{r}{Propensity score} & \multicolumn{1}{r}{0.569} & \multicolumn{1}{l}{(0.429)} & \multicolumn{1}{r}{4.402***} & \multicolumn{1}{l}{(0.491)} & \multicolumn{1}{l}{-1.0385***} & \multicolumn{1}{l}{(0.3011)} & \multicolumn{1}{l}{0.3256} & \multicolumn{1}{l}{(0.3591)} \\
    \multicolumn{1}{r}{Propensity score square} & \multicolumn{1}{r}{0.069} & \multicolumn{1}{l}{(0.506)} & \multicolumn{1}{r}{-5.287***} & \multicolumn{1}{l}{(0.582)} & \multicolumn{1}{l}{1.4218***} & \multicolumn{1}{l}{(0.2901)} & \multicolumn{1}{l}{0.5093} & \multicolumn{1}{l}{(0.3689)} \\
    \multicolumn{1}{r}{Propensity score cubic} & \multicolumn{1}{r}{-0.223} & \multicolumn{1}{l}{(0.196)} & \multicolumn{1}{r}{1.626***} & \multicolumn{1}{l}{(0.227)} & \multicolumn{1}{l}{-0.4558***} & \multicolumn{1}{l}{(0.0903)} & \multicolumn{1}{l}{-0.3255***} & \multicolumn{1}{l}{(0.1209)} \\
    \midrule
    \textit{Fixed effects} &       &       &       &       &       &       &       &  \\
    \multicolumn{1}{r}{Bank x County} & \multicolumn{2}{c}{YES} & \multicolumn{2}{c}{YES} & \multicolumn{2}{c}{YES} & \multicolumn{2}{c}{YES} \\
    \multicolumn{1}{r}{Time} & \multicolumn{2}{c}{YES} & \multicolumn{2}{c}{YES} & \multicolumn{2}{c}{NO} & \multicolumn{2}{c}{NO} \\
    \multicolumn{1}{r}{Country x Time} & \multicolumn{2}{c}{NO} & \multicolumn{2}{c}{NO} & \multicolumn{2}{c}{YES} & \multicolumn{2}{c}{YES} \\
    \multicolumn{1}{r}{Bank x Time} & \multicolumn{2}{c}{NO} & \multicolumn{2}{c}{NO} & \multicolumn{2}{c}{YES} & \multicolumn{2}{c}{YES} \\
    \midrule
    Number of observations & \multicolumn{2}{c}{194,267} & \multicolumn{2}{c}{194,267} & \multicolumn{2}{c}{196,090} & \multicolumn{2}{c}{196,090} \\
    R-square & \multicolumn{2}{c}{0.926} & \multicolumn{2}{c}{0.915} & \multicolumn{2}{c}{} & \multicolumn{2}{c}{} \\
    \bottomrule
    \end{tabular}%
    \begin{tablenotes}
        \scriptsize
        \item Note: Robust standard errors of serial correlation and heteroscedasticity are reported in parentheses. * means p-value < 0.05; ** means p-value < 0.01; *** means p-value < 0.001.
    \end{tablenotes}
  \label{tab:addlabel}%
  \end{threeparttable}
\end{table}%

\newpage

Tables \ref{tab:alt_IV_deposits} and \ref{tab:alt_IV_loans}  present results demonstrating the robustness of our findings to different combinations of time series and spatial instruments. In addition, we have also considered specification in which we incorporate BLP-type instruments and lagged-by-3-period values into the set of instruments used in the main specifications. The main findings remain consistent.

\begin{table}[htbp]
  \centering
  \scriptsize
  \begin{threeparttable}
  \caption{Alternative set of instruments for EOS: Deposit equation} \label{tab:alt_IV_deposits}
    \begin{tabular}{llcccccc}
    \toprule
    \multicolumn{1}{r}{\textbf{Variables}} &       & \multicolumn{2}{c}{Spec. 1} & \multicolumn{2}{c}{Spec. 2} & \multicolumn{2}{c}{Spec. 3} \\
    \midrule
    \textit{Number of branches} &       &       &       &       &       &       &  \\
    \multicolumn{1}{r}{Second branch} &       & \multicolumn{1}{l}{0.6803***} & \multicolumn{1}{l}{(0.0089)} & \multicolumn{1}{l}{0.6792***} & \multicolumn{1}{l}{(0.0087)} & \multicolumn{1}{l}{0.6673***} & \multicolumn{1}{l}{(0.0090)} \\
    \multicolumn{1}{r}{Third branch} &       & \multicolumn{1}{l}{0.3801***} & \multicolumn{1}{l}{(0.0076)} & \multicolumn{1}{l}{0.3785***} & \multicolumn{1}{l}{(0.0074)} & \multicolumn{1}{l}{0.3722***} & \multicolumn{1}{l}{(0.0077)} \\
    \multicolumn{1}{r}{Fourth branch} &       & \multicolumn{1}{l}{0.2904***} & \multicolumn{1}{l}{(0.0079)} & \multicolumn{1}{l}{0.2889***} & \multicolumn{1}{l}{(0.0078)} & \multicolumn{1}{l}{0.2825***} & \multicolumn{1}{l}{(0.0081)} \\
    \multicolumn{1}{r}{Fifth branch} &       & \multicolumn{1}{l}{0.3645***} & \multicolumn{1}{l}{(0.0100)} & \multicolumn{1}{l}{0.3649***} & \multicolumn{1}{l}{(0.0097)} & \multicolumn{1}{l}{0.3525***} & \multicolumn{1}{l}{(0.0100)} \\
    \multicolumn{1}{r}{> Fifth} &       & \multicolumn{1}{l}{0.0335***} & \multicolumn{1}{l}{(0.0024)} & \multicolumn{1}{l}{0.0336***} & \multicolumn{1}{l}{(0.0023)} & \multicolumn{1}{l}{0.0310***} & \multicolumn{1}{l}{(0.0022)} \\
    \midrule
    \textit{Securitization} &       &       &       &       &       &       &  \\
    \multicolumn{1}{r}{\% of loan resold} &       & \multicolumn{1}{l}{-0.0354***} & \multicolumn{1}{l}{(0.0116)} & \multicolumn{1}{l}{-0.0448***} & \multicolumn{1}{l}{(0.0114)} & \multicolumn{1}{l}{-0.0573***} & \multicolumn{1}{l}{(0.0114)} \\
    \midrule
    \textit{Econ of scope and Qd} &       &       &       &       &       &       &  \\
    \multicolumn{1}{r}{log own local loans} &       & \multicolumn{1}{l}{0.0692***} & \multicolumn{1}{l}{(0.0102)} & \multicolumn{1}{l}{0.0762***} & \multicolumn{1}{l}{(0.0099)} & \multicolumn{1}{l}{0.1146***} & \multicolumn{1}{l}{(0.0101)} \\
    \multicolumn{1}{r}{log own total deposit} &       & \multicolumn{1}{l}{0.0452***} & \multicolumn{1}{l}{(0.0049)} & \multicolumn{1}{l}{0.0497***} & \multicolumn{1}{l}{(0.0048)} & \multicolumn{1}{l}{0.0378***} & \multicolumn{1}{l}{(0.0050)} \\
    \midrule
    Number of observations &       & \multicolumn{2}{c}{241,911} & \multicolumn{2}{c}{241,911} & \multicolumn{2}{c}{241,911} \\
    \bottomrule
    \end{tabular}%
    \begin{tablenotes}
        \scriptsize
        \item Note: All specifications include Bank x County, County x Time, and Bank x Time fixed effects. The instrumental variables used for the EOS in each specification are as follows. Spec.1, value of loans in the neighbouring counties, and value of loans in the neighbours of the neighbouring counties; Spec. 2, value of loans in the neighbouring counties, and value of loans in the neighbouring counties lagged by 2 periods; Spec. 3, value of loans in the neighbouring counties, and value of loans lagged by 2 periods. 
    \end{tablenotes}
  \end{threeparttable}
\end{table}%

\begin{table}[htbp]
  \centering
  \scriptsize
    \begin{threeparttable}
  \caption{Alternative set of instruments for EOS: Loan equation}\label{tab:alt_IV_loans}
    \begin{tabular}{llcccccc}
    \toprule
    \multicolumn{1}{r}{\textbf{Variables}} &       & \multicolumn{2}{c}{Spec. 1} & \multicolumn{2}{c}{Spec. 2} & \multicolumn{2}{c}{Spec. 3} \\
    \midrule
    \textit{Number of branches} &       &       &       &       &       &       &  \\
    \multicolumn{1}{r}{Second branch} &       & \multicolumn{1}{l}{0.0957***} & \multicolumn{1}{r}{(0.0298)} & \multicolumn{1}{l}{0.0945***} & \multicolumn{1}{r}{(0.0296)} & \multicolumn{1}{l}{0.1433**} & \multicolumn{1}{r}{(0.0618)} \\
    \multicolumn{1}{r}{Third branch} &       & \multicolumn{1}{l}{0.0619***} & \multicolumn{1}{r}{(0.0191)} & \multicolumn{1}{l}{0.0611***} & \multicolumn{1}{r}{(0.0190)} & \multicolumn{1}{l}{0.0879**} & \multicolumn{1}{r}{(0.0355)} \\
    \multicolumn{1}{r}{Fourth branch} &       & \multicolumn{1}{l}{0.0692***} & \multicolumn{1}{r}{(0.0167)} & \multicolumn{1}{l}{0.0694***} & \multicolumn{1}{r}{(0.0167)} & \multicolumn{1}{l}{0.0866***} & \multicolumn{1}{r}{(0.0285)} \\
    \multicolumn{1}{r}{Fifth branch} &       & \multicolumn{1}{l}{0.0899***} & \multicolumn{1}{r}{(0.0201)} & \multicolumn{1}{l}{0.0889***} & \multicolumn{1}{r}{(0.0200)} & \multicolumn{1}{l}{0.1133***} & \multicolumn{1}{r}{(0.0348)} \\
    \multicolumn{1}{r}{> Fifth} &       & \multicolumn{1}{l}{0.0092***} & \multicolumn{1}{r}{(0.0022)} & \multicolumn{1}{l}{0.0094***} & \multicolumn{1}{r}{(0.0022)} & \multicolumn{1}{l}{0.0105***} & \multicolumn{1}{r}{(0.0036)} \\
    \midrule
    \textit{Securitization} &       &       &       &       &       &       &  \\
    \multicolumn{1}{r}{\% of loan resold} &       & \multicolumn{1}{l}{0.6623***} & \multicolumn{1}{r}{(0.0304)} & \multicolumn{1}{l}{0.6629***} & \multicolumn{1}{r}{(0.0304)} & \multicolumn{1}{l}{0.6634***} & \multicolumn{1}{r}{(0.0307)} \\
    \midrule
    \textit{Econ of scope and Qd} &       &       &       &       &       &       &  \\
    \multicolumn{1}{r}{log own local deposit} &       & \multicolumn{1}{l}{0.3571***} & \multicolumn{1}{r}{(0.0437)} & \multicolumn{1}{l}{0.3614***} & \multicolumn{1}{r}{(0.0432)} & \multicolumn{1}{l}{0.2820***} & \multicolumn{1}{r}{(0.0907)} \\
    \multicolumn{1}{r}{log own total deposit} &       & \multicolumn{1}{l}{0.1672***} & \multicolumn{1}{r}{(0.0157)} & \multicolumn{1}{l}{0.1582***} & \multicolumn{1}{r}{(0.0119)} & \multicolumn{1}{l}{0.1737***} & \multicolumn{1}{r}{(0.0148)} \\
    \textit{Selection - Control function} &       &       &       &       &       &       &  \\
    \multicolumn{1}{r}{Propensity score} &       & \multicolumn{1}{l}{-0.6647**} & \multicolumn{1}{r}{(0.3373)} & \multicolumn{1}{l}{-0.6403*} & \multicolumn{1}{r}{(0.3441)} & \multicolumn{1}{l}{-0.6357*} & \multicolumn{1}{r}{(0.3553)} \\
    \multicolumn{1}{r}{Propensity score square} &       & \multicolumn{1}{l}{0.9122***} & \multicolumn{1}{r}{(0.3288)} & \multicolumn{1}{l}{0.8727***} & \multicolumn{1}{r}{(0.3330)} & \multicolumn{1}{l}{1.0109***} & \multicolumn{1}{r}{(0.3568)} \\
    \multicolumn{1}{r}{Propensity score cubic} &       & \multicolumn{1}{l}{-0.3001***} & \multicolumn{1}{r}{(0.1004)} & \multicolumn{1}{l}{-0.2863***} & \multicolumn{1}{r}{(0.1003)} & \multicolumn{1}{l}{-0.3630***} & \multicolumn{1}{r}{(0.1255)} \\
    \midrule
    Number of observations &       & \multicolumn{2}{c}{196,090} & \multicolumn{2}{c}{196,090} & \multicolumn{2}{c}{196,090} \\
    \bottomrule
    \end{tabular}%
    \begin{tablenotes}
    \scriptsize
        \item Note: All specifications include Bank x County, County x Time, and Bank x Time fixed effects. The instrumental variables used for the EOS in each specification are as follows. Spec.1, value of deposits lagged by 2 periods; Spec. 2, value of deposits lagged by 2 periods, and value of deposits in the neighours of the neighbouring counties; Spec.  3, number of branches lagged by 2 periods, and value of deposits in the neighbouring counties lagged by 2 periods. 
    \end{tablenotes}
  \end{threeparttable}
\end{table}%

\clearpage

\section{Description of an equilibrium and algorithm \label{appendix_equilibrium_algorithm}}
\footnotesize
Let $\mathbf{s}$ be the $2MJ \times 1$ vector of local market shares for deposits and loans for every bank and county. An equilibrium is a vector $\mathbf{s}$ that satisfies the following system of equations:
\begin{equation*}
    \left\{
\begin{array}{rrcl}
    (A) & s_{jm}^{d} & = &  0, 
    \text{ } \forall \text{ } (j,m), m \notin \mathcal{M}_{j}^{d} \\ 
    &  &  &  \\   
    (B) & s_{jm}^{\ell} & = &  0, 
    \text{ } \forall \text{ } (j,m), m \notin \mathcal{M}_{j}^{\ell} \\ 
    &  &  &  \\   
    (C) & s_{jm}^{d} & = & 
    \left(1 - \displaystyle \sum_{i=1}^{J} s_{im}^{d} \right)
    \exp \left\{ 
        e^{d}_{jm} - \dfrac{1}{1-s_{jm}^{d}} +
        \theta_{\ell}^{d} \text{ } 
        \ln(1 + H_{m}^{\ell} s_{jm}^{\ell}) + 
        \theta_{Q}^{d} \text{ } \ln Q_{j}^{d}    
        \right\},
    \text{ } \forall \text{ } (j,m), m \in \mathcal{M}_{j}^{d} \\ 
    &  &  &  \\   
    (D) & s_{jm}^{\ell} & = & 
    \left(1 - \displaystyle \sum_{i=1}^{J} s_{im}^{\ell} \right)
    \exp \left\{ 
        e^{\ell}_{jm} - \dfrac{1}{1-s_{jm}^{\ell}} + 
        \theta_{d}^{\ell} \text{ } 
        \ln(1 + H_{m}^{d} s_{jm}^{d}) +
        \theta_{Q}^{\ell} \text{ } \ln Q_{j}^{d}
        \right\},
    \text{ } \forall \text{ } (j,m), m \in \mathcal{M}_{j}^{\ell} \\ 
    &  &  &  \\   
    (E) & Q_{j}^{d} & = & 
    \displaystyle \sum_{j=1}^{J} s_{jm}^{d} \text{ } H_{m},
    \text{ } \forall \text{ } j
\end{array}
    \right.
\label{equilibrium mapping}
\end{equation*}
where $e^{d}_{jm}$ and $e^{\ell}_{jm}$ are exogenous terms that we define in the paper. For the description of the algorithms, it is convenient to represent this system of equations in the following compact form:
\begin{equation*}
    \left\{
\begin{array}{rrcl}
    (A') & \mathbf{s}_{m} & = &  
    f_{m} \left( \mathbf{s}_{m}, \mathbf{Q}^{d} \right), 
    \text{ } \forall \text{ } m = 1,2, ...M \\ 
    &  &  &  \\   
    (B') & \mathbf{Q}^{d} & = &  
    F \left( \mathbf{s} \right)
\end{array}
    \right.
\label{compact mapping}
\end{equation*}

The system of equations in (A') represents equations (A) to (D) for market $m$, and $f_{m}$ is a vector-valued function with $2J$ elements. The system of equations in (B') is simply the system in (E) in vector form, and $F$ is a vector-valued function with $J$ elements. 

The system of equations (A') possesses two properties that significantly ease the computation of an equilibrium. First, for a fixed value of the vector $\mathbf{Q}^{d}$, solving for the market shares in (A') is separable across counties. This means that the system of $2JM$ equations and unknowns can be divided into $M$ separate systems, each with a dimension of $2J$. Second, for fixed $\mathbf{Q}^{d}$ and deposit shares $s_{jm}^{d}$, there exists a unique vector of loan shares that solves the system (A'). Likewise, for fixed $\mathbf{Q}^{d}$ and loan shares $s_{jm}^{\ell}$, there is a unique vector of deposit shares that solves the system (A').

Taking these properties into account, our algorithm for computing an equilibrium follows these steps:
\begin{itemize}
    \item [1.] Initialization. We start by initializing the vector \(\mathbf{Q}^{d}\). Typically, in most counterfactual experiments, this initial value is set to the banks' total deposits observed in the data. However, if the counterfactuals involve changes to branch networks, we adjust this initial value accordingly.

    \item [2.] Iterative Process. During each iteration, we execute the following four steps sequentially:
    \begin{itemize}
        \item Step 1. Compute the unique equilibrium for the vector of loan shares $s_{jm}^{\ell}$ in each county, given $Q_{j}^{d}$ and local deposit shares $s_{jm}^{d}$ for every $j$. This unique local equilibrium is computed using a Bisection algorithm.

        \item Step 2. Compute the unique equilibrium of deposit shares $s_{jm}^{d}$, given $Q_{j}^{d}$ and the local loan shares $s_{jm}^{\ell}$ computed in Step 1. We again use a Bisection algorithm for this computation.
    
        \item Step 3. Using the market shares from Step 2, aggregate over counties to obtain $Q_{j}^{d}$ for each $j$, thereby updating the vector $\mathbf{Q}^{d}$.

        \item Step 4. Check for convergence by calculating the distance between the values of the vector $\mathbf{Q}^{d}$ at the beginning and the end of the iteration.
\end{itemize}
\end{itemize}
This process can be repeated until the algorithm converges.

\clearpage
\newpage

\section{Additional Counterfactual results}\label{sec:CF_results_app}

\subsection{Counterfactual National Imbalance Index}

	\begin{figure}[!h]
\caption{Evolution of the National Imbalance Index -- Counterfactual scenarios}\label{fig:II_DF_vs_DS}
\centering
\begin{minipage}{\linewidth}
\centering
\includegraphics[width=0.99\linewidth]{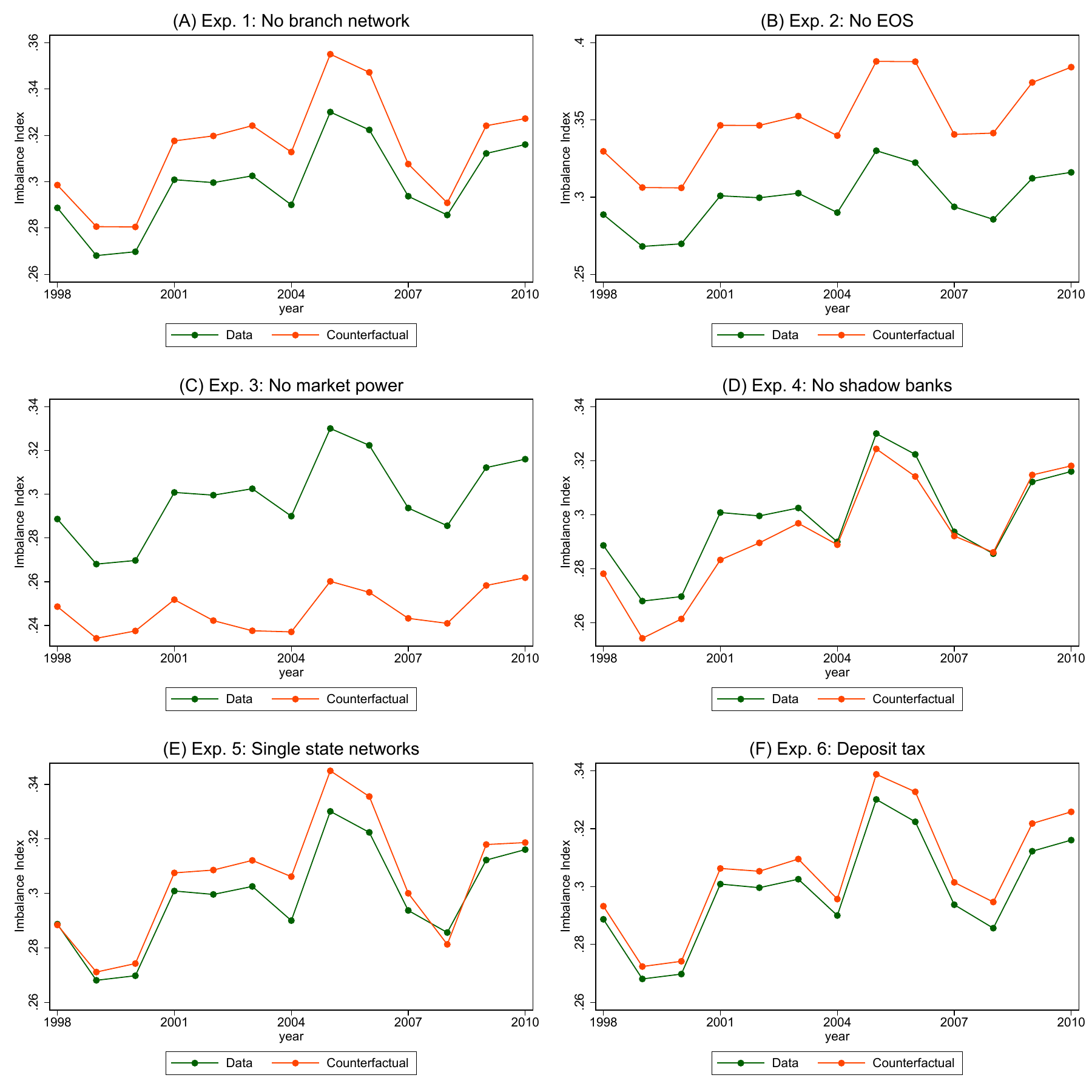}

\end{minipage}
\end{figure}

\subsection{Validation of the Riegle-Neal Counterfactual}\label{app:RN_test}

Here we test our Riegle-Neal counterfactual against the data. To do so we adopt the following strategy. We look for mergers involving two single-state banks.\footnote{That is, banks operating branches in only one state although they may make loans in multiple states.} Specifically, we are interested in cases where in period $t$ bank $A$ operates only in state 1 and acquires bank $B$ that operates only in state 2, such that in period $t+1$, the survivor $A'$  operates in states 1 and 2. There are 71 such cases throughout our sample period.

We then run our Counterfactual in which we effectively undo Riegle-Neal to see how outcomes change as we split $A'$ into separate single-state banks $A$ and $B$ in $t+1$. 

The first rows of Panels A and B of Table \ref{tab:RN_test} present the Imbalance Index for the counterfactual unmerged survivor and non-survivor banks respectively.  

Next, following the referee's suggestion we compare these counterfactual separate entities from $t+1$ to the data for $A$ and $B$ in $t$.\footnote{The referee had suggested using data from before 1998, but in fact any merger that takes place across state lines is the result of Riegle-Neal and so we do not need to go back to pre-1998. This is useful, since not many banks had actually taken advantage of the expansion opportunities afforded them by Riegle-Neal by this time (see \citeauthor{Aguirregabiria_Clark_2016} \citeyear{Aguirregabiria_Clark_2016}).} The Imbalance Index scores for $A$ and $B$ in period $t$ are presented in the second rows of Panels A and B of Table \ref{tab:RN_test}, respectively.  

Finally, the third rows of Panels A and B present statistics on the differences between the counterfactual Imbalance Index in $t+1$ and the Imbalance Index constructed using data from $t$, for survivor and non-survivor banks respectively.

Importantly, we would not expect the counterfactual outcomes for $A$ and $B$ from $t+1$  to be identical to data outcomes for $A$ and $B$ from $t$, since naturally there could have been some other changes in the market. However, should outcomes be similar, then this would validate our approach.\footnote{The reason we look at expansion through mergers, instead of expansion through denovo entry is that, for the latter cases, there is no bank entity in state 2 in year t. By contrast, for expansion through mergers, for survivors, we can compare their CF II in t+1 in state 1 with their real II in t (Panel A of Table \ref{tab:RN_test}); and for non-survivor, we can compare the survivor's CF II in t+1 in state 2 with their real II in t (Panel B of Table \ref{tab:RN_test}).} 

Results:

1. For both survivor and non-survivor banks, the mean Imbalance Index score of the year $t+1$ counterfactual  is very close to the mean Imbalance Index score of in the data in year $t$, as are their 25, 50, and 75 percentiles. 

2. The difference between these two measures (the third row of each panel) is also very small at the mean, and the corresponding percentiles. 

Overall, we believe that this test validates our Riegle-Neal counterfactual and allows us to learn about the importance of the ability to expand across states for the flow of credit. 

\begin{table}[htbp]
  \centering
  \caption{Test of Riegle-Neal Counterfactual: Mergers between Two Single-state Banks} \label{tab:RN_test}
    \begin{tabular}{llrrrrr}
    \toprule
    \toprule
          & \# obv & Mean  & Std   & p25   & p50   & p75 \\
    \midrule
    \textit{Panel A: Survivor bank} &       &       &       &       &       &  \\
    \quad CF II in t+1 & 71    & 0.304  & 0.170  & 0.180  & 0.292  & 0.403  \\
    \quad Real II in t & 71    & 0.314  & 0.159  & 0.174  & 0.309  & 0.427  \\
    \quad $CF_{t+1} - Real_t$ & 71    & -0.009  & 0.102  & -0.046  & 0.001  & 0.042  \\
    \quad $Diff/Real_t$ & 71    & 3.4\% & 51.7\% & -18.8\% & 0.7\% & 16.2\% \\
          &       &       &       &       &       &  \\
    \textit{Panel B: Non-survivor bank} &       &       &       &       &       &  \\
    \quad CF II in t+1 & 71    & 0.302  & 0.221  & 0.124  & 0.300  & 0.471  \\
    \quad Real II in t & 71    & 0.302  & 0.226  & 0.138  & 0.272  & 0.512  \\
    \quad $CF_{t+1} - Real_t$ & 71    & 0.000  & 0.223  & -0.129  & -0.020  & 0.129  \\
    \quad $Diff/Real_t$ & 71    & 68.3\% & 370.3\% & -41.6\% & -10.9\% & 33.2\% \\
    \bottomrule
    \end{tabular}%
  \label{tab:addlabel}
\end{table}

\clearpage
\newpage

\baselineskip14pt

\bibliographystyle{econometrica}

\bibliography{references}

\end{document}